\documentclass[preprint,authoryear,12pt]{elsarticle}

\usepackage{amsmath}
\usepackage{subfigure}
\usepackage{graphicx}

\begin{document}
\title{
Vascular phyllotaxis transition and 
an evolutionary mechanism of phyllotaxis
}

\author{Takuya Okabe}
\address{
Faculty of Engineering, Shizuoka University, 3-5-1 Johoku, 
Hamamatsu 432-8561,Japan}
\ead{ttokabe@ipc.shizuoka.ac.jp}

\begin{abstract}

Leaves of vascular plants are arranged regularly around stems,  
a phenomenon known as phyllotaxis. 
A constant angle between two successive leaves is called divergence angle. 
On the one side, 
the divergence angle $\alpha_0$ 
of an initial pattern of leaf primordia at a shoot apex is 
most commonly
 an irrational number of about 137.5 degrees, called limit divergence. 
On the other side, 
the divergence $\alpha$
of a final pattern of leaf traces in the vascular system 
of a mature stem is expressed 
in terms of a sequence of 
rational numbers, 
$\frac{1}{2}$, $\frac{1}{3}$, $\frac{2}{5}$, $\frac{3}{8}$,
 $\frac{5}{13}$,  $\frac{8}{21}$, 
 called phyllotactic fractions. 
The mathematical relationship between 
 the initial divergence $\alpha_0$, the final divergence $\alpha$, 
 and the number of internodes traversed by the leaf traces $n_c$ is
 investigated by means of a theoretical model of vascular phyllotaxis. 
It is shown that continuous changes of the trace length $n_c$ 
induce transitions between the fractional orders 
in  the vascular structure.  
The vascular phyllotaxis transition suggests 
an evolutionary mechanism 
for the phenomenon of phyllotaxis.    
To provide supporting evidence for the model and mechanism, 
available experimental results 
for fossil remains of {\it Lepidodendron} and the vascular structure of
 {\it Linum} and {\it Populus} are analyzed with the model. 
%


\end{abstract}


\maketitle

%
%
%
%

\section{Introduction}

\subsection{Review, background, and motivation}

Astonishing regularity 
manifested in plant architecture 
has fascinated various fields of scientists for centuries.
The regular arrangement of leaves, flowers and floral organs of higher plants is called phyllotaxis.
A constant angle of rotation between two successive organs is called divergence angle, 
on which two apparently irreconcilable concepts have been in general use
since the inception of quantitative investigations on phyllotaxis. 
%

%
%
%
%
%
%

%

\begin{figure}
  \begin{center}
  \subfigure[]{
\includegraphics[width= .5\textwidth]{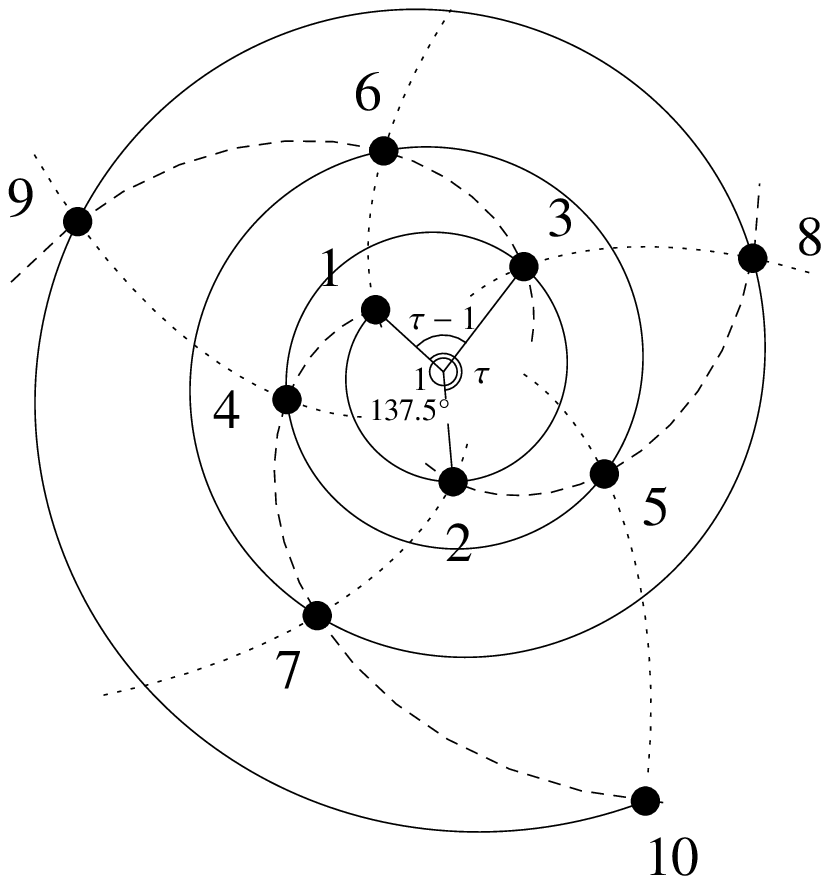}  
    \label{leftfig0}
 }
  \hfill
  \subfigure[]{
\includegraphics[width= .3\textwidth]{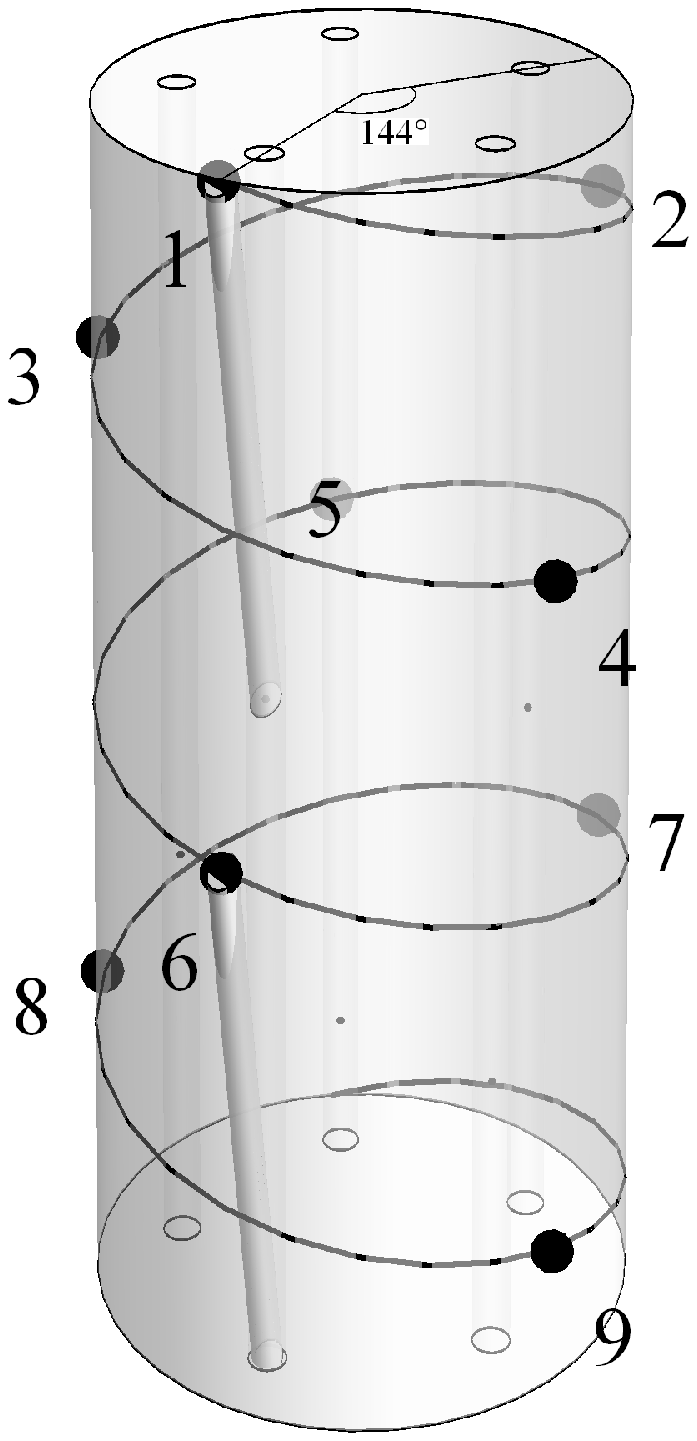}  
    \label{rightfig0}
  }
  \end{center}
  \vspace{-0.5cm}
  \caption{
(a) 
A typical pattern of leaf primordia (points) on a shoot apex
 with the initial divergence of $\alpha_0=1/(1+\tau)$, 
or 
$360 \alpha_0 \simeq 137.5$ in degrees.  
The irrational number $\tau\simeq 1.618$ is 
the golden ratio defined by the proportion equation $1:\tau=\tau-1:1$. 
The primordia are numbered in the reverse order of production. 
A solid spiral connecting all the primordia in the numerical order 
is the genetic spiral.
Three dashed spirals (clockwise inward) and five dotted spirals
 (counterclockwise inward) 
are 3 and 5 parastichies, respectively. 
This pattern has 
a parastichy pair $(3, 5)$. 
(b) 
A typical pattern of leaves on a mature stem 
characterized with a divergence fraction of 
 $\alpha=\frac{2}{5}$ ($ 360 \alpha=144 ^\circ$). 
Oblique strands diverging to 
leaves 1 and 6 are leaf traces.
A solid spiral surrounding the stem is the genetic spiral.
}
  \label{apexstemp}
\end{figure}
\cite{braun31,braun1835} and \cite{schimper1835beschreibung}
noticed that divergence angle is various but not arbitrary. 
It is 
a fraction, or a {\it rational number}, 
a number that can be expressed as the quotient 
$\frac{n}{m}$ of two integers $n$ and $m$. 
The most widespread is the helical phyllotaxis, also called spiral or alternate phyllotaxis, 
in which stems bear a leaf per node. 
In the helical phyllotaxis, 
the numerator $n$ and denominator $m$ of the fraction 
normally are 
two alternate terms of a Fibonacci sequence, 1, 
2, 3, 5, 8, 13, 21, 34, 55, 89, $\cdots$. 
It is generated by the Fibonacci recurrence relation that each number after the first two terms 
is the sum of the previous two numbers. 
The phyllotactic fractions 
$\frac{1}{2}$, $\frac{1}{3}$, $\frac{2}{5}$, $\frac{3}{8}$, $\frac{5}{13}$,
$\frac{8}{21}$, $\frac{13}{34}$,  $\cdots$
comprise what is called the main sequence of phyllotaxis.  
A $\frac{2}{5}$ phyllotaxis
 is schematically shown in Fig.~\ref{rightfig0}. 
In multijugate, 
verticillate or whorled phyllotaxis, where more than two leaves are
borne at each node, 
the divergence angle is divided by the number of leaves in a whorl. 
In general, a plant stem is partitioned into nodes and internodes. 
A node is a point at which a leaf or leaves are attached, 
and an internode is a section of the stem between two successive nodes. 
In the fractional phyllotaxis, there are leaves 
aligned vertically above 
each other along a stem, as represented by leaves 1 and 6 in Fig.~\ref{rightfig0}. 
A straight line connecting the superposed leaves 
is called an orthostichy. 
In the helical phyllotaxis, the denominator of the phyllotactic fraction
is equal to the number of orthostichies. 
It is also the number of
internodes between two adjacent leaves on an orthostichy.  
Thus, the $\frac{2}{5}$ phyllotaxis in Fig.~\ref{rightfig0} has
five orthostichies, 1-6, 2-7, 3-8, 4-9 and 5-10, and 
five internodes separate leaves on each orthostichy. 
%
An imaginary spiral connecting all the leaves in the order of production 
is called the genetic, fundamental, generative, or ontogenetic spiral.
The numerator of the fraction refers to the number of turns 
of the genetic spiral 
between the two adjacent leaves on an orthostichy.
In Fig.~\ref{rightfig0}, a solid spiral is the genetic spiral. 
From the leaf 6 to 1,  
the genetic spiral winds around the stem twice,  
the number two being the numerator of $\frac{2}{5}$. 
%
As remarked below, the phyllotactic fraction does not lose its
significance 
even though vertical alignment is actually not exact but approximate. 

In contrast, 
\cite{bravais1837}
suggested that divergence angle is uniquely and invariably 
given by an {\it irrational number}, that is, 
a number which cannot be expressed as a fraction.
The most typical angle of $360/(1+\tau)$ degrees is called the golden angle, where 
the irrational number $\tau$, 
called the golden ratio, golden mean, golden section, or extreme and mean ratio,  
is defined by the proportional relation $1:\tau=\tau-1:1$. 
%
%
%
%
As the positive solution of the quadratic equation $\tau(\tau-1)=1$, 
it is given by 
\begin{equation}
 \tau=\frac{\sqrt{5}+1}{2} \simeq 1.61803399 \cdots. 
\label{tau}
\end{equation}
The defining equation is transformed to $\tau^{-1}=1/(1+\tau^{-1})$.  
Recursive substitution of $\tau^{-1}$ in the left-hand side 
to the right-hand side gives
an infinite continued fraction
representation, 
\[
\tau^{-1}= \dfrac{1}{1+\dfrac{1}{1+\dfrac{1}{1+\ddots}}}, 
\qquad
\tau=1+\tau^{-1}. 
\]
The golden angle $360/(1+\tau)=360/\tau^2$ is approximately $137.50776$
degrees. 
By definition, 
the golden angle is the smaller angle created by sectioning the
circumference of a circle (360 degrees)
according to the golden ratio $1: \tau$,  the golden section. 
A phyllotactic pattern with divergence equal to the golden angle is
shown in Fig.~\ref{leftfig0}. 
The ratio of the angle subtended by 1 and 2 to the 
angle between
1 and 3 is $\tau$, 
or $\angle 1O2: \angle 1O3 =1: \tau-1= \tau: 1$, where $O$ is the origin. 
Similarly, 
$\angle 1O4: \angle 2O4 = \angle 1O9: \angle 4O9 =1: \tau$, and so on. 
Thus,  phyllotactic patterns with constant divergence equal to the golden angle have harmonious proportions. 
Patterns with an irrational divergence angle have 
no orthostichy in a strict sense, as no two leaves align vertically or radially. 
Instead, therefore, 
attention is directed to secondary spirals connecting positionally nearby leaves,  called parastichies.  
Like an orthostichy, 
a parastichy is characterized by a difference in number of leaves on it. 
In Fig.~\ref{leftfig0}, 
the genetic spiral, 
three parastichies 
and five parastichies are drawn with a solid curve, 
dashed curves and dotted curves, respectively. 
Each of three parastichies 1-4-7-10, 2-5-8 and 3-6-9, is called a 3-parastichy.  
Hence there are three 3-parastichies and five 5-parastichies in
Fig.~\ref{leftfig0}, and the pattern in Fig.~\ref{leftfig0} is said to have a parastichy pair
of $(3,5)$, which is also denoted as $(3+5)$ or $3: 5$. 
As a remarkable fact, 
parastichy numbers are almost always given by Fibonacci numbers.  
This is a mathematical consequence of the fact that
divergence angle is 
almost always the special irrational number, the golden angle. 
The golden angle is also called the Fibonacci angle, for 
it is the limit angle of divergence for 
the phyllotactic fractions
belonging to the main sequence;  
\[
\begin{array}{c}
\textstyle
360 \times \frac{2}{5} =144,\ \ \ \\
360\times \frac{3}{8}=135,\ \ \ \\
360\times \frac{5}{13}\simeq 138.46,  \\
360\times \frac{8}{21} \simeq 137.14.
\end{array}
\]
%
The rational angles beyond $\frac{5}{13}$
are practically indistinguishable
from 
the `ideal' irrational angle of 
137.507764$\cdots$ degrees. 
Therefore, it is argued 
that 
what appear to be different rational angles are 
nothing but 
a single irrational angle disturbed by 
inevitable random errors. 
%



The seemingly conflicting views on the divergence angle,
whether rational numbers or an irrational number, 
are a source of inspiration and confusion. 
%
%
%
%
In effect, they are not only compatible but both indispensable. 
On the one hand, 
the irrational number applies to 
the divergence angle of phyllotactic patterns of undifferentiated tissues at shoot tips or
 apical meristems (\cite{church04,hirmer22,hirmer31}). 
Let us call it the initial divergence angle. 
It is commonly referred to as the ideal or limit divergence angle 
for the reason mentioned above. 
%
On the other hand,  the rational (fractional) divergence 
applies to phyllotaxis of leaves, or primary vascular architecture 
on a mature stem  (\cite{lestiboudois, naegeli58}). 
%
%
In the literature, 
the majority of studies discuss 
the former, i.e., the process of organ initiation, 
positioning of the leaf primordia from which leaves will develop, 
and transitions of patterns at the shoot apical meristem. 
In recent years, substantial  
progress has been made 
in understanding plant hormonal factors 
that influence or control the formation of leaf primordia and their arrangement 
on the apical meristem (\cite{reinhardt05,kuhlemier07}). 
In striking contrast, the fractional phyllotaxis of the mature stem
have received less scholarly attention, unfortunately.
This is not because the latter is less important than the former.  
As a matter of fact, 
experimental findings on the close relationship between phyllotactic fraction and vascular
organization have been accumulated without being theorized from a general perspective 
(\cite{sterling45b,girolami53,jensen68, nb68, larson77,bsr82,kirchoff84}). 

%

\begin{figure}
  \begin{center}
\includegraphics[width= .7\textwidth,angle=90]{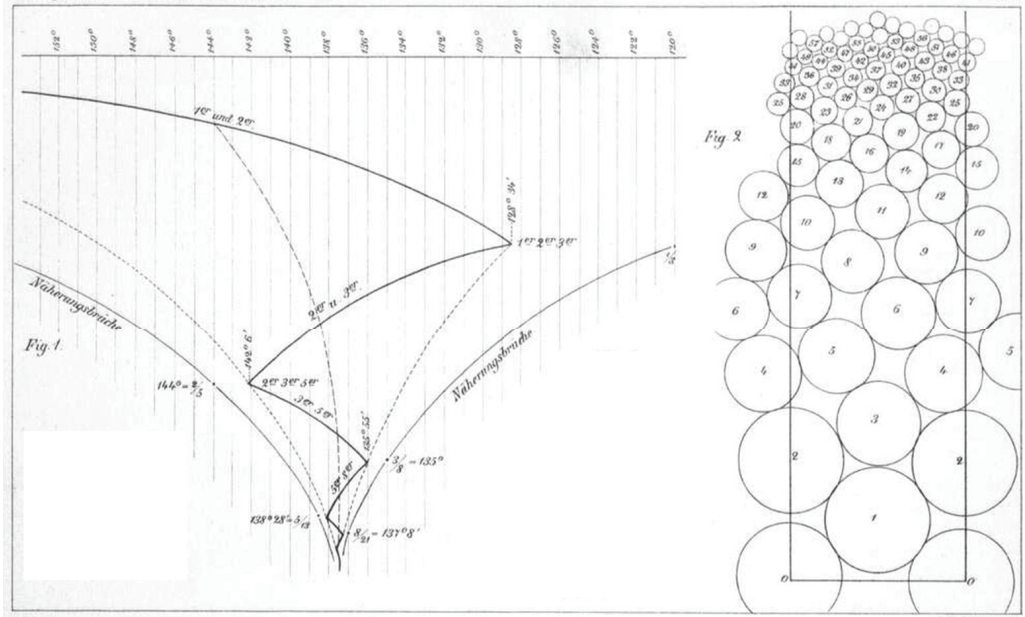}  
  \end{center}
  \caption{
Schwendener's causal model.  
Right: 
Contiguous circles with decreasing radius are stacked 
on an unrolled surface of a stem cylinder.
Contact parastichy numbers, or differences in the numbers of the circles in contact, 
change from (1,1) at the bottom to (5,8) at the top. 
Left: A mathematical relation 
between the divergence angle (the horizontal axis) and the
 radius of the contiguous circles (the vertical axis) is indicated with
 a solid zigzag curve
starting from the top left corner (divergence of 180$^\circ$, 
corresponding to the bottom part of the right figure) 
 down to the golden angle 137.5$^\circ$ (the top of the right figure). 
The zigzagging is due to shifts in the contact parastichy numbers 
from $(1,1)$ through $(1,2)$, $(2,3)$,  $(3,5)$, $(5,8)$, $(8,13)$, $(13,21)$ to
 $(21,34)$. 
The top branch for $(1,2)$ extends from 180$^\circ$ to 
$128^\circ 34'$. 
Adapted from \cite{schwendener83}.
}
  \label{fig:schwendener}
\end{figure}

Since the influential text by \cite{hofmeister68}, 
research into causal or dynamical mechanisms of 
primordia initiation 
has been the central pillar in the study of phyllotaxis. 
The empirical observation that new leaf primordia arise 
in the largest space between the older primordia is called Hofmeister's rule. 
%
What was originally a rule of thumb of botanists 
has been refined and developed into causal or dynamical models. 
%
%
\cite{airy1873} speculated on a causal 
mechanism 
in terms of
geometrical objects in mechanical action.  
%
\cite{schwendener78} put a similar idea 
on a more solid mathematical basis
by regarding leaves on a stem as solid disks 
contiguously covering a cylinder surface of infinite length (Fig.~\ref{fig:schwendener}). 
In Schwendener's model, 
contiguous circles of a constant radius are 
arranged in a periodic pattern 
characterized with a given set of contact parastichy numbers. 
Then it is a purely geometrical problem to derive 
various mathematical relations for the divergence angle, 
the radius of the circles, 
the girth of the cylinder, 
and 
the set of parastichy numbers. 
The radius of the circles is regarded as an independent variable, or a control parameter of the model. 
%
By letting the radius 
change  continuously along the stem cylinder, 
the divergence angle varies concomitantly with the contact parastichy numbers 
according to the mathematical relations.  
As a remarkable result, 
the divergence angle 
converges toward
the golden angle $137.5^\circ$ 
by decreasing the radius sufficiently slowly from a large initial value to a small constant value. 
Before attaining to the golden angle, 
the model predicts that 
the divergence angle 
oscillates with decreasing amplitude  (Fig.~\ref{fig:schwendener}). 
The decrease in the radius 
corresponds to 
decrease in relative size of leaf primordia on the stem or apex. 
%
This is a brief summary of Schwendener's causal mechanism for the golden
angle. 
The model is referred to as a mechanical or causal model of phyllotaxis. 
%


%
%

%
%
%
%
%
%
Related causal models were discussed in depth by
\cite{delpino1883teoria} and \cite{vaniterson07}. 
In recent decades, models of 
Schwendener and van Iterson 
have been elaborated on and 
developed further mathematically (\cite{adler74,rk89b,levitov91b,kunz01,agh02}) 
or numerically (\cite{wb84, hed06}), 
and even realized dynamically in a physics laboratory experiment (\cite{dc96a}). 
%
%
%
The causal models are founded on the basic assumption of causal determinism
that a phyllotactic pattern is 
a result of causal interaction of pattern units. 
In particular,  
the position of an initiated leaf primordium is 
determined 
by the position of the older primordia according to (supposedly simple) causal rules. 
%
The manner in which the units are arranged 
depends on the dynamic history of growth, 
or particularly on
the course of changes in size of leaf primordia. 
Thus, a common key factor of the causal models is a gradual change in size
of leaf primordia under mutual repulsion. 
%
%
%
%
%
Accordingly, there is a variety of causal models 
in which 
 the repulsive interaction is ascribed 
not to the mechanical contact pressure as supposed by Airy and Schwendener,  
but 
 to a chemical diffusion process
(\cite{schoute13,thornley75,mitchison77,veenlindenmayer77,young78, 
mk83, sc84, 
cp87,
roberts87,
ss89, 
yotsumoto93,km94,mkb98}). 
There are another 
causal models based on physical (\cite{hofmeister68,gss96,nss08})
and chemical (\cite{cs98}) instabilities.  
%
Recently, more intricate models based on molecular-genetic experiments have been
discussed (\cite{smith06,smith06b, jhsmm06, sspn11}), while 
geometrical models 
have been used 
to interpret patterns of real systems 
(\cite{malygin06,hjwzagd06, zms08}).
All these 
causal models are based on the assumption that 
{\it divergence angle is intrinsically variable and determined causally}. 

In the recent literature, 
we have had few opportunities of finding phyllotactic fractions in use. 
Most theoretical and experimental works 
attach importance to parastichy numbers instead, 
and the phyllotactic fraction is not mentioned or  regarded merely as an approximation 
even if mentioned
(\cite{williams74, ss89, lyndon90, jean94}). 
The fractional phyllotaxis is the original problem.
%
%
%
%
%
There are some reasons for this trend. 
First, early researchers 
did not appreciate the structural significance of the phyllotactic
fraction (\cite{hofmeister68, candolle81, church20, hirmer22, 
richards51,snow55}). 
Second, 
the studies of vascular structure organization 
are comparatively so few in number
that they are
overshadowed by intensive research interests directed
towards the shoot apical meristems. 
%
%
Third, Schwendener's causal model and its descendants are at variance
with the fractional divergence. 
According to the model, 
the divergence angle varies
depending on size of leaves, 
the vertical coordinate of Fig.~\ref{fig:schwendener}. 
%
%
\cite{schwendener83} 
guessed that a fractional pattern 
 would be made secondarily 
as a result of mechanical straightening of 
parastichous bundles connecting initiated leaves.
\cite{teitz88} confirmed indeed that 
the fractional phyllotaxis is accomplished by 
secondary torsion occurring in the vascular system during growth of the
stem.  
When there is little or no secondary 
distortion for lack of subsequent growth or internodal elongation, 
the original pattern at the apex may grow to 
a similar pattern of mature organs. 
%
This holds true for the most eye-catching patterns of closely packed reproductive organs, 
which, therefore, are often compared favorably with 
numerically simulated outputs of causal models.
Even then, 
the basic concept of the phyllotactic fraction may remain significant 
internally in vascular connections (\cite{wc84}). 
In fact, a stem with short internodes takes a high-order
fraction, which can be indistinguishable from the limit divergence of the undistorted stem. 
%
Thus, the phyllotactic fraction 
may not be judged by the external appearance alone.  
In vascular plants, 
each leaf is connected to the main stem vascular system through a strand 
of fluid-carrying vascular tissue called a {leaf trace}.  
A leaf may have several to many leaf traces. 
Leaf traces diverge from the stem vascular system some distance
below or very near the nodes at which they enter the leaves
(\cite{lestiboudois, naegeli58,beck10}). 
At the level of the shoot apex,
leaf traces 
form parastichous strands winding
obliquely round the stem axis. 
As the stem elongates, the leaf traces align up along the stem to make 
orthostichous 
bundles by forcing the whole stem to twist slightly 
from the original pattern (Fig.~\ref{rightfig0}), 
thereby a phyllotactic pattern characterized by a phyllotactic fraction
is established.
In the final pattern, there is a definite relationship between the
denominator of the phyllotactic fraction and the number of vascular orthostichies (\cite{kirchoff84}). 
The term orthostichy might be misleading,  
because the straightened bundles still may maintain their tilted course. 
The fraction neatly represents
geometrical arrangement of leaf traces, 
and the fractional order need not 
mean that leaves are positioned exactly vertically. 
%
Developmental sequences in differentiation and vascularization of leaf primordia are numerically 
correlated with the phyllotactic fraction of the shoot (\cite{ps36,girolami53, esau65}). 
%
%
%
Seemingly irregular 
rhythmical variations 
in various lengths of the external structure of a mature plant may be understood 
as a consequence of a hidden phyllotactic order in  the vascular system 
(\cite{unruh50, kumazawa71}). 
There is evidence for restricted pathways of translocation 
of photosynthetic assimilates related to phyllotaxis (\cite{wc84}). 
The patterns of translocation are sectorial, 
or the phyllotactic fraction has biological significance. 
The observation most pertinent to the present work is 
the significant correlation existing between the phyllotactic fraction and 
the number of internodes traversed by leaf traces: 
The higher phyllotactic fractions are associated with the longer leaf traces (\cite{girolami53,esau65}). 
While leaf traces of plants with helical phyllotaxis typically traverse more
than one internode, 
in distichous phyllotaxis,  a $\frac{1}{2}$ phyllotaxis of two-ranked
leaf arrangement, and in verticillate phyllotaxis, 
leaf traces are approximately one internode or less (\cite{bsr82}).  
Accordingly, 
low-order systems of a $\frac{1}{2}$ and $\frac{1}{3}$ phyllotaxis
are seen on the stems of plants with long internodes,  
while plants with short internodes show high-order fractions, 
as remarked above.

The stem vascular bundles, or axial bundles, and associated
leaf traces comprise sympodia, 
on the nature of which there are two perspectives (\cite{beck10}). 
In one view, 
the sympodia are 
of cauline origin, or derived from stem vascular tissue (\cite{bsr82}). 
In the other view, 
they are 
of foliar origin, or derived from leaf traces (\cite{esau65}). 
%
%
%
There are two different views on the causal relation between 
initiation of primordia and development of leaf traces or procambial strands, 
the strands differentiating into vascular bundles of  xylem and phloem. 
%
%
In one view, 
the initiation of leaf primordia brings about the differentiation of
the leaf traces. 
Hence the initiation of the leaf traces occurs basipetally, or  
in the direction from the leaf primordia 
toward the vascular system of the stem. 
%
In the opposing view, phyllotaxis of leaf primordia 
is dictated by the vascular organization that has been established
before the leaf primordia are initiated (\cite{larson77,larson83}). 
The latter is consistent with the observation that 
an incipient leaf trace develops acropetally,  
or in the direction toward the leaf primordium it serves
(\cite{esau65,nd97}). 
\cite{ps33} was criticized by \cite{snows34}. 
They both do not cast doubt on 
Hofmeister's rule, that is, 
they share  the causal view that phyllotaxis is a natural consequence of 
growth and development of an individual plant. 
They differ in what they regard as a basic unit of phyllotaxis. 
The former adopts growth units including leaf traces, 
while the latter places primary emphasis on 
leaf primordia at the apex. 
Accordingly, the former and the latter attach little importance to the irrational and
rational divergence, respectively. 
Thus, 
the causal view has been the paradigm of phyllotaxis. 

On the whole, causal models are successful in deriving indefinitely
continuing stable systems, resembling actual phyllotactic patterns. 
From a computational point of view, 
they are particularly appealing in that they provide us with programmable
protocols leading to the golden angle.
%
%
Irrespective of detailed mechanisms, however, 
realistic phyllotactic patterns are derived based on 
the following observational facts 
(\cite{vogel79, ropl84, pl91}): 
(i) Divergence angle is constant. 
(ii) The constant is the golden angle.  
For the sake of argument, 
the former is often taken so broadly 
that 
the constant 
may take any value. 
On this 
premise,  phyllotaxis is rendered to a geometrical playground of mathematics. 
There are mathematical arguments for (ii) 
based on the generalized hypothesis (i) (\cite{candolle81,coxeter72,leigh72,ridley82a,mk83}). 
It is often stated in this regard that the golden angle is a special
angle at which optimal packing is achieved. 
As a matter of fact, this is not true literally,  
for it is only under the constraint (i) that the golden
angle may be said optimal and there is no {\it a priori} reason for the constancy. 
For living organisms,  the property (i) is far from obvious and no less astounding than (ii), 
especially because the angular regularity may persist 
in spite of temporal irregularity. 
A time interval between the formation of successive leaves is called a plastochron, 
which is used as a morphological or developmental time scale. 
Plants grown in different environmental conditions
may be compared in plastochron units but not in physical time units. 
The fact that unit of time is a plastochron and duration of 
a plastochron is not constant in physical time 
poses a problem for realistic causal models based on physical time. 

%
%
%
%

Phyllotactic patterns at a shoot apex are more regular than those on a mature stem, 
because internodes tend to be elongated less regularly on the mature stem.   
As a matter of fact, the exact level in the stem at which a 
bifurcation or recombination of  vascular bundles 
takes place is not an important morphological constant (\cite{dormer72}). 
Accordingly, trace lengths may vary arbitrarily along the stem.  
For this reason, it is often argued that 
one should devote oneself exclusively to the study of the growing apex (\cite{church04,snow55}). 
%
Nonetheless, 
further mathematical relations for the spiral patterns at the apex can
be derived 
by assuming a stronger mathematical constraint of exponential
growth, according to which the leaf primordia are arranged
on logarithmic spirals in a centric representation
(Fig.~\ref{leftfig0})
(\cite{church04, richards51, thomas75, jean94}). 
In the exponential growth, 
the ratio of the distances from the center of the apex to two
successively numbered primordia 
is a constant, called the plastochron ratio. 
For a fixed value of divergence angle, 
\cite{richards51} has advocated the use of a phyllotaxis index 
defined in terms of the plastochron ratio (cf. (\ref{P.I.})). 
The index is used to designate two sets of parastichies intersecting orthogonally. 
For instance, for Fig.~\ref{leftfig0}, the plastochron ratio is 1.2,  the phyllotaxis
index is 3, and $(3,5)$ parastichies cross at right angles. 
A fractional value of the index, such as 2.7, means that no two parastichies are orthogonal. 
In this geometrical model,  a shift in parastichy numbers, 
e.g. from $(3,5)$ to $(5,8)$, a phenomena called rising phyllotaxis, 
is related to a variation of the plastochron ratio, or the exponential growth rate. 
The model has been generalized to allow for other constant divergence angles
than the golden angle (\cite{richards51,thomas75,jean94}). 
In contrast to these geometrical models 
based on {\it constant} divergence angle, there exist 
geometrical causal models 
in line with Schwendener's model, which 
aim at deriving the limit divergence angle 
by assuming {\it variable} divergence angles 
depending on plastochron, the plastochron ratio, and their own rules 
(\cite{vaniterson07, williams74, erickson83, wb84}). 
For a vegetative shoot, 
a plastochron index is defined in terms of length of leaves, 
and a leaf on a shoot is labeled with a leaf plastochron index (\cite{em57}). 
A developmental index of this kind is indispensable for the
systematic study of plant development (\cite{meicenheimer06}). 
The exponential growth is a practically useful 
approximation in dealing with 
young organs and early stages of development, 
although 
it is 
neither essential nor peculiar 
to phyllotaxis.

%
%



%
%

Despite the apparent success of causal models, neither their intrinsic
mechanisms nor predictions have yet been subjected to experimental tests
specifically.  
To name several problems on a descriptive level,  
existing causal models that explain all types of observed patterns
cannot help predicting also a multiplicity of unreal or too rare patterns. 
Even when they are capable of deriving normal patterns, 
they are not free from instabilities apparently irrelevant to
living organs. 
%
%
%
%
%
%
%
%
Causal models in general are confronted with a subtle trade-off. 
Normal phyllotactic patterns 
must be stable enough to account for the current prevalence in nature, 
while they cannot be quite 
stable in order to allow for 
many other exceptional ideal angles 
just to such a degree that they are actually existent. 
In short,  rare patterns should be neither too common nor too rare. 
It is not clear how and why this subtle balance between stability and instability is maintained universally, 
since fine control of relative size of phyllotactic units depends not
only on species but individual plants or even on parts of the individual plant. 
We get puzzled all the more by the observations of more frequent occurrence
of rare patterns among fossil plants.  
%
%

%
%
%
%
%

%


Causal models, whether physical or chemical, 
 provide  dynamical schemes of self-adjusting the system
under the influence of the older leaf primordia. 
On the premise that divergence angles between
successive leaves are freely variable by nature, 
they aim to derive a special angle, normally the golden angle,  
toward which the variable divergence angles tend ultimately. 
They do not assume any special constant divergence {\it a priori}.  
For the very reasons, they are likely to be 
beset with a fundamental difficulty in protecting the system against disturbance. 
%
%
In this regard, Hofmeister's empirical rule is often 
overestimated. 
Observed patterns satisfy Hofmeister's rule, but 
Hofmeister's rule is not sufficient for 
observed patterns. 
Hofmeister's rule 
does not imply the {\it periodic} appearance of new primordia 
(\cite{kirchoff03}), 
nor does it ensure precise regulation of 
the divergence of 137.5$^\circ$ (cf. Fig.~\ref{fig:fujitaL}). 
It is not difficult to draw an unreal pattern 
according to Hofmeister's rule.
The remarkable empirical fact 
is rather that 
divergence angle 
during steady growth seems always regulated stably to one of 
special angles closely related to the golden ratio. 
%
In fact, 
if a causal rule is to be strictly applied throughout, 
fluctuations in size of the domain of influence of a 
leaf primordium 
should inevitably 
leave behind everlasting irregularities propagated in 
the developing pattern (\cite{snow62}). 
Mathematically, 
the instability is a general consequence of the fact that the number of
possible phyllotactic configurations proliferates 
as relative size of phyllotactic units decreases. 
According to causal interpretations, 
higher phyllotaxis becomes more vulnerable. 
The difficulty may not be obvious if one were interested only in 
 low-order patterns like a $(2,3)$ and $(3,5)$ phyllotaxis, 
but it should become conspicuous when dealing with a higher order
pattern which requires higher precision maintenance. 
Besides this stability problem for high-order patterns, 
causal models have another difficulty for low-order patterns (Sec.~\ref{sec:3}). 
There are apparent geometric correlations between parastichy numbers 
and relative size of primordia on the apex (\cite{church04, richards51,
kirchoff03}) 
and between leaf arcs and 
the plastochron ratio (\cite{rutishauser98}). 
%
%
Causal models implement them as causal relationships
with the intention of proving that
a phyllotactic pattern, especially the divergence angle of 137.5$^\circ$, 
 is a necessary consequence of changes in the causal agent, relative size of leaf primordia. 
According to this interpretation, divergence angles and contact parastichies must 
 depend not only on the shape of primordia but on the geometry of the
surface on which they are located. 
The dependence has been investigated by \cite{vaniterson07}
on the assumption that 
all the primordia keep a common shape while they are allowed to change their sizes. 
%
%
%
%
%
%
%
%
%
%
%
So far, however, 
no direct evidence has been provided to support the presumed causal relationship. 
%
As a matter of fact, 
there are very few studies in which sufficiently detailed data 
are obtained 
to make a close comparison with the models possible or useful (\cite{erickson83}). 
In particular, 
%
the prediction of causal models that 
rising phyllotaxis, or change in parastichy numbers, 
should accompany 
wide variations and abrupt turns of divergence angle, 
as indicated in Fig.~\ref{fig:schwendener}, 
has not been supported unequivocally. 
On the contrary, 
%
the success of Richards' model indicates 
the exponential growth with {\it constant} divergence angle 
irrespective of whether parastichy numbers rise or fall. 
\cite{church04} refuted Schwendener's model 
by counterexamples showing normal spiral patterns of 
circular primordia whose positions are widely separated. 
In comparing treated plants, 
\cite{me77} found no significant change in divergence angle 
in a correlation diagram for 
the plastochron ratio and divergence angle.  
Statistical analysis of \cite{fujita39} has revealed that 
divergence angles do not depend so much on parastichy numbers
as expected from causal models (\cite{jean84}).  
%
There is clear evidence against the basic assumption that
the primordia size is the causal factor of divergence angle. 
A plant appears to accomplish 
geometrical correlations in a phyllotactic pattern by adapting the size and shape of leafy organs 
as if it knows the end pattern at which it aims. 
%
%
\cite{snow62} observe that the secondary extension of a leaf base 
adjusts itself 
so that 
divergence angle 
is little affected in spite of artificial disturbances.  
This observation, despite the authors' claim,  undermines 
their space-filling mechanism that the leaf base extension regulates the phyllotactic pattern. 
%
%
%
%
To the contrary, 
 apparent causal changes in the position, size and shape of leaves or scales in chemical or
physical contact 
 may be just incidental 
phenomena 
(\cite{church04, richards48, mh91}). 
No doubt there are cases in which physical or chemical 
contact pressure 
may induce secondary displacement of compactly packed lateral organs.  

Natural selection plays no role 
in causal interpretations of phyllotaxis. 
%
%
If one supposes to the contrary that natural selection holds the key to
understanding the golden angle at the shoot apex, 
then one should 
investigate a special {effect} of the special angle, instead of its {cause}. 
%
In other words, one should look for the distal or ultimate cause of the
special angle, instead of the proximity cause. 
This sort of theory intends to explain special traits 
not in terms of immediate physiological factors,
but in terms of evolutionary forces acting on them. 
It aims at 
a full understanding of the phenomena at a phenomenological level, 
independently of whatever physiological mechanisms may be involved. 
There is a long history 
of investigations into 
selective advantage of the observed divergences 
based on the external structure. 
It goes as follows:  
common phyllotactic patterns
distribute leaves 
as evenly as possible 
and maximize exposure of leaves to enhance the
capacity to intercept sunlight (\cite{wright1873}).  
Such an argument is unpromising because 
leaves are aligned in vertical ranks. 
Indeed, 
changes in leaf shape 
and stem length can compensate
for the negative effects of leaf overlap (\cite{niklas88, niklas98}). 
%
For this obvious reason, 
it is often argued to the contrary in favor of  the `most irrational'
divergence angle; 
no two leaves lie precisely under one another when divergence angle is equal to the golden angle 
 (\cite{candolle81, wiesner75,wiesner07, coxeter72,leigh72, 
takenaka94, py98, vb04, kbl04,bryntsev04}). 
%
%
%
%
%
There is also a long history of criticism of this view (\cite{dt17}). 
In the first place, 
the golden angle is not a general rule for mature shoots, 
and the light-capture mechanism deepens the riddle of 
the common occurrence of a $\frac{2}{5}$ phyllotaxis.
%
%
In general, existing theories 
tend to 
argue for 
the uses of irrational angles 
without regard to the uses of rational angles or vice versa.

\subsection{Aim and scope of this paper}
\label{aimandscope}

Let us direct attention to the internal structure,  the vascular system. 
Mathematical interrelationship between the initial (apical) and the mature (vascular) phyllotactic pattern 
seems to have not been discussed experimentally nor theoretically. 
This paper develops a theory of vascular phyllotaxis 
to fill in the gap between the two distinct but intimately related
phenomena. 
A physical model has been described mathematically 
in the previous paper (\cite{okabe11}). 
However, the model was abstract 
and its relevance to real phenomena was not clearly elucidated. 
The aim of this paper is to develop the model to show its experimental validity and relevance. 
This is relevant 
to a fundamental problem of phyllotaxis:  
{\it Is phyllotaxis determined causally or genetically?}
In contrast to numerous models holding the causal view, 
the present model is
based on the genetic perspective that 
special numbers in phyllotaxis are primarily of genetic origin, 
so that it is assumed that 
constant primordial divergence angle during steady growth is genetically
determined.  
According to the model, the effect of constant divergence angle is investigated, 
and what value of the constant is advantageous is settled. 
This work is not concerned about 
transient fluctuations 
of divergence angle during ontogeny.  
%
%
%
Therefore, 
the model is compatible with any physical or chemical causal models  for
the positioning of leaf initiation at the shoot apex, 
although  the limit divergence angle at the apex is interpreted totally differently. 
The special angle is not an inevitable consequence of ontogenetic
dynamics, whether physical or chemical. 
It is regarded as a heritable trait of a plant. 
It is supposed that once there was a wide variation in the traits of individuals, or 
there have formerly been wide variations of divergence angles.  
The special limit divergences found in nature have survived natural selection. 
This conforms with the traditional view that 
biological features that are under tight genetic control
and that have very narrow 
ranges of variation are believed to be adaptive (\cite{niklas97}). 
Although the author believes that the premise of the model,
divergence angle as a trait of a plant, is 
not only plausible but supported by circumstantial evidence, 
 it has not been unanimously accepted at present. 
It may be verified or refuted experimentally in the future.


%
%
For the efficient transport of materials 
throughout 
an indefinite 
number of leaves 
attached to a stem 
of a finite cross section, 
the leaves should be aligned along 
a finite number of `orthostichious' bundles. 
At this point, a whole number enters the theory. 
%
%
There are modes of orthostichous order depending on the initial
arrangement and length of leaf traces.  
The number of vascular orthostichies may be increased or decreased,  
but not arbitrarily. 
By regarding a leaf primordium {and} the leaf trace(s) associated with it 
as the fundamental unit of phyllotactic patterns, 
a mathematical correspondence is derived 
between the divergence angle of the initial phyllotactic pattern, 
 $360 \alpha_0$ degrees, 
and the phyllotactic fraction $\alpha$ of a mature pattern, 
where 
the number of internodes traversed by the leaf traces, $n_c$,
plays a pivotal role.
As a general rule,
it has been known 
that phyllotactic fraction of a vascular plant may 
vary sequentially during growth 
(\cite{braun1835,skutch27, allard42, pulawska65, larson77}). 
By means of the mathematical relation 
between $\alpha_0$ (an irrational number) and $\alpha$ (rational numbers), 
it is shown that 
changes in $n_c$ cause the phyllotactic transitions in $\alpha$. 
As a natural consequence, 
an evolutionary mechanism for the phenomenon of phyllotaxis is suggested. 
Supporting evidence for the model 
and 
the evolutionary mechanism 
is presented 
by 
analyzing 
experimental results.  

%
%
%
%
%

%




In Sec.~\ref{sec:model}, 
 a model and results used in the following sections 
are presented 
by means of figures and tables without using mathematics. 
Tables~\ref{tab:2}$\sim$\ref{tab:2112} have not been presented before.
%
%

In Sec.~\ref{sec:3}, 
 observed precision of 
the initial divergence $\alpha_0$ is explained 
by means of a correlation predicted between the range of 
 $\alpha_0$ and the highest-order fraction $\alpha$. 
In short, 
divergence angle $\alpha_0$ of a system 
with a high phyllotactic fraction $\alpha$
should be accurately controlled 
in order to avoid unnecessary changes in vascular structure. 



%

%

%

%

In Sec.~\ref{sec:fossilrecord}, 
phyllotaxis of 
{\it Lepidodendron} by \cite{dickson71} 
is analyzed. 
Diversity of phyllotaxis is discussed 
as a result of ineffective selective pressures.

In Sec.~\ref{sec:girolami}, 
the vascular structure of {\it Linum usitatissimum} by \cite{girolami53}
is investigated. 
Various relations between phyllotactic fraction and parastichy numbers, 
the phyllotactic fraction $\alpha$ and 
the length per internode of leaf traces $n_c$,  
and directions of parastichies and the genetic spiral
are pointed out.

In Sec.~\ref{sec:larson}, 
the phyllotactic transition of {\it Populus deltoides} by
\cite{larson77} is analyzed. 
It is shown that 
a continuous change in length of leaf traces causes 
the discontinuous effect of the phyllotactic transition in the vascular structure. 

In the appendix, 
a relation between the trace length $n_c$ 
and the plastochron ratio $a$ 
is discussed 
to indicate that the former serves as a useful developmental index for the
mature stem as the latter is used for the apex. 


%
%
%

\section{Model}
\label{sec:model}

\begin{figure}
  \begin{center}
  \subfigure[]{
\includegraphics[width= .47\textwidth]{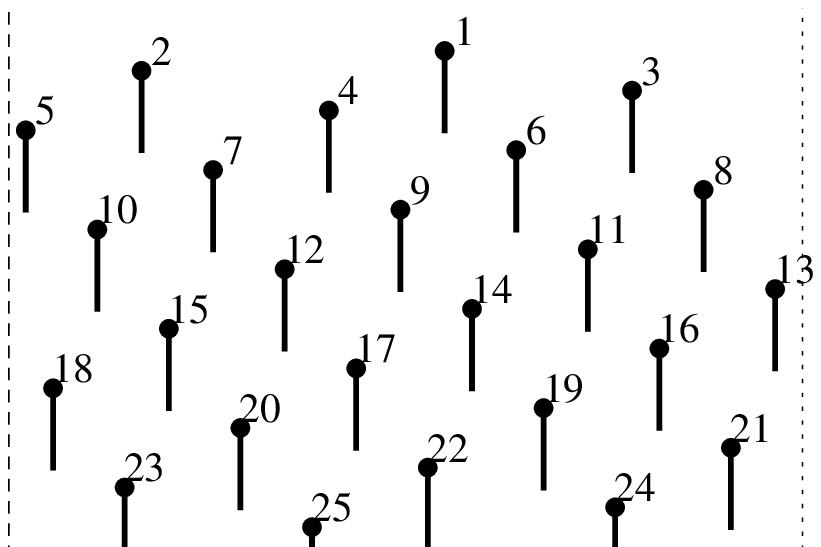}  
    \label{leftfig1}
  }
  \hfill
  \subfigure[]{
\includegraphics[width= .47\textwidth]{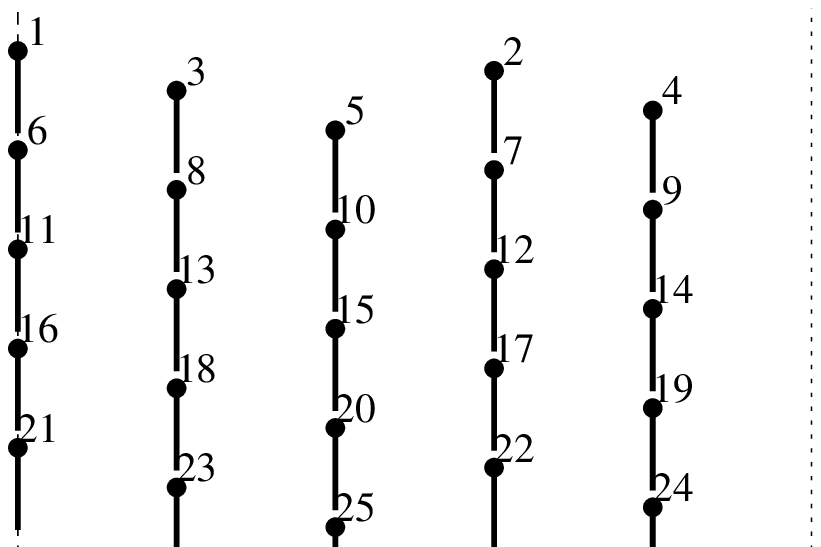}  
    \label{rightfig1}
  }
  \end{center}
  \vspace{-0.5cm}
  \caption{
Phyllotactic patterns of leaf traces with a length of $n_c=4$
before and after secondary torsion 
are arranged side by side.  
A dotted and dashed line 
of each figure 
represent a vertical cut of a cylinder surface unrolled. 
(a) A pattern with initial divergence of 
$360\alpha_0\simeq 137.5^\circ$ 
($\alpha_0=1/(1+\tau)\simeq  0.382$).  
(b) The final pattern of
a fractional divergence $\alpha=\frac{2}{5}$ resulting from (a). 
Leaf traces in the upper part 
move rightward 
while the pattern (a) becomes (b), 
thereby
five 5-parastichies in (a), such as 1-6-11-16-21, 
align themselves to make five orthostichies in (b).  
}
  \label{unrolledcyl1}
\end{figure}
\begin{figure}[t]
  \begin{center}
\includegraphics[width=.8 \textwidth]{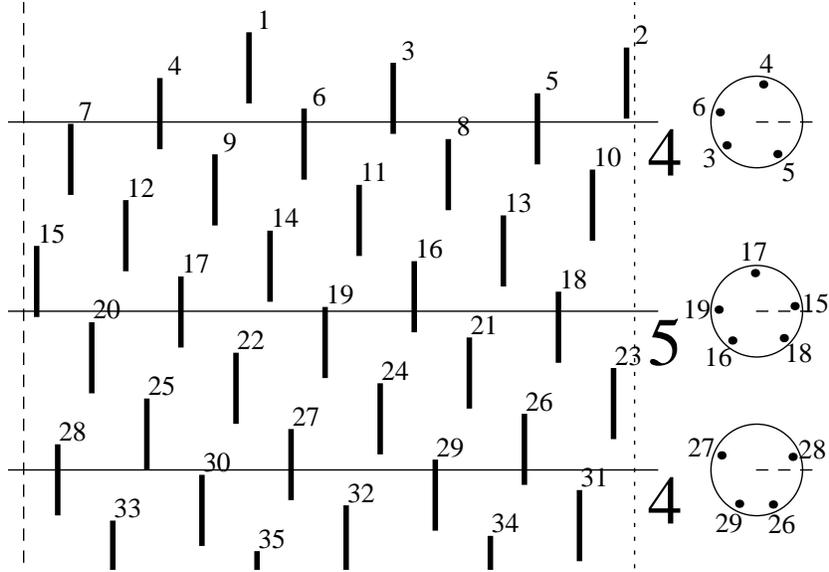}  
  \end{center}
  \caption{
For three transverse sections of 
a pattern of leaf traces with a length of $n_c=4.3$,  
the number of the traces in each section 
is indicated on the right-hand side  below the cut line.  
The number averaged over sections should approach $n_c$. 
}
  \label{nc4.3}
\end{figure}

A regular helical pattern of leaf traces 
is 
schematically plotted 
as a lattice of line segments on an unrolled surface of
a cylinder.
The divergence angle of the initial pattern is denoted as $360\alpha_0$
in degrees, which is assumed to be less than 180 degrees, i.e., 
$0\le \alpha_0 \le \frac{1}{2}$ without loss of generality. 
Fig.~\ref{leftfig1} presents a typical pattern 
for 
$360\alpha_0\simeq 137.5^\circ$ ($\alpha_0\simeq 0.382$). 
The length of leaf traces measured in internodes is denoted as $n_c$ 
in accordance with the previous notation (\cite{okabe11}).  
As in Fig.~\ref{rightfig0}, $n_c=4$ in Fig.~\ref{unrolledcyl1}. 
The trace length $n_c$ need not be an integer; 
 $n_c$ is the average number of leaf traces cut by 
a transverse section (Fig.~\ref{nc4.3}). 
As the number in a section is an integer,  
this method gives a good estimate of $n_c$ 
particularly for $n_c\gg 1$. 
%
%
%
%
The model comprises two parameters $\alpha_0$ and $n_c$. 
For the sake of argument, 
patterns with constant values of them are considered below.
%
Effects of their fluctuations may be discussed based on results to be obtained.

The leaf traces repel with each other laterally 
to arrange themselves in an orthostichous pattern. 
The mutual interaction is likely to be regulated by the plant hormone auxin (\cite{beck10}). 
Fig.~\ref{rightfig1} is the final pattern resulting from Fig.~\ref{leftfig1}.  
Divergence of the final pattern is expressed in terms of the phyllotactic fraction $\alpha$. 
The pattern of Fig.~\ref{rightfig1} is characterized with $\alpha=\frac{2}{5}$. 
In Fig.~\ref{leftfig1}, there are five parastichies of 
1-6-11-16-21,  2-7-12-17-22, 3-8-13-18-23, 4-9-14-19-24 and
5-10-15-20-25, each of which is called a 5-parastichy.  
The five 5-parastichies are lined up vertically to make five orthostichies
of the $\frac{2}{5}$ phyllotaxis in Fig.~\ref{rightfig1}. 
In the patterns of Fig.~\ref{unrolledcyl1},  
the next visible parastichies 
are 3-parastichies (1-4-7-10-13-16-19-22-25, 
2-5-8-11-14-17-20-23 and 3-6-9-12-15-18-21-24) 
and 2-parastichies (1-3-5-7-9-11-13-15-17-19-21-23-25 and 
2-4-6-8-10-12-14-16-18-20-22-24). 
As these parastichies remain conspicuous in the two patterns, 
both patterns 
may be referred to as having a parastichy pair of $(2,3)$. 
Thus, 
for $n_c=4$,   
there is one-to-one correspondence 
between $\alpha_0\simeq 0.382$ (angle of $360\alpha_0\simeq 137.5^\circ$)  
of the initial pattern and $\alpha=\frac{2}{5}$ of the final pattern. 
In a similar manner, 
 $\alpha$ is obtained for arbitrary values of $\alpha_0$ and $n_c$. 
Indeed, 
we get $\alpha=\frac{2}{5}$ insofar as $3\le
n_c<5$ and $\frac{1}{3}<\alpha_0<\frac{1}{2}$ 
( see \cite{okabe11} for the mathematical implementation).  
Below we discuss phyllotactic changes in $\alpha$ 
that occur when $n_c$ and $\alpha_0$ are set out of their respective ranges.


\begin{figure}
  \begin{center}
  \subfigure[
]{
\includegraphics[width= .47\textwidth]{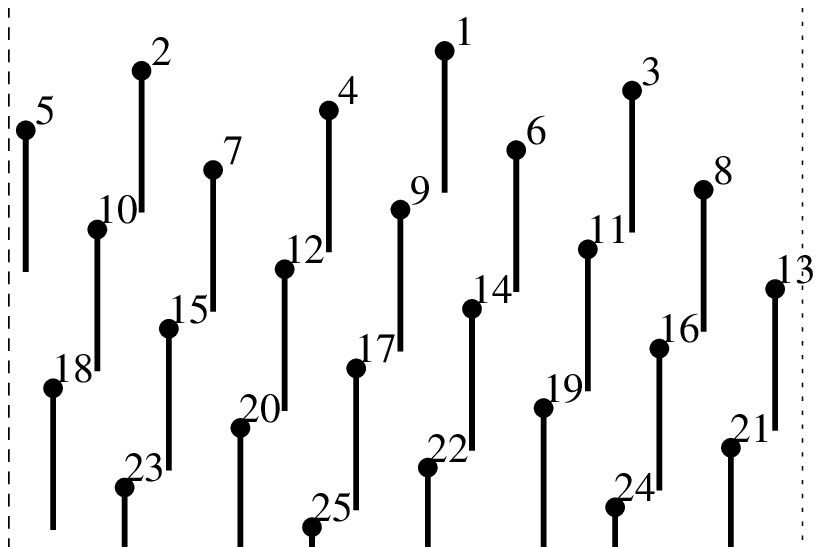}  
    \label{leftfig}
  }
  \hfill
  \subfigure[]{
\includegraphics[width= .47\textwidth]{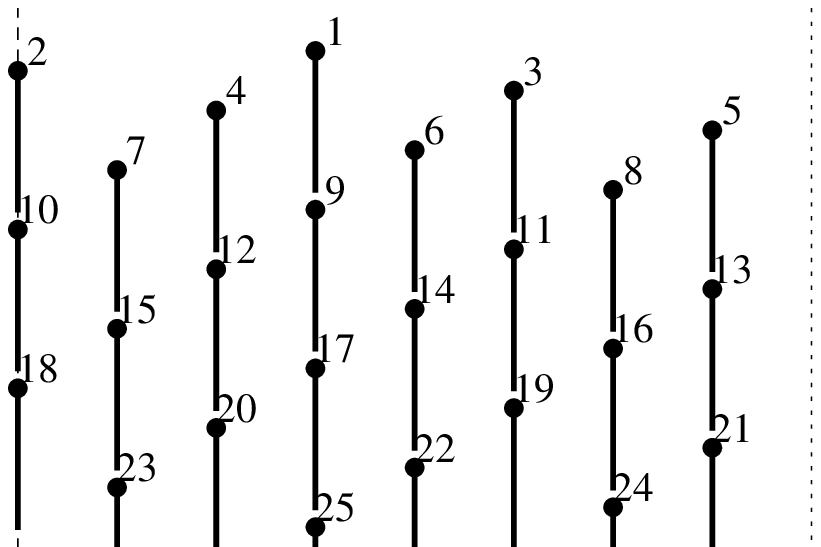}  
    \label{rightfig}
  }
  \end{center}
  \vspace{-0.5cm}
  \caption{
Change in a phyllotactic pattern of leaf traces with a length of $n_c=7$
(cf. Fig.~\ref{unrolledcyl1}). 
(a) The initial pattern with $\alpha_0=1/(1+\tau)=1/(2+\tau^{-1})$ 
($360 \alpha_0 \simeq 137.5 ^\circ$). 
(b) The final pattern with $\alpha=\frac{3}{8}$. 
}
  \label{unrolledcyl}
\end{figure}
For a fixed value of $\alpha_0\simeq 0.382$, 
Fig.~\ref{leftfig} is for $n_c=7$ in comparison with Fig.~\ref{leftfig1} for $n_c=4$. 
As the traces of length longer than five internodes 
cannot be aligned in five orthostichies, 
we obtain $\alpha=\frac{3}{8}$ for Fig.~\ref{leftfig},  
while $\alpha=\frac{2}{5}$ in Fig.~\ref{leftfig1}. 
Thus, 
it is explained that 
a higher phyllotactic fraction is obtained for 
a longer length of leaf traces. 
Phyllotactic transition from $\alpha=\frac{2}{5}$ to $\alpha=\frac{3}{8}$
occurs when $n_c$ increases past a threshold value of $n_c=5$. 
Experimental evidence of this transition is presented below in Fig.~\ref{fig:larsonnc}.

\begin{figure}
  \begin{center}
  \subfigure[
]{
\includegraphics[width= .47\textwidth]{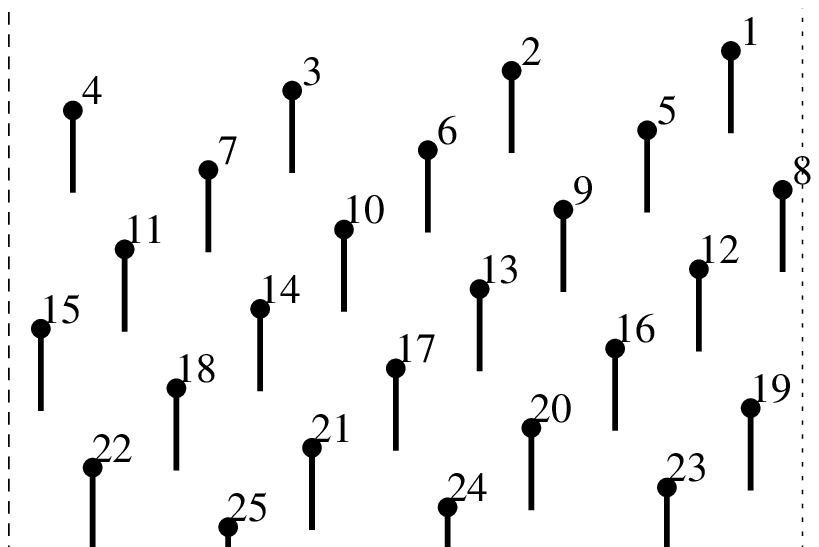}  
    \label{leftfig3}
  }
  \hfill
  \subfigure[]{
\includegraphics[width= .47\textwidth]{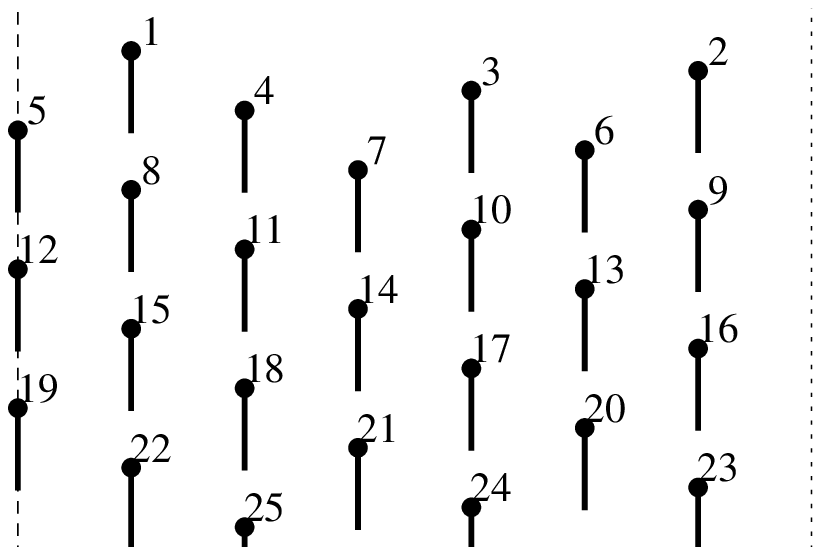}  
    \label{rightfig3}
  }
  \end{center}
  \vspace{-0.5cm}
  \caption{
Change in a phyllotactic pattern of leaf traces with a length of $n_c=4$
(cf. Fig.~\ref{unrolledcyl1}). 
(a) The initial pattern with $\alpha_0=1/(3+\tau^{-1})\simeq 0.276$ 
($360 \alpha_0 \simeq 99.5 ^\circ$). 
(b) The final pattern with $\alpha=\frac{2}{7}$. 
}
  \label{unrolledcyl3}
\end{figure}
For a fixed value of $n_c$, 
the phyllotactic fraction $\alpha$
depends
on the initial divergence $\alpha_0$.
For $n_c=4$,  Fig.~\ref{unrolledcyl3} is for 
$\alpha_0\simeq 0.276$ (angle of $99.5 ^\circ$), 
which is compared with   
Fig.~\ref{unrolledcyl1} for 
$\alpha_0\simeq 0.382$ $(137.5 ^\circ)$. 
The former leads to a final pattern of $\alpha=\frac{2}{7}$ in Fig.~\ref{rightfig3}, 
while the latter gives $\alpha=\frac{2}{5}$ in Fig.~\ref{rightfig1}.
In fact, there are three fractional patterns conceivable for $n_c =4$, namely 
(a) $\alpha=\frac{1}{5}$ for $0 < \alpha_0< \frac{1}{4}$, 
(b)  $\alpha=\frac{2}{7}$  for $\frac{1}{4}< \alpha_0< \frac{1}{3}$  and 
(c) $\alpha=\frac{2}{5}$  for $\frac{1}{3} < \alpha_0< \frac{1}{2}$.


\begin{figure}
\begin{center}
\includegraphics[width= \textwidth, height=.77\textheight
]{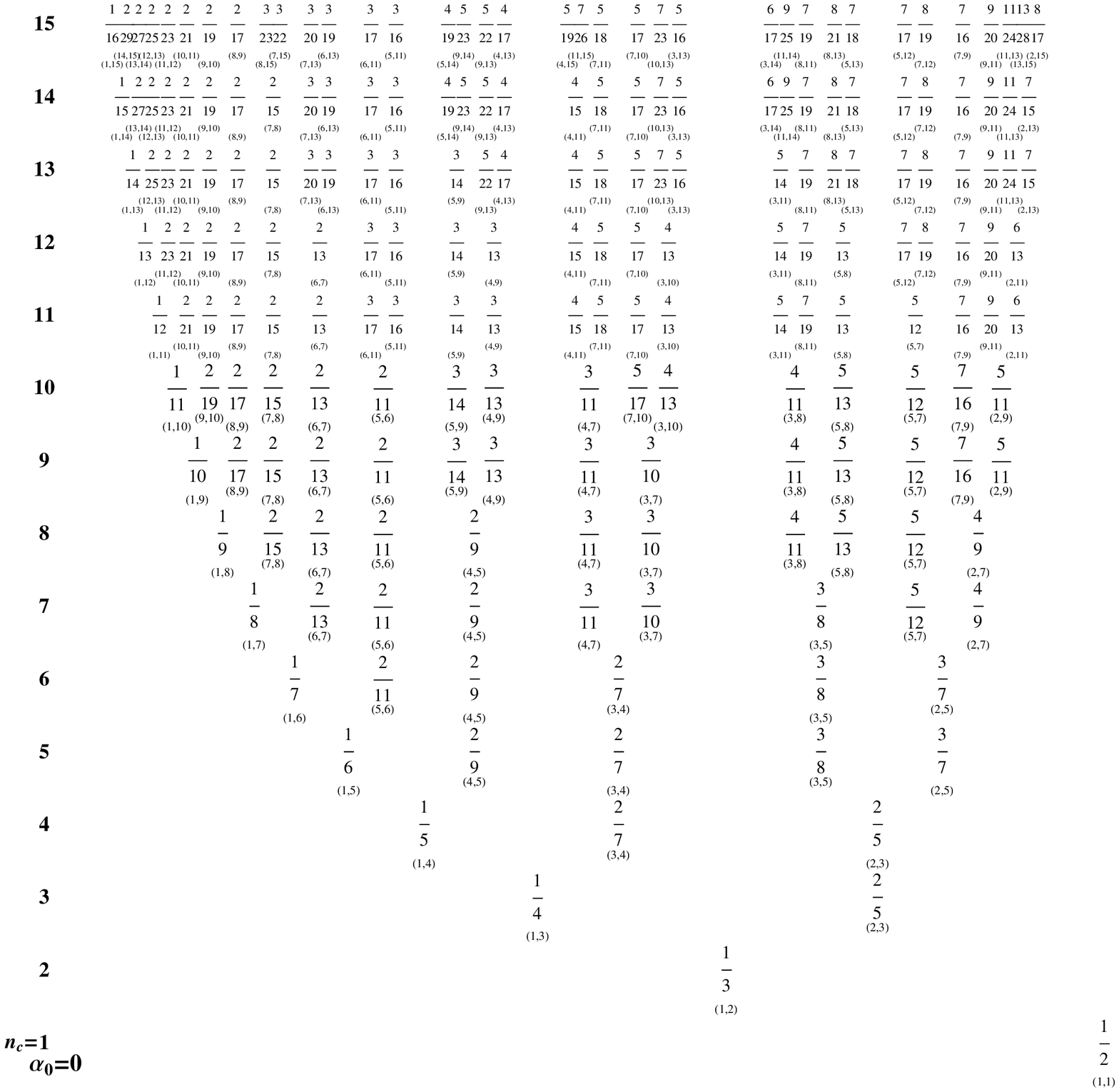}
\caption{
Tree diagram for the phyllotactic fraction $\alpha$. 
The horizontal axis is the initial divergence $\alpha_0$, 
and the vertical axis is the number of internodes traversed by leaf traces $n_c$. 
Numbers in parentheses below each fraction are 
the parastichy pair corresponding to the fraction.
%
By way of explanation, let us take $\frac{3}{8}$ in the right-bottom
 quoter as an example. 
The fraction $\frac{3}{8}$ with the parastichy pair $(3,5)$ is in three different positions 
at $n_c=5,6$ and 7. 
By means of lower order fractions lying below them, 
the fraction $\frac{3}{8}$ is bracketed 
between $\frac{1}{3}$ and $\frac{2}{5}$. 
Therefore, we obtain 
$\alpha=\frac{3}{8}$ with the parastichy pair $(3,5)$ 
insofar as $\frac{1}{3}\le \alpha_0 < \frac{2}{5}$ and $5\le n_c < 8$ (Table~\ref{tab:2}). 
Similarly,  we find 
$\alpha=\frac{5}{12}$ with $(5,7)$ 
for $\frac{2}{5}\le \alpha_0 < \frac{3}{7}$ and $7\le n_c < 12$ (Table~\ref{tab:22}). 
}
\label{SBtree}
\end{center}
\end{figure}
Every phyllotactic fraction for $\alpha$ has its own ranges of values for $\alpha_0$ and $n_c$. 
The mathematical correspondence is presented succinctly 
as a tree diagram in Fig.~\ref{SBtree}. 
For instance, Fig.~\ref{SBtree} gives 
the conditions $\frac{1}{4}< \alpha_0< \frac{1}{3}$ and $4\le n_c<7$ for
$\alpha=\frac{2}{7}$. 
For the former inequalities, 
the boundary fractions $\frac{1}{4}$ and $\frac{1}{3}$ lie below
$\frac{2}{7}$ in Fig.~\ref{SBtree}. 
The latter condition $4\le n_c<7$ is reasoned 
from the vertical coordinate $n_c=4, 5$ and $6$ of three $\frac{2}{7}$'s
in Fig.~\ref{SBtree}. 
The phyllotactic sequence of fractions derived from an arbitrary value
of initial divergence $\alpha_0$
may be traced 
by climbing up the tree of Fig.~\ref{SBtree} along the vertical line at $\alpha_0$. 
For $\alpha_0\simeq 0.382$ $(137.5 ^\circ)$, 
the main sequence 
$\frac{1}{2}, \frac{1}{3}, \frac{2}{5}, \frac{3}{8}, \frac{5}{13},
\frac{8}{21}, \cdots$ 
is obtained in the increasing order of $n_c$. 
The tree diagram extended for all values of $n_c$ includes all conceivable phyllotactic fractions. 

For the sake of convenience, let us introduce shorthand notation 
for the irrational numbers found in nature, 
\begin{eqnarray}
\ [n] &\equiv & \frac{1}{n+ \tau^{-1}},  
\nonumber\\ 
\ 
\qquad 
 [n,m] &\equiv &
\frac{1}{n+ \dfrac{1}{m+\tau^{-1}}},  
\nonumber\\
\ [n,m,l] &\equiv &
\frac{1}{n+ \dfrac{1}{m+\dfrac{1}{l+\tau^{-1}}}}, 
\label{[nml]}
\end{eqnarray}
and so on, 
%
where $n$, $m$, $l$ are positive integers.
With this notation, 
$\alpha_0=[2]=1/(2+\tau^{-1})=1/(1+\tau)$ gives the main sequence. 
The last equality holds by the definition of $\tau$ in (\ref{tau}). 
Note that the pattern with $\alpha_0=[1]$ 
($360\alpha_0= 222.5^\circ$)
is nothing but the mirror image of
$\alpha_0=[2]$ ($137.5^\circ$), 
because $[1]=1-[2]$ or $222.5^\circ=360^\circ-137.5^\circ $. 
For future reference, 
Tables \ref{tab:2}$\sim$\ref{tab:2112} are provided
for the initial divergence $\alpha_0$ given by typical irrational numbers. 
These are not exhaustive, but they include almost all
phyllotactic fractions observed in nature. 
The main sequence is presented in Table \ref{tab:2}.
In the second column for $\alpha=\frac{2}{5}$,  
 `$(2,3)$' in the second row represents the parastichy pair
 corresponding to the fraction $\frac{2}{5}$, 
`$3\sim$' in the third row abbreviates $3\le n_c <5$, 
where 5 for the upper limit is taken from the next column, 
and  `$\frac{1}{3}\sim \frac{1}{2}$' in the fourth row 
indicates $\frac{1}{3}< \alpha_0 < \frac{1}{2}$. 
As an example, let us take a fraction $\alpha=\frac{21}{76}$. 
It is found in the eighth column of Table \ref{tab:3}, 
from which  the conditions $47\le n_c < 76$ and
$\frac{8}{29}<\alpha_0<\frac{13}{47}$ are read. 
These results are used in the next section (Table~\ref{tab:dickson}).

The tables show that the denominator of a fraction $\alpha$ and the
parastichy numbers are correlated with the threshold numbers for $n_c$. 
The numbers comprise a characteristic sequence of integers. 
The main sequence in Table~\ref{tab:2} is
characterized with the Fibonacci sequence of 1, 2, 3, 5, 8, 13, 21,
$\cdots$, while Table~\ref{tab:22} has a sequence of 
1, 2, 3, 2, 5, 7, 12, 19, $\cdots$, which is sometimes called the lateral sequence. 
Three consecutive numbers of a sequence 
satisfy the Fibonacci recurrence relation ($2+5=7$, $5+7=12$, $7+12=19$), 
except for the first several numbers (like 1, 2, 3 in the lateral sequence). 
Therefore, each phyllotactic sequence is referred to by a pair of seed
integers for the Fibonacci recurrence formula.  
The seed pair of each table, such as $(2,5)$ in Table~\ref{tab:22}, is highlighted in boldface.

\begin{table}[t]
\begin{center}
 \begin{tabular}{ccccccccc}\hline
$\alpha$   &$\frac{2}{5}$ & $\frac{3}{8}$ & $\frac{5}{13}$ &
  $\frac{8}{21}$ & $\frac{13}{34}$ & $\frac{21}{55}$ &$\frac{34}{89}$
  &$\frac{55}{144}$ 
\\
 &(2,3) &(3,5) &(5,8) &(8,13) &(13,21) &(21,34) &(34,55) &(55,89) 
\\\hline
$n_c$      &$3\sim$ &$5\sim$ &$8\sim$ &$13\sim$ &$21\sim$ &$34\sim$
			  &$55\sim$ &$89\sim$ 
\\
$\alpha_0$ &$\frac{1}{3}\sim\frac{1}{2}$ &$\frac{1}{3}\sim\frac{2}{5}$
	  &$\frac{3}{8}\sim\frac{2}{5}$ &$\frac{3}{8}\sim\frac{5}{13}$
		  &$\frac{8}{21}\sim\frac{5}{13}$
		      &$\frac{8}{21}\sim\frac{13}{34}$
			  &$\frac{21}{55}\sim\frac{13}{34}$
			      &$\frac{21}{55}\sim\frac{34}{89}$ 
\\\hline
 \end{tabular}
\caption{
Parastichy numbers (in parentheses) and ranges of $n_c$ and $\alpha_0$  
for the phyllotactic fractions $\alpha$ 
belonging to the main sequence with 
the limit divergence of 
$\alpha_0=[2]=1/(2+\tau^{-1})\simeq 0.3820$
 ($360\alpha_0  \simeq 137.5^\circ$). 
The parastichy pairs are generated 
from the seed pair $(1,2)$ for $\alpha=\frac{1}{3}$ (not shown) by a Fibonacci recurrence relation. 
The golden angle $\alpha_0=[2]$ and the main sequence
are called the Fibonacci angle and the Fibonacci sequence. 
}
\label{tab:2}
\end{center}\end{table}
\begin{table}[t]
\begin{center}
 \begin{tabular}{ccccccccc}\hline
$\alpha$  & $\frac{1}{4}$& $\frac{2}{7}$& $\frac{3}{11}$&
  $\frac{5}{18}$& $\frac{8}{29}$& $\frac{13}{47}$&
  $\frac{21}{76}$&$\frac{34}{123}$\\
  &{\bf (1,3)}   &(3,4) &(4,7) &(7,11) &(11,18) &(18,29) &(29,47)&(47,76) \\\hline
$n_c$     &3$\sim$    &4$\sim$ &7$\sim$ &11$\sim$ &18$\sim$ &29$\sim$
			  &47$\sim$&76$\sim$ \\
$\alpha_0$  
&$0\sim\frac{1}{3}$
&$\frac{1}{4}\sim\frac{1}{3}$
	  &$\frac{1}{4}\sim\frac{2}{7}$ &$\frac{3}{11}\sim\frac{2}{7}$
		  &$\frac{3}{11}\sim\frac{5}{18}$
		      &$\frac{8}{29}\sim\frac{5}{18}$
			  &$\frac{8}{29}\sim\frac{13}{47}$
			      &$\frac{21}{76}\sim\frac{13}{47}$\\
\hline
 \end{tabular}
\caption{
Table for the limit divergence of 
$\alpha_0=[3]=1/(3+\tau^{-1})\simeq 0.2764$ ($360\alpha_0
 \simeq 99.5^\circ$). 
The parastichy pairs 
are generated from the seed pair $(1,3)$
by a Fibonacci recurrence relation. 
The sequence 1,3,4,7,11,$\cdots$ is called 
the Lucas sequence or the first accessory sequence. 
}
\label{tab:3}
\end{center}\end{table}
\begin{table}[t]
\begin{center}
 \begin{tabular}{ccccccccc}\hline
$\alpha$ & $\frac{1}{4}$ & $\frac{1}{5}$  &$\frac{2}{9}
  $&$\frac{3}{14}$&$\frac{5}{23}$&$\frac{8}{37}$&$\frac{13}{60}$&$\frac{21}{97}$\\
&(1,3)& {\bf (1,4)}  &(4,5) &(5,9) &(9,14) &(14,23) &(23,37)&(37,60) \\\hline
$n_c$  &3$\sim$  &4$\sim$      &5$\sim$ &9$\sim$ &14$\sim$ &23$\sim$
		      &37$\sim$&60$\sim$ \\
$\alpha_0$    &$0\sim\frac{1}{3}$
&$0\sim\frac{1}{4}$
    &$\frac{1}{5}\sim\frac{1}{4}$
      &$\frac{1}{5}\sim\frac{2}{9}$ &$\frac{3}{14}\sim\frac{2}{9}$
	      &$\frac{3}{14}\sim \frac{5}{23}$ & $\frac{8}{37}\sim
			      \frac{5}{23}$ & $\frac{8}{37}\sim
				  \frac{13}{60}$\\ \hline
 \end{tabular}
\caption{
Table for the limit divergence of 
$\alpha_0=[4]=1/(4+\tau^{-1})\simeq 0.2165$ ($360\alpha_0
 \simeq 78.0^\circ$). 
The parastichy pairs except (1,3)
are generated 
from the seed pair (1,4) by a Fibonacci recurrence relation. 
The sequence 1,4,5,9,14,$\cdots$ is called 
the second accessory sequence. 
}
\label{tab:4}
\end{center}\end{table}
\begin{table}[t]
\begin{center}
 \begin{tabular}{ccccccccc}\hline
$\alpha$   &$\frac{1}{4}$ &$\frac{1}{5}$
  &$\frac{1}{6}$&$\frac{2}{11}$&$\frac{3}{17}
  $&$\frac{5}{28}$&$\frac{8}{45}$&$\frac{13}{73}$\\
  &(1,3)  &(1,4)  &{\bf (1,5)}  &(5,6) &(6,11) &(11,17) &(17,28)&(28,45)\\\hline
$n_c$   &3$\sim$   &4$\sim$   &5$\sim$      &6$\sim$ &11$\sim$ &17$\sim$
			  &28$\sim$&45$\sim$ \\
$\alpha_0$  & $0\sim\frac{1}{3}$& $0\sim\frac{1}{4}$& $0\sim\frac{1}{5}$
& $\frac{1}{6}\sim\frac{1}{5}$& $\frac{1}{6}\sim\frac{2}{11}$&
			  $\frac{3}{17}\sim\frac{2}{11}$&
			      $\frac{3}{17}\sim\frac{5}{28}$&
				  $\frac{8}{45}\sim\frac{5}{28}$\\  \hline
 \end{tabular}
\caption{$\alpha_0=[5]=1/(5+\tau^{-1})\simeq 0.1780$ ($360\alpha_0
 \simeq 64.1^\circ$). 
}
\label{tab:5}
\end{center}\end{table}
%
\begin{table}[t]
\begin{center}
 \begin{tabular}{ccccccccc}\hline
$\alpha$   &$\frac{1}{4}$ &$\frac{1}{5}$  &$\frac{1}{6}$&$\frac{1}{7}$&$\frac{2}{13}
  $&$\frac{3}{20}$&$\frac{5}{33}$&$\frac{8}{53}$\\
  &(1,3)  &(1,4)  &{(1,5)}  &{\bf (1,6)} & (6,7) &(7,13) &(13,20)&(20,33)\\\hline
$n_c$   &3$\sim$   &4$\sim$   &5$\sim$      &6$\sim$ &7$\sim$ &13$\sim$
			  &20$\sim$&33$\sim$ \\
$\alpha_0$  & $0\sim\frac{1}{3}$& $0\sim\frac{1}{4}$& $0\sim\frac{1}{5}$
& $0\sim\frac{1}{6}$& $\frac{1}{7}\sim\frac{1}{6}$&
			  $\frac{1}{7}\sim\frac{2}{13}$&
			      $\frac{3}{20}\sim\frac{2}{13}$&
				  $\frac{3}{20}\sim\frac{5}{33}$\\  \hline
 \end{tabular}
\caption{$\alpha_0=[6]=1/(6+\tau^{-1})\simeq 0.15112$ ($360\alpha_0 \simeq 54.4^\circ$). 
}
\label{tab:6} 
\end{center}\end{table}
%
%
%
%
\begin{table}[t]
\begin{center}
 \begin{tabular}{ccccccccc}\hline
$\alpha$   &$\frac{2}{5}$ &$\frac{3}{7 }$&$\frac{5}{12 }$&$\frac{8}{19
  }$&$\frac{13}{31 }$&$\frac{21}{50}$&$\frac{34}{81}$&$\frac{55}{131}$
\\
 & (2,3) & {\bf (2,5)} &(5,7) &(7,12) &(12,19) &(19,31) &(31,50)&(50,81) 
\\\hline
$n_c$       &3$\sim$    &5$\sim$ &7$\sim$ &12$\sim$ &19$\sim$ &31$\sim$
		      &50$\sim$&81$\sim$ 
\\

$\alpha_0$   &$\frac{1}{3}\sim\frac{1}{2}$&$\frac{2}{5}\sim\frac{1}{2}$ &$\frac{2}{5}\sim\frac{3}{7}$
	  & $\frac{5}{12}\sim\frac{3}{7}$
	      &$\frac{5}{12}\sim\frac{8}{19}$
		  &$\frac{13}{31}\sim\frac{8}{19}$
		      &$\frac{13}{31}\sim\frac{21}{50}$
			  &$\frac{34}{81}\sim\frac{21}{50}$ 
\\
  \hline

 \end{tabular}
\caption{$\alpha_0=[2,2]=1/(2+1/(2+\tau^{-1}))\simeq 0.4198$
 ($360\alpha_0 \simeq 151.1^\circ$). 
The parastichy pairs except (2,3)
are generated from the seed pair (2,5).  
The sequence 2,5,7,12,19,$\cdots$ is called the first lateral sequence. 
}
\label{tab:22}
\end{center}\end{table}
\begin{table}[t]
\begin{center}
 \begin{tabular}{ccccccccc}\hline
$\alpha$    &$\frac{ 1}{4}$  &$\frac{2}{7}$  &$\frac{ 3}{10}$&$\frac{
  5}{17}$&$\frac{ 8}{27}$&$\frac{ 13}{44}$&$\frac{21}{71}$&$\frac{
  34}{115}$\\
  &(1,3)  &(3,4)  &{\bf (3,7)} &(7,10) &(10,17) &(17,27) &(27,44)&
				  (44,71)\\\hline
$n_c$ &3$\sim$ &4$\sim$       &7$\sim$ &10$\sim$ &17$\sim$ &27$\sim$ &44$\sim$&71$\sim$\\
$\alpha_0$  &$0\sim \frac{1}{3}$ &$\frac{1}{4}\sim \frac{1}{3}$
 &$\frac{2}{7}\sim \frac{1}{3}$ & $\frac{2}{7}\sim \frac{3}{10}$ &
		      $\frac{5}{17}\sim \frac{3}{10}$ &$\frac{5}{17}\sim
			  \frac{8}{27}$ &$\frac{13}{44}\sim
			      \frac{8}{27}$ &$\frac{13}{44}\sim
				  \frac{21}{71}$\\ \hline
 \end{tabular}
\caption{$\alpha_0=[3,2]=1/(3+1/(2+\tau^{-1}))\simeq 0.2957$
 ($360\alpha_0 \simeq 106.4^\circ$). 
}
\label{tab:32}
\end{center}\end{table}
\begin{table}[t]
\begin{center}
 \begin{tabular}{ccccccccc}\hline
$\alpha$   &$\frac{2}{5 }$ &$\frac{3}{7 }$  &$\frac{4}{9 }$&$\frac{7}{16
  }$&$\frac{11}{25 }$&$\frac{18}{41 }$&$\frac{29}{66}$&$\frac{47}{107
  }$\\
 &(2,3) &(2,5) &{\bf (2,7)} &(7,9) &(9,16) &(16,25) &(25,41)&(41,66) \\\hline
$n_c$       &3$\sim$     &5$\sim$     &7$\sim$ &9$\sim$ &16$\sim$ &25$\sim$ &41$\sim$&66$\sim$\\
$\alpha_0$   &$\frac{1}{3}\sim\frac{1}{2}$  &$\frac{2}{5}\sim\frac{1}{2}$ 
&$\frac{3}{7}\sim\frac{1}{2}$ &$\frac{3}{7}\sim\frac{4}{9}$
		  &$\frac{7}{16}\sim\frac{4}{9}$
		      &$\frac{7}{16}\sim\frac{11}{25}$
			  &$\frac{18}{41}\sim\frac{11}{25}$
			      &$\frac{18}{41}\sim\frac{29}{66}$\\ \hline
 \end{tabular}
\caption{$\alpha_0=[2,3]=1/(2+1/(3+\tau^{-1}))\simeq 0.4393$
 ($360\alpha_0 \simeq 158.1^\circ$). 
The sequence 2,7,9,16,$\cdots$ is called the second lateral sequence. 
}
\label{tab:23}
\end{center}\end{table}
\begin{table}[t]
\begin{center}
 \begin{tabular}{ccccccccc}\hline
$\alpha$  &$\frac{2}{7 }$&$\frac{3}{10 }$ &$\frac{4}{13 }$&$\frac{7}{23 }$&$\frac{11}{36 }$&$\frac{18}{59}$&$\frac{ 29}{95}$&$\frac{47}{154}$ \\
 &(3,4) &(3,7) &{\bf (3,10)} &(10,13) &(13,23) &(23,36)&(36,59) &(59,95) \\\hline
$n_c$      &4$\sim$    &7$\sim$     &10$\sim$ &13$\sim$ &23$\sim$ &36$\sim$&59$\sim$
		      &95$\sim$ \\

$\alpha_0$  &
$\frac{1}{4}\sim\frac{1}{3}$ 
&
$\frac{2}{7}\sim\frac{1}{3}$
 &$\frac{3}{10}\sim\frac{1}{3}$ & $\frac{3}{10}\sim\frac{4}{13}$ &$\frac{7}{23}\sim\frac{4}{13}$ & $\frac{7}{23}\sim\frac{11}{36}$& $\frac{18}{59}\sim\frac{11}{36}$& $\frac{18}{59}\sim\frac{29}{95}$\\ \hline
 \end{tabular}
\caption{$\alpha_0=[3,3]=1/(3+1/(3+\tau^{-1}))\simeq 0.3052$
 ($360\alpha_0 \simeq 109.9^\circ$). }
\label{tab:33}
\end{center}
\end{table}
\begin{table}[t]
\begin{center}
 \begin{tabular}{ccccccccc}\hline
$\alpha$  &$\frac{1}{4 }$ &$\frac{1}{5 }$&$\frac{2}{9 }$ &$\frac{3}{13
  }$&$\frac{5}{22 }$&$\frac{8}{35 }$&$\frac{13}{57}$&$\frac{21}{92}$\\
&(1,3) &(1,4)  &(4,5)  &{\bf (4,9)} &(9,13) &(13,22) &(22,35)&(35,57)
				  \\\hline
$n_c$    &3$\sim$  &4$\sim$    &5$\sim$       &9$\sim$ &13$\sim$ &22$\sim$ &35$\sim$&57$\sim$\\
$\alpha_0$ 
&$0\sim\frac{1}{3}$
&$0\sim\frac{1}{4}$ &$\frac{1}{5}\sim\frac{1}{4}$ 
 &$\frac{2}{9}\sim\frac{1}{4}$ & $\frac{2}{9}\sim\frac{3}{13}$
		      &$\frac{5}{22}\sim\frac{3}{13}$ &
			      $\frac{5}{22}\sim\frac{8}{35}$&
				  $\frac{13}{57}\sim\frac{8}{35}$\\
  \hline
 \end{tabular}
\caption{$\alpha_0=[4,2]=1/(4+1/(2+\tau^{-1}))\simeq 0.2282$
 ($360\alpha_0 \simeq 82.2^\circ$). }
\label{tab:42}
\end{center}
\end{table}
\begin{table}[t]
\begin{center}
 \begin{tabular}{ccccccccc}\hline
$\alpha$   &$\frac{2}{5 }$ &$\frac{3}{8}$
  &$\frac{4}{11 }$&$\frac{7}{19 }$&$\frac{11}{30 }$&$\frac{18}{49
  }$&$\frac{29}{79 }$&$\frac{47}{128}$\\
  &(2,3)   &(3,5)   &{\bf (3,8)} &(8,11) &(11,19) &(19,30) &(30,49)&(49,79)
  \\\hline
$n_c$   &3$\sim$  &5$\sim$     &8$\sim$ &11$\sim$ &19$\sim$ &30$\sim$ &49$\sim$&
			      79$\sim$\\
  $\alpha_0$
&$\frac{1}{3}\sim\frac{1}{2}$
&$\frac{1}{3}\sim\frac{2}{5}$
&$\frac{1}{3}\sim\frac{3}{8}$
 &$\frac{4}{11}\sim\frac{3}{8}$ &$\frac{4}{11}\sim\frac{7}{19}$ &
			      $\frac{11}{30}\sim\frac{7}{19}$&
				  $\frac{11}{30}\sim\frac{18}{49}$&
  $\frac{29}{79}\sim\frac{18}{49}$\\ \hline
 \end{tabular}
\caption{$\alpha_0=[2,1,2]
=1/(2+1/(1+1/(2+\tau^{-1})))
\simeq 0.3672$ ($360\alpha_0 \simeq 132.2^\circ$). }
\label{tab:212}
\end{center}\end{table}
\begin{table}[t]
\begin{center}
 \begin{tabular}{ccccccccc}\hline
$\alpha$   &$\frac{2}{5 }$ &$\frac{3}{7 }$ &$\frac{5}{12 }$
  &$\frac{7}{17 }$&$\frac{12}{29 }$&$\frac{19}{46 }$&$\frac{31}{75
  }$&$\frac{50}{121}$\\
 &(2,3) &{(2,5)} &(5,7) &{\bf (5,12)} &(12,17) &(17,29) &(29,46)&(46,75)\\\hline
$n_c$      &3$\sim$    &5$\sim$    &7$\sim$     &12$\sim$ &17$\sim$ &29$\sim$ &46$\sim$&75$\sim$\\
  $\alpha_0$
&$\frac{1}{3}\sim\frac{1}{2}$&$\frac{2}{5}\sim\frac{1}{2}$&$\frac{2}{5}\sim\frac{3}{7}$
&$\frac{2}{5}\sim\frac{5}{12}$ & $\frac{7}{17}\sim\frac{5}{12}$
		      &$\frac{7}{17}\sim\frac{12}{29}$
			  &$\frac{19}{46}\sim\frac{12}{29}$
			      &$\frac{19}{46}\sim\frac{31}{75}$ \\ \hline
 \end{tabular}
\caption{$\alpha_0=[2,2,2]
=1/(2+1/(2+1/(2+\tau^{-1})))
\simeq 0.4133$ ($360\alpha_0 \simeq 148.8^\circ$). }
\label{tab:222}
\end{center}\end{table}
\begin{table}[t]
\begin{center}
 \begin{tabular}{ccccccccc}\hline
$\alpha$  &$\frac{1}{4 }$&$\frac{2}{7}$&$\frac{3}{11 }$  &$\frac{4}{15
  }$&$\frac{7}{26 }$&$\frac{11}{41 }$&$\frac{18}{67}$&$\frac{
  29}{108}$\\
 &(1,3) &(3,4) &(4,7)  &{\bf (4,11)} &(11,15) &(15,26) &(26,41)&(41,67)
				  \\\hline
$n_c$      &3$\sim$    &4$\sim$    &7$\sim$    &11$\sim$ &15$\sim$ &26$\sim$ &41$\sim$&
			  67$\sim$\\
  $\alpha_0$
&$0\sim\frac{1}{3}$&$\frac{1}{4}\sim\frac{1}{3}$&$\frac{1}{4}\sim\frac{2}{7}$
&$\frac{1}{4}\sim\frac{3}{11}$ & $\frac{4}{15}\sim\frac{3}{11}$
		      &$\frac{4}{15}\sim\frac{7}{26}$ &
			      $\frac{11}{41}\sim\frac{7}{26}$&
				  $\frac{11}{41}\sim\frac{18}{67}$\\
  \hline
 \end{tabular}
\caption{$\alpha_0=[3,1,2]
=1/(3+1/(1+1/(2+\tau^{-1})))
\simeq 0.2686$ ($360\alpha_0 \simeq 96.7^\circ$). }
\label{tab:312}
\end{center}\end{table}
\begin{table}[t]
\begin{center}
 \begin{tabular}{ccccccccc}\hline
$\alpha$ 
 &$\frac{1}{4 }$ &$\frac{2}{7}$ &$\frac{3}{10}$ &$\frac{5}{17 }$
  &$\frac{7}{24 }$&$\frac{12}{41 }$&$\frac{19}{65 }$&$\frac{31}{106}$\\
 &(1,3) &(3,4) &(3,7) &(7,10)
 &{\bf (7,17)} &(17,24) &(24,41)&(41,65)\\\hline
$n_c$  &3$\sim$  &4$\sim$  &7$\sim$  &10$\sim$ 
      &17$\sim$ &24$\sim$ &41$\sim$& 65$\sim$\\

$\alpha_0$ 
 &$0\sim\frac{1}{3}$ &$\frac{1}{4}\sim\frac{1}{3}$ &$\frac{2}{7}\sim\frac{1}{3}$ &$\frac{2}{7}\sim\frac{5}{17}$
 &$\frac{2}{7}\sim\frac{5}{17}$ &$\frac{7}{24}\sim\frac{5}{17}$
			  &$\frac{7}{24}\sim\frac{12}{41}$ &
				  $\frac{19}{65}\sim\frac{12}{41}$\\
  \hline
 \end{tabular}
\caption{$\alpha_0=[3,2,2]
=1/(3+1/(2+1/(2+\tau^{-1})))
\simeq 0.2924$ ($360\alpha_0 \simeq 105.3^\circ$). }
\label{tab:322}
\end{center}\end{table}
\begin{table}[t]
\begin{center}
 \begin{tabular}{ccccccccc}\hline
$\alpha$  &$\frac{2}{5}$&$\frac{3}{8 }$&$\frac{4}{11}$ &$\frac{5}{14
  }$&$\frac{9}{25 }$&$\frac{14}{39 }$&$\frac{23}{64}$&$\frac{
  37}{103}$ \\
 &(2,3) &(3,5) &(3,8)  &{\bf (3,11)} &(11,14) &(14,25) &(25,39)&(39,64)
  \\\hline
$n_c$   &3$\sim$   &5$\sim$   &8$\sim$ 
     &11$\sim$ &14$\sim$ &25$\sim$ &39$\sim$&64$\sim$ \\
  $\alpha_0$
&$\frac{1}{2} \sim\frac{1}{3}$&$\frac{1}{3}\sim\frac{2}{5}$&$\frac{1}{3}\sim\frac{3}{8}$
&$\frac{1}{3}\sim\frac{4}{11}$ &$\frac{5}{14}\sim\frac{4}{11}$
		      &$\frac{5}{14}\sim\frac{9}{25}$ &
			      $\frac{14}{39}\sim\frac{9}{25}$&
				  $\frac{14}{39}\sim\frac{23}{64}$\\
  \hline
 \end{tabular}
\caption{$\alpha_0=[2,1,3]
=1/(2+1/(1+1/(3+\tau^{-1})))
\simeq 0.3593$ ($360\alpha_0 \simeq 129.3^\circ$). }
\label{tab:213}
\end{center}\end{table}
\clearpage
\begin{table}[t]
\begin{center}
 \begin{tabular}{ccccccccc}\hline
$\alpha$   
 &$\frac{2}{5}$ &$\frac{3}{7}$ &$\frac{4}{9}$ &$\frac{7}{16}$
 &$\frac{10}{23 }$&$\frac{17}{39 }$&$\frac{27}{62}$&$\frac{ 44}{101}$\\
 &(2,3) &(2,5) &(2,7) &(7,9)
 &{\bf (7,16)} &(16,23) &(23,39)&(39,62) \\\hline
$n_c$     &3$\sim$   &5$\sim$   &7$\sim$   &9$\sim$ 
    &16$\sim$ &23$\sim$ &39$\sim$& 62$\sim$\\
  $\alpha_0$
&$\frac{1}{3}\sim\frac{1}{2}$ &$\frac{2}{5}\sim\frac{1}{2}$ &$\frac{3}{7}\sim\frac{1}{2}$ &$\frac{3}{7}\sim\frac{4}{9}$ 
&$\frac{3}{7}\sim\frac{7}{16}$ &$\frac{10}{23}\sim\frac{7}{16}$
			      &$\frac{10}{23}\sim\frac{17}{39}$ &
  $\frac{27}{62}\sim\frac{17}{39}$\\ \hline
 \end{tabular}
\caption{$\alpha_0=[2,3,2]
=1/(2+1/(3+1/(2+\tau^{-1})))
\simeq 0.4356$ ($360\alpha_0 \simeq 156.8^\circ$). }
\label{tab:232}
\end{center}\end{table}
\begin{table}[t]
\begin{center}
 \begin{tabular}{ccccccccc}\hline
$\alpha$   
&$\frac{2}{5}$&$\frac{3}{7}$&$\frac{5}{12}$&$\frac{7}{17}$
&$\frac{ 9}{22}$&$\frac{16}{39 }$&$\frac{25}{61}$&$\frac{41}{100 }$\\
 &(2,3)  &(2,5)  &(5,7)  &(5,12) 
 &{\bf (5,17)} &(17,22) &(22,39)&(39,61) \\\hline
$n_c$   &3$\sim$&5$\sim$&7$\sim$&12$\sim$    &17$\sim$ &22$\sim$
			  &39$\sim$&61$\sim$\\
  $\alpha_0$
&$\frac{1}{3}\sim\frac{1}{2}$&$\frac{2}{5}\sim\frac{1}{2}$&$\frac{2}{5}\sim\frac{3}{7}$&$\frac{2}{5}\sim\frac{5}{12}$
&$\frac{2}{5}\sim\frac{7}{17}$ &$\frac{9}{22}\sim\frac{7}{17}$ &
				  $\frac{9}{22}\sim\frac{16}{39}$ &
  $\frac{25}{61}\sim\frac{16}{39}$ \\ \hline
 \end{tabular}
\caption{$\alpha_0=[2,2,3]=1/(2+1/(2+1/(3+\tau^{-1})))\simeq 0.4100$ 
($360\alpha_0 \simeq 147.6^\circ$). }
\label{tab:223}
\end{center}
\end{table}
\begin{table}[t]
\begin{center}
 \begin{tabular}{ccccccccc}\hline
$\alpha$   &$\frac{2}{5}$ & $\frac{3}{8}$ & $\frac{5}{13}$ &
  $\frac{7}{18}$ & $\frac{12}{31}$ & $\frac{19}{49}$ &$\frac{31}{80}$
  &$\frac{50}{129}$ \\
 &(2,3) &(3,5) &(5,8) & {\bf (5,13)} & (13,18) &(18,31) &(31,49) &(49,80) 
\\\hline
$n_c$     &$3\sim$ &$5\sim$ &$8\sim$ &$13\sim$ &$18\sim$ &$31\sim$
			  &$49\sim$ &$80\sim$ 
\\
$\alpha_0$ &$\frac{1}{3}\sim\frac{1}{2}$ &$\frac{1}{3}\sim\frac{2}{5}$
	  &$\frac{3}{8}\sim\frac{2}{5}$ &$\frac{5}{13}\sim\frac{2}{5}$
		  &$\frac{5}{13}\sim\frac{7}{18}$
		      &$\frac{12}{31}\sim\frac{7}{18}$
			  &$\frac{12}{31}\sim\frac{19}{49}$
			      &$\frac{31}{80}\sim\frac{19}{49}$ 
\\\hline
 \end{tabular}
\caption{$\alpha_0=[2,1,1,2]
=1/(2+1/(1+1/(1+1/(2+\tau^{-1}))))\simeq 0.3876$ ($360\alpha_0 \simeq 139.5^\circ$). }
\label{tab:2112}
\end{center}\end{table}
\begin{table}
\begin{center}
 \begin{tabular}{ccccccccc}\hline
$\alpha$   &$\frac{1}{4}$ &$\frac{1}{5}$  &$\frac{1}{6}$&$\frac{1}{7}$&
$\frac{2}{13}  $&$\frac{3}{19}$&$\frac{5}{32}$&$\frac{8}{51}$\\
  &(1,3)  &(1,4)  &{(1,5)}  &{(1,6)} & {(6,7)} & {\bf (6,13)} &(13,19)&(19,32)\\\hline
$n_c$   &3$\sim$   &4$\sim$   &5$\sim$      &6$\sim$ &7$\sim$ &13$\sim$
			  &19$\sim$&32$\sim$ \\
$\alpha_0$  & $0\sim\frac{1}{3}$& $0\sim\frac{1}{4}$& $0\sim\frac{1}{5}$
& $0\sim\frac{1}{6}$& $\frac{2}{13}\sim\frac{1}{6}$&
			  $\frac{2}{13}\sim\frac{3}{19}$&
			      $\frac{5}{32}\sim\frac{3}{19}$&
				  $\frac{5}{32}\sim\frac{8}{51}$\\  \hline
 \end{tabular}
\caption{$\alpha_0=[6,2]=1/(6+1/(2+\tau^{-1}) \simeq 0.1567$ ($360\alpha_0 \simeq 56.4^\circ$).}
\label{tab:62} 
\end{center}
\end{table}

\begin{figure}
  \begin{center}
   \includegraphics[width= \textwidth]{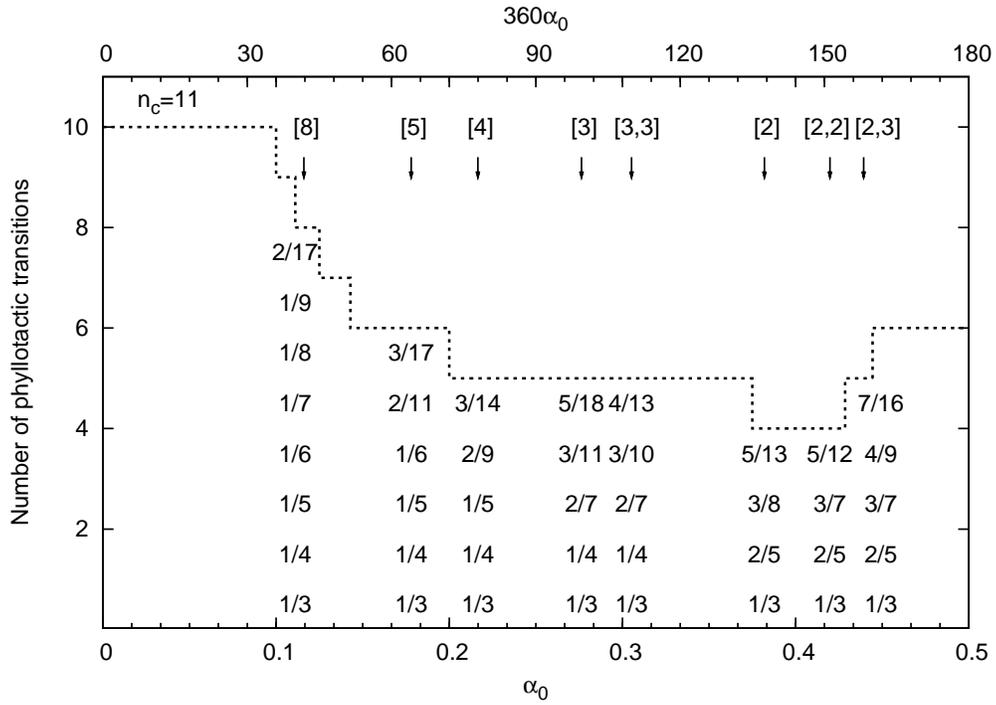}  
  \caption{
Phyllotactic fractions resulting 
while $n_c$ increases to eleven are arranged vertically 
for eight representative values of the limit divergence $\alpha_0$. 
A dashed line is the number of phyllotactic transitions 
counted from a $\frac{1}{2}$ phyllotaxis.
Initial divergences 
within a narrow range around the golden angle $\alpha_0=[2] \simeq
   0.382$ ($137.5^\circ$)
are most likely to survive. 
}
  \label{ntfraction}
  \end{center}
\end{figure}

Having prepared the mathematical relationship between 
the initial divergence $\alpha_0$, the final divergence $\alpha$
and the trace length $n_c$,  
we are in a position to give an account of what is special about the
golden angle. 
As shown below in  Figs.~\ref{fig:larson} and \ref{fig:larsonfig4},   
discontinuous change in phyllotactic fraction $\alpha$, 
or phyllotactic transition, 
involves reconstruction of the vascular structure.  
Therefore, it is advantageous for a plant to suppress the transitions as few as possible. 
As internodes vary in length during growth, 
the trace length per internode $n_c$ may change accordingly.  
For instance, $n_c$ may depend on the plastochron ratio $a$ (\ref{appendix}). 
Patterns with a fraction that appears in many places of
Fig.~\ref{SBtree} are stable against occasional changes in $n_c$. 
%
The lowest fraction that appears more than once is $\frac{2}{5}$.
Thus, 
systems with initial divergence angle giving rise to stable fractions are most likely to survive. 
Among all possible values of $\alpha_0$, 
the initial divergence angle which suffers the least number of phyllotactic transitions is 
the golden angle $\alpha_0=[2]$ (137.5 degrees). 
This is a summary of the evolutionary mechanism for the golden angle (\cite{okabe11}).  
%
Fig.~\ref{ntfraction} shows phyllotactic fractions resulting from 
various representative values of $\alpha_0$ while $n_c$ increases up to eleven.  
The number of phyllotactic transitions is indicated by a dashed line. 
In this example, initial divergence angles from 135$^\circ$ to 
154$^\circ$ are most likely to be naturally selected.  

\section{Precision of initial divergence angle}
\label{sec:3}

As the evolutionary mechanism relies on statistical screening processes, 
it does not predict a limit divergence angle with unlimited precision.
It is an empirical fact that 
divergence angles at the level of the shoot apex are regulated
toward a mean value comparable with an ideal angle given by 
the formula (\ref{[nml]}) after some transient fluctuations 
(\cite{davies39,snow62,bbyl10}).  
Excepting initial fluctuations, 
the precision with which leaves are organized on the apical meristems is remarkable. 
It is undoubtedly controlled by genetics,  though it may be slightly 
affected by light stimuli depending on the orientation (\cite{kumazawa71}). 
Twenty samples of young shoots of {\it Erigeron sumatrensis} (Sumatran fleabane)
show mean divergence angles from 137.23$^\circ$ to 137.97$^\circ$ 
with the sample average of $137.499 \pm 0.212^\circ$ (\cite{kumazawa71}). 
The mean divergence angle of the individual plant 
may deviate statistically significantly from the ideal limit angle (\cite{me77}). 
Sometimes there occur other ideal divergence angles than the normal
golden angle of 137.5$^\circ$.  
Phyllotaxis of {\it Musa sapientum} (banana) 
changes with the age of the plant from 
$\frac{2}{5}$ through 
$\frac{3}{7}$ 
to $\frac{4}{9}$ (\cite{skutch27}). 
This is consistent with a unique initial divergence of $\alpha_0=[2,3]$ (Table~\ref{tab:23}), 
which seems to be true for all species of {\it Musa} propagated vegetatively. 
\cite{rutishauser98} has presented 
a remarkably exotic pattern of {\it Picea abies} (Norway spruce)
showing a $(6,13)$ phyllotaxis (Table~\ref{tab:62}). 
%

The evolutionary mechanism predicts a 
correlation between the range of values of the initial divergence $\alpha_0$
and the highest-order fraction $\alpha$ 
attained in evolutionary or phylogenetic processes. 
The correlation may seem strange at first glance, 
as it appears as an advanced correlation 
in developmental or ontogenetic processes of a plant; 
the precision of divergence angles on a young shoot 
is determined by the phyllotactic form at its maturity. 
This phylogenetic correlation is contrasted with 
the instantaneous correlation that causal models predict between 
the divergence angle and parastichies of the standing pattern.  
In general, divergence angles of a $(3,5)$ phyllotaxis are widely variable within
$\frac{1}{3}<\alpha_0<\frac{2}{5}$,  
whereas the range is narrowed to 
$\frac{3}{8}<\alpha_0<\frac{2}{5}$
when the parastichy pair is raised to $(5,8)$. 
Remark that these are general results drawn from regularity of phyllotactic patterns.  
The ranges may be restricted further depending on 
specific assumptions of models. 
%
%
%
%
%
%
%
For instance, consider a regular pattern with a parastichy pair $(1,2)$,
which is realized for any value of divergence angle. 
According to Schwendener's model, however, 
$(1,2)$ patterns for $0< \alpha_0 <0.36$, i.e., from 0 to 128.6 degrees,
are not realized, 
for a transition to a $(2,3)$ phyllotaxis 
intervenes at $\alpha_0=0.36$ (\cite{adler74,levitov91b,dc96a}). 
See the top branch of the zigzag path in Fig.~\ref{fig:schwendener}. 
The threshold angle $\alpha_0=0.36$ specifically depends on 
geometrical assumptions, e.g. the circular shape of `leaves' on the stem cylinder surface. 
Accordingly, the divergence angles for the parastichy pair $(1,2)$
is predicted to vary continuously 
 within $0.36 <\alpha_0 <\frac{1}{2}$, i.e., from 128.6$^\circ$ to
 180$^\circ$. 
The range is narrowed substantially but still so wide that it is incompatible with observations that 
divergence angles are very close to 137.5$^\circ$ even in systems of low phyllotaxis. 
Causal models attain a target pattern with 137.5$^\circ$ 
by way of an almost opposite $(1,2)$ pattern with divergence of about 180$^\circ$. 
Therefore, they cannot but allow the wide latitude of divergence angles for the $(1,2)$ pattern, 
in disagreement with precise control of actual systems (cf. Fig.~\ref{fig:fujitaL}). 
This is a very old problem which \cite{vaniterson07} (p.~247) was well aware of. 
Nonetheless, it has been left unnoticed 
despite a marginal rise of various causal models in recent years. 
With reference to experimental evidence,  
\cite{church04} (p.~340) remarks that already at a $(2, 3)$ system 
the ideal angle is attained within an error of about one degree. 
The present model explains
the non-correspondence between 
divergence angle $\alpha_0$ and parastichy numbers
by relating the allowed range of $\alpha_0$  
not with the parastichy numbers but 
with the highest order fraction $\alpha$ that the plant would attain 
in its mature state.

%
%

%
%
%
%
%
%
%
%
%
%
By measuring initial divergence angles for thirty species of plants,  
\cite{fujita39} found that frequency distributions of the divergence
angles are almost independent of the parastichy numbers. 
The divergence angles cluster in a narrow range.  
The width of the range quantifies 
the remarkable constancy of the divergence angle  (Fig.~\ref{fig:fujitaL}). 
This results look puzzling from a causal viewpoint (\cite{jean84,jean94}). 
By contrast, they are consistent with the evolutionary mechanism 
in that the initial angle $\alpha_0$ is independent of the parastichy numbers. 
According to \cite{fujita39}, initial divergence angles for the main sequence 
fall within $138 \pm 7 ^\circ$ (\cite{fujita39}), irrespective of the parastichy pair.  
This corresponds to $\frac{3}{8}<\alpha_0<\frac{2}{5}$ (135 to 144 degrees),  
which is as expected for the highest-order fraction of $\alpha=\frac{5}{13}$ (Table~\ref{tab:2}). 
Similarly, an estimate of 
 $99 \pm 4 ^\circ$ for {\it Cunninghamia lanceolata} (China fir) (\cite{fujita39})
is consistent with 
$\frac{1}{4}<\alpha_0<\frac{2}{7}$ for $\alpha=\frac{3}{11}$
in Table \ref{tab:3}, and a narrow scattering of
$151 \pm 3 ^\circ$ for $(2,5)$ phyllotaxis at the apex of {\it
Cephalotaxus drupacea} (Japanese plum yew)
(\cite{fujita37}) is consistent with 
$\frac{5}{12}<\alpha_0<\frac{3}{7}$ for $\alpha=\frac{8}{19}$ 
(Table~\ref{tab:22}). 

Let us make a general remark that parastichy does not substitute for
divergence angle.   
The former depends on size and shape of the pattern unit or on a radial or internodal length scale. 
Therefore, several different parastichy pairs may be arbitrarily related to a single divergence angle. 
Parastichy numbers given in Tables~\ref{tab:2}$\sim$\ref{tab:2112} are 
the simplest pairs, which normally represent contact parastichies. 

\begin{figure}
  \begin{center}
   \includegraphics[width= \textwidth]{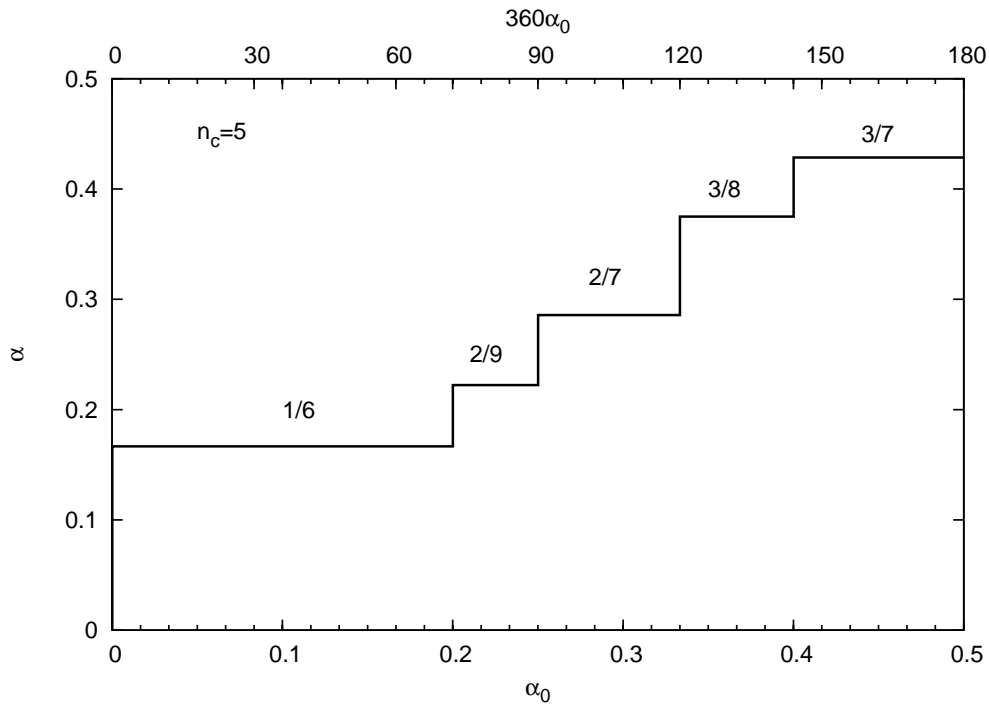}  
  \caption{
Phyllotactic fraction $\alpha$ versus 
initial divergence angle $\alpha_0$ for a fixed length $n_c=5$ of leaf traces.  
The fractional order changes discontinuously while $\alpha_0$ changes continuously. 
There are plateaus for 
five phyllotactic orders with 
$\frac{1}{6}$, $\frac{2}{9}$, $\frac{2}{7}$, $\frac{3}{8}$ and $\frac{3}{7}$. 
The initial divergence $\alpha_0$ is `quantized' within a  plateau 
to avoid the discontinuous transition. 
In other words, leaves are initiated regularly 
with a given angular precision.
The wide plateau for $\alpha=\frac{1}{6}$ is the most unstable against changes in $n_c$, 
while the plateau at $\alpha=\frac{3}{8}$ is the most stable. 
}
  \label{alpha0-alpha}
  \end{center}
\end{figure}

Large fluctuations in the initial divergence $\alpha_0$ may
cause the phyllotactic transition in the vascular structure, 
 even if the trace length $n_c$ is fixed constant. 
To suppress the transition that could happen, 
the divergence $\alpha_0$ has to be restricted within 
one of the ranges determined by $n_c$. 
For a fixed length of $n_c=5$, 
the fraction $\alpha$ is plotted against the initial divergence
$\alpha_0$ in Fig.~\ref{alpha0-alpha}.  
To maintain a $\frac{3}{8}$ phyllotaxis, 
the initial divergence $\alpha_0$ must stay within 
$\frac{1}{3}<\alpha_0<\frac{2}{5}$ (from $120^\circ$ to $144^\circ$); 
otherwise one would observe occasional excursions to 
$\frac{2}{7}$ (for $\alpha_0<\frac{1}{3}$) 
or $\frac{3}{7}$ (for $\frac{2}{5}<\alpha_0$) 
in the midst of a steady course of the $\frac{3}{8}$ phyllotaxis.  
Similarly, to maintain a $\frac{5}{13}$ phyllotaxis, 
the initial divergence $\alpha_0$ has to be kept within
$\frac{3}{8}<\alpha_0<\frac{2}{5}$ (from $135^\circ$ to $144^\circ$);  
otherwise one would find 
$\frac{4}{11}$ (for $\alpha_0<\frac{3}{8}$) 
or $\frac{5}{12}$ (for $\frac{2}{5}<\alpha_0$) 
within the mature state of the $\frac{5}{13}$ phyllotaxis 
(cf. Fig.~\ref{ntfraction}). 
%
Thus, it is explained why the initial divergence angle has to be `quantized' or 
fixed around a special constant with precision determined by the length of
leaf traces.
For this mechanism to work, 
stepwise changes in the fraction $\alpha$ of the vascular order,  
which are presumed to occur 
if the initial divergence angle $\alpha_0$ were not optimum, 
should incur penalties of extra energy. 
%
%
Thus, efficiency of the mechanism depends on the energy cost per
transition, which should depend on species.
By and large, however, 
the number of transition may be used as a good measure of the total cost, 
at least as a first approximation (Fig.~\ref{ntfraction}).

%
%
%

%
%
%
%
%
%
%
%
%
%
%
%

%
%
%
%
%

\section{Fossil record and diversity of phyllotaxis}
\label{sec:fossilrecord}

\cite{dickson71} found that nine among thirteen specimens 
of fossil remains of {\it Lepidodendron} (scale tree) show helical phyllotaxis,
of which only three belong to the main sequence.
This is in striking contrast to the current 
dominance of the main sequence in existing species 
(\cite{fujita38,zm85,jean94}). 
Therefore, 
Dickson concluded that 
the phyllotaxis of {\it Lepidodendron} is extremely variable, 
as much so as that of those most variable plants like cacti. 
His results provide us with important information
when they are analyzed in terms of the model.  

\begin{figure}
  \begin{center}
  \subfigure[
]{
\includegraphics[width=.45 \textwidth]{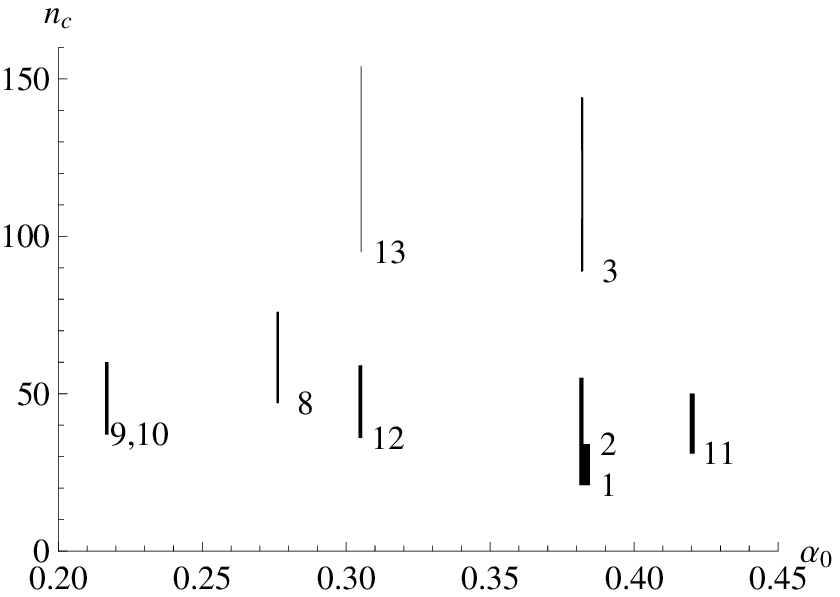}
    \label{fig:dickson}
  }
  \hfill
  \subfigure[]{
\includegraphics[width= .48\textwidth]{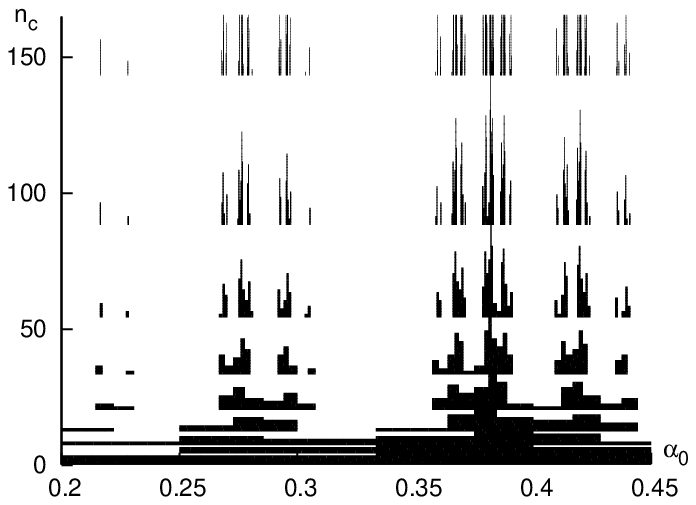}
    \label{fig:okabe}
}
  \end{center}
  \vspace{-0.5cm}
  \caption{
(a) 
Ranges of $\alpha_0$ and $n_c$ for 
the specimens of \cite{dickson71} (Table~\ref{tab:dickson}) 
are painted black in the $\alpha_0$-$n_c$ plane. 
(b) 
A theoretical result for the regions in which the number of phyllotactic
 transitions is minimal (adapted from Fig.~13 of \cite{okabe11}).  
The golden angle $\alpha_0\simeq 0.382$ (137.5$^\circ$) is singled out 
for $n_c$ below Fibonacci numbers such as 34, 55, 89 and 144. 
}
  \label{dickson}
\end{figure}

In the second and third line of Table~\ref{tab:dickson}, 
the phyllotactic fractions and the parastichy pairs for the nine
specimens are presented after Dickson. 
The fourth and fifth line are 
the corresponding ranges of $n_c$ and $\alpha_0$ 
according to Tables~\ref{tab:2}, \ref{tab:3}, \ref{tab:4}, \ref{tab:22},
and \ref{tab:33}. 
The last line is the limit divergence in terms of the bracket notation 
defined by (\ref{[nml]}) in the last section.  
For instance, in the second column, 
the specimen No.~1 
has a $\frac{13}{34}$ phyllotaxis
$(\alpha=\frac{13}{34})$ with the parastichy pair of $(13,21)$, 
for which 
$21\le n_c< 34$ and $\frac{8}{21} < \alpha_0 <\frac{5}{13}$. 
The limit divergence of $\alpha_0=[2]$ (137.5$^\circ$) satisfies the latter condition. 
The specimens Nos.~1-3 belong to the main sequence $\alpha_0=[2]$. 
Fig.~\ref{fig:dickson} represents graphically 
the parameter regions allowed for $n_c$ and $\alpha_0$.  
By comparison,  Fig.~\ref{fig:okabe} gives 
a theoretical result for the most favored regions 
in which the number of phyllotactic transitions is minimal 
(\cite{okabe11}).  

\begin{table}[t]
\begin{center}
\footnotesize
\begin{tabular}{lccccccccc}
No. &1 &2 &3 & 8 & 9,10& 11& 12& 13 \\\hline
$\alpha$ & $\frac{13}{34}$ & $\frac{21}{55}$ & $\frac{55}{144}$ &
		 $\frac{21}{76}$
& $\frac{13}{60}$
& $\frac{21}{50}$& $\frac{18}{59}$&
				     $\frac{47}{154}$ \\
  &\footnotesize (13,21) &\footnotesize(21,34) &\footnotesize(55,89) &\footnotesize (29,47)
&\footnotesize (23,37) 
&\footnotesize (19,31)&
			 \footnotesize (23,36)&\footnotesize (59,95) \\
\hline
$n_c$&\footnotesize [21, 34)&\footnotesize [34,55)&\footnotesize
	     [89,144)&\footnotesize [47,76)&\footnotesize [37,60)
&\footnotesize [31,50)&\footnotesize [36,59)&\footnotesize [95,154)\\
$\alpha_0$ 
 & $\frac{8}{21}\sim \frac{5}{13}$ & $\frac{8}{21}\sim \frac{13}{34}$
	 &$\frac{21}{55}\sim \frac{34}{89}$  &$\frac{8}{29}\sim
		 \frac{13}{47}$ & $\frac{8}{37}\sim \frac{5}{23}$ &
			     $\frac{13}{31}\sim \frac{8}{19}$&
				 $\frac{7}{23}\sim
				 \frac{11}{36}$&$\frac{18}{59}\sim
				     \frac{29}{95}$ \\
& [2] & [2] &[2] & [3]& [4]
& [2,2]& [3,3]&[3,3] \\
\hline
\end{tabular}
\caption{
Ranges of $n_c$ and $\alpha_0$  for the phyllotactic fraction $\alpha$ and the contact parastichy pair $(n,m)$
of  the nine specimens of {\it Lepidodendron} by \cite{dickson71}. 
Abbreviations $[21,34)$ and $\frac{8}{21}\sim   \frac{5}{13}$ mean 
  $21\le n_c<34$ 
and 
$\frac{8}{21}< \alpha_0< \frac{5}{13}$. 
The bracket notation in (\ref{[nml]}) is used 
for the limit divergence in the last row. 
Only the first three specimens
belong to the main sequence $\alpha_0=[2]$ (137.5$^\circ$). 
}
\label{tab:dickson}
\end{center}\end{table}
According to Fig.~\ref{fig:dickson}, 
the trace length $n_c$ appears to be independent of the initial
divergence $\alpha_0$. 
Moreover,  $n_c$ is not as variable as $\alpha_0$. 
As the order of phyllotaxis is very high,  
there is considerable uncertainty in $n_c$, while $\alpha_0$ is quite accurate. 
The specimens may be divided
into two groups in terms of $n_c$, 
i.e., one with $n_c\sim 50$ and the other with $n_c> 100$.
The fact that the fossil specimens show various but accurate values of $\alpha_0$
%
strongly suggests the evolutionary origin of the special divergence angles.  
It is impossible to tabulate all phyllotactic fractions 
for such a large value as $n_c=50$ due to lack of space, but 
it is mentioned only 
that 
the number of possible phyllotactic fractions at $n_c=50$ amounts to 387 
$(\simeq 3n_c^2/\pi^2/2)$. 
Among them, 
only the single fraction $\alpha=\frac{21}{55}$ of the main sequence 
falls in the optimum regions depicted in Fig.~\ref{fig:okabe}, 
while the specimens Nos.~8-12 do not meet the optimum condition. 
Nevertheless, all the reported specimens 
possess 
the irrational numbers 
expressed in the form of (\ref{[nml]}), 
as expected in the evolutionary mechanism (cf. Table~2 of \cite{okabe11}). 
Phyllotactic patterns for $\alpha_0=[3,2]$, $[2,3]$, 
$[2,1,2]$ and others are not reported, 
presumably because 
of  lack of enough samples. 
Thus, anomalous patterns 
are regarded as relics of evolutionary processes. 



\begin{figure}
  \begin{center}
\includegraphics[width= \textwidth]{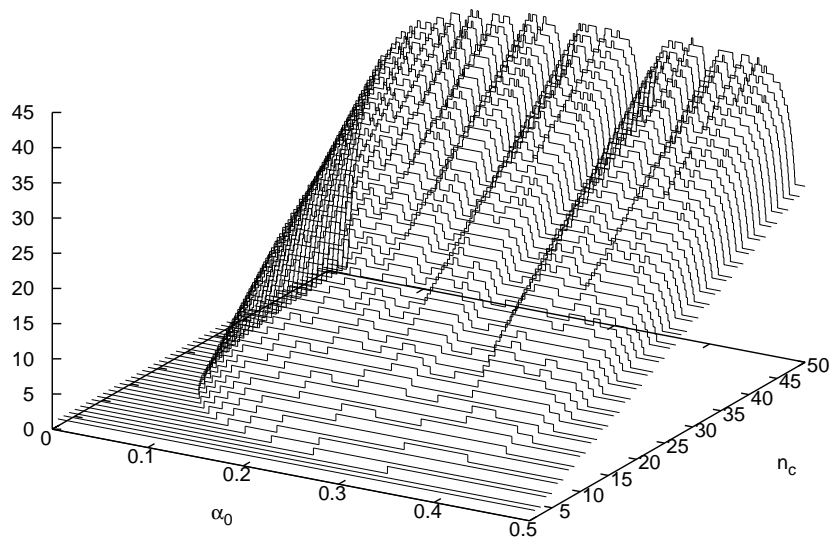}
  \end{center}
  \caption{
Three-dimensional fitness landscape. 
The vertical axis 
representing `fitness' 
is $n_c$ minus the number of phyllotactic transition. 
%
%
The variables on the base plane are $\alpha_0$ and $n_c$. 
The number of transition increases with $n_c$. 
The fitness has a flat bottom minimum in the worst case of $\alpha_0\simeq 0$, 
whereas there are `fitness peaks' at $\alpha_0=[2]\simeq 0.38$ 
(the main sequence), $\alpha_0=[3]\simeq 0.28$ (an accessory sequence) 
and others, 
whose widths decrease as $n_c$ increases (\cite{okabe11}). 
}
  \label{fitness}
\end{figure}

\begin{figure}
  \begin{center}
  \subfigure[]{
\includegraphics[width= .46\textwidth]{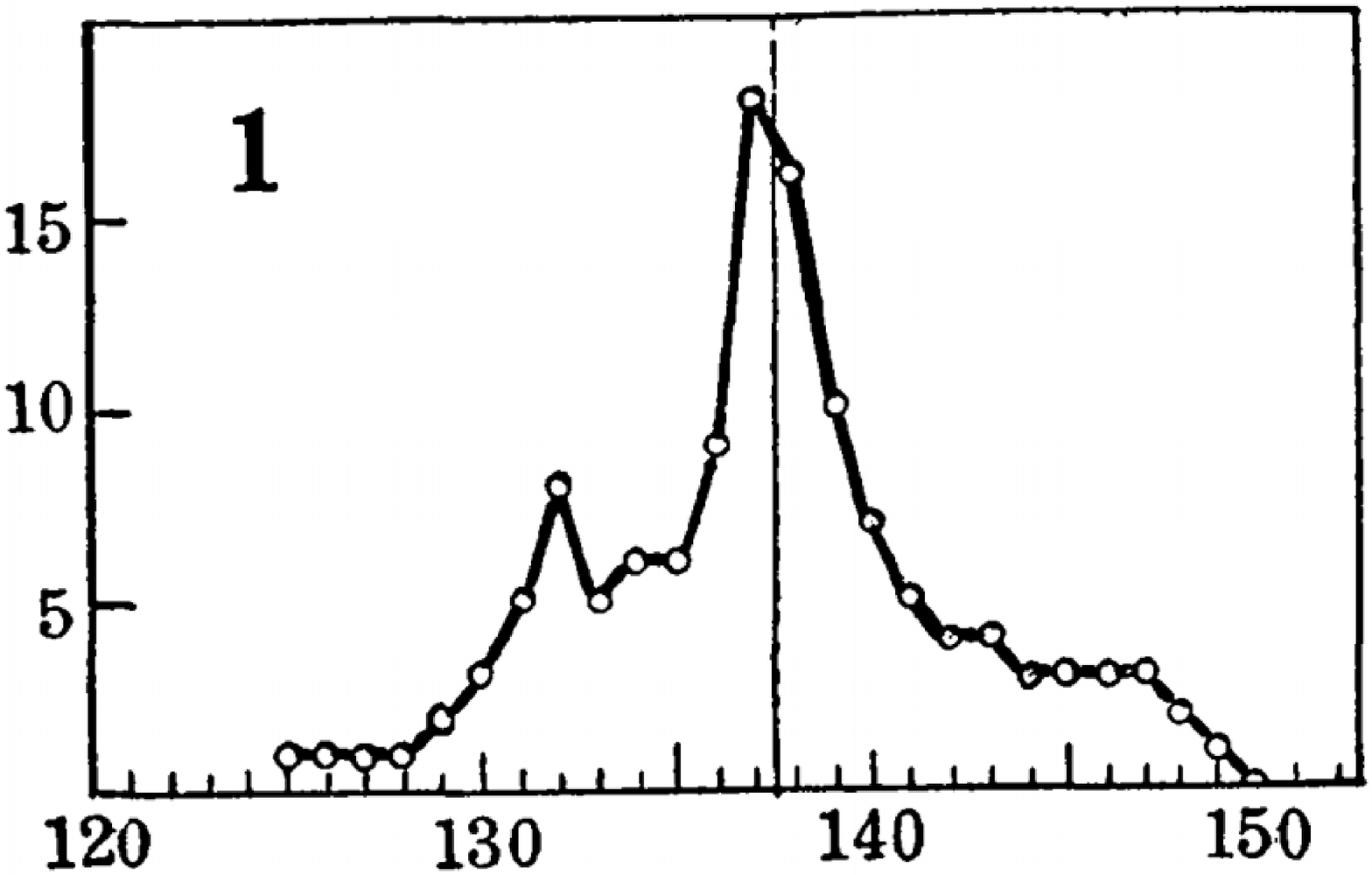}  
    \label{fig:fujitaL}
 }
  \hfill
  \subfigure[]{
\includegraphics[width= .49\textwidth,]{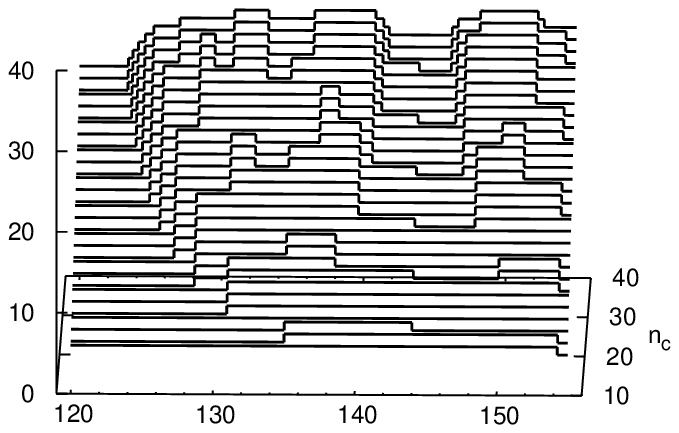}
    \label{fig:fujitaR}
  }
  \end{center}
  \vspace{-0.5cm}
  \caption{
(a) Frequency distribution of initial divergence angles for
 a $(1,2)$ phyllotaxis of {\it Lysimachia clethroides} by \cite{fujita39}. 
(b) 
An enlarged view of a relative `fitness'
in Fig.~\ref{fitness} 
is plotted against the divergence angle in degrees, $360\alpha_0$.
The peak plateau 
extends from 135 to 144 degrees at $n_c=12$, 
and from 135 to 138 degrees at $n_c=19$.  
}
  \label{fig:fujita}
\end{figure}

It has been an unresolved problem in what quantitative terms normal and
anomalous phyllotaxis are differentiated. 
The number of phyllotactic transition during a steady growth
provides us with a numerical measure of relative fitness in evolution. 
The most fit divergence angles 
are indicated in Fig.~\ref{fig:okabe}. 
They are peaks of a `fitness landscape' (\cite{niklas97}),  
shown in Fig.~\ref{fitness} (\cite{okabe11}). 
A close inspection of 
the frequency distribution curves of \cite{fujita39} indicates that 
a primary peak 
accompanies small subsidiary peaks at anomalous angles. 
In Fig.~\ref{fig:fujita}, 
Fujita's result for {\it Lysimachia clethroides} (gooseneck
loosestrife)  
is arranged 
along with transections of the fitness landscape in Fig.~\ref{fitness}. 
\cite{roberts84} has discussed that his chemical contact pressure model explains 
the anomalous subsidiary peaks. 
However, 
his conclusion is based on circular reasoning 
that anomalous systems are less frequent because they are anomalous. 
%
%
Similar fitness curves are obtained for light absorption efficiency
of rosette plants (\cite{niklas88,niklas98,py98, kbl04}). 


Let us remark incredible precision 
of the divergence angle. 
As already mentioned, it is no less astonishing than the widely noticed fact 
that divergence angles converge on one of the special irrational numbers.  
Let us take the specimen No.~13 as an example. 
The divergence angle of the $\frac{47}{154}$ phyllotaxis 
is a rational number $360\alpha\simeq 109.870^\circ$. 
This is very close to 
an irrational, ideal angle of $\alpha_0=[3,3]$, or $360\alpha_0\simeq 109.877^\circ$. 
According to Table~\ref{tab:dickson}, 
the range of $\alpha_0$ 
for the $\frac{47}{154}$ phyllotaxis is very narrow, that is, 

\begin{equation}
109.831^\circ < 360\alpha_0 < 109.895^\circ, 
\label{inequalities33}
\end{equation}
or $360 \alpha_0\simeq 109.863\pm 0.032$ degrees.  
The relative precision is less than about a part per three thousand. 
For reference, 
we present results that would be obtained 
if 
$\alpha_0$ happens to be off the narrow range of
(\ref{inequalities33}). 
Instead of $\frac{47}{154}$ and the parastichy
pair $(59,95)$ for (\ref{inequalities33}), 
we would have obtained $\frac{43}{141}$ and $(59,82)$ 
if $\alpha_0$ were slightly below the lower limit of (\ref{inequalities33}),  
or $\alpha=\frac{40}{131}$ and $(36,95)$
if $\alpha_0$ were above the upper limit of (\ref{inequalities33}).  
Neither of the last two cases is listed in Tables \ref{tab:2}$\sim$\ref{tab:2112}, 
for they are hardly ever likely to occur. 
The plants' ability to distinguish  $\frac{47}{154}$ from $\frac{43}{141}$ and $\frac{40}{131}$ 
is due to high precision regulation of initial divergence angle. 
%
The range width of 
$\alpha_0$ depends not so much on $\alpha_0$ as on $n_c$.  
Indeed, we find $\Delta \alpha_0\simeq (\tau/n_c)^2$ 
according to Eq.~(B.39) in \cite{okabe11}. 
The precision as high as the above cannot be attained by a limited
number of cells on the apex (\cite{kbr98, mkb98, smith06}).
It seems very unlikely that existing causal models can explain 
this anomalous phyllotaxis with this precision 
in this probability of one out of thirteen specimens. 



Diversities of phyllotaxis is considered as a result of 
selective pressures being ineffective.  
In extant plants, the main Fibonacci phyllotaxis is dominant 
while some species specifically show very diverse phyllotaxis (\cite{zm94}). 
In general, the trait diversity will be reduced 
if there is selective pressure acting on it. 
Strength of selective pressure 
depends specifically on extra cost required while 
rearranging phyllotactic patterns of leaf traces during growth of individual plants.
Accordingly, the diversity may be preserved 
for some reason or other, e.g., when 
leaf traces are so fragile that the energy cost 
of rearrangement is insignificant. 
This view is consistent with recent research on {\it Licopodium} 
revealing a link between variability of leaf traces and diversity of phyllotaxis (\cite{gjz07}). 
In contrast, 
the diversity in phyllotaxis of scale trees is considered as a result of
strong selective pressure of insufficient time durations, 
strong because divergence angles are highly accurate 
whereas insufficient because various angles besides 137.5$^\circ$ 
are still in existence. 
In discussing diversity of phyllotaxis, 
one should make a clear distinction between  
the variance, or standard deviation, of divergence angle of an individual 
and varieties of divergence angles of individuals. 
This section was devoted to the latter, while 
the former was discussed in the last section. 

\section{Phyllotaxis and vascular organization}
\label{sec:girolami}

\begin{figure}
\begin{center}
\includegraphics[width=0.65 \textwidth,angle=-90]{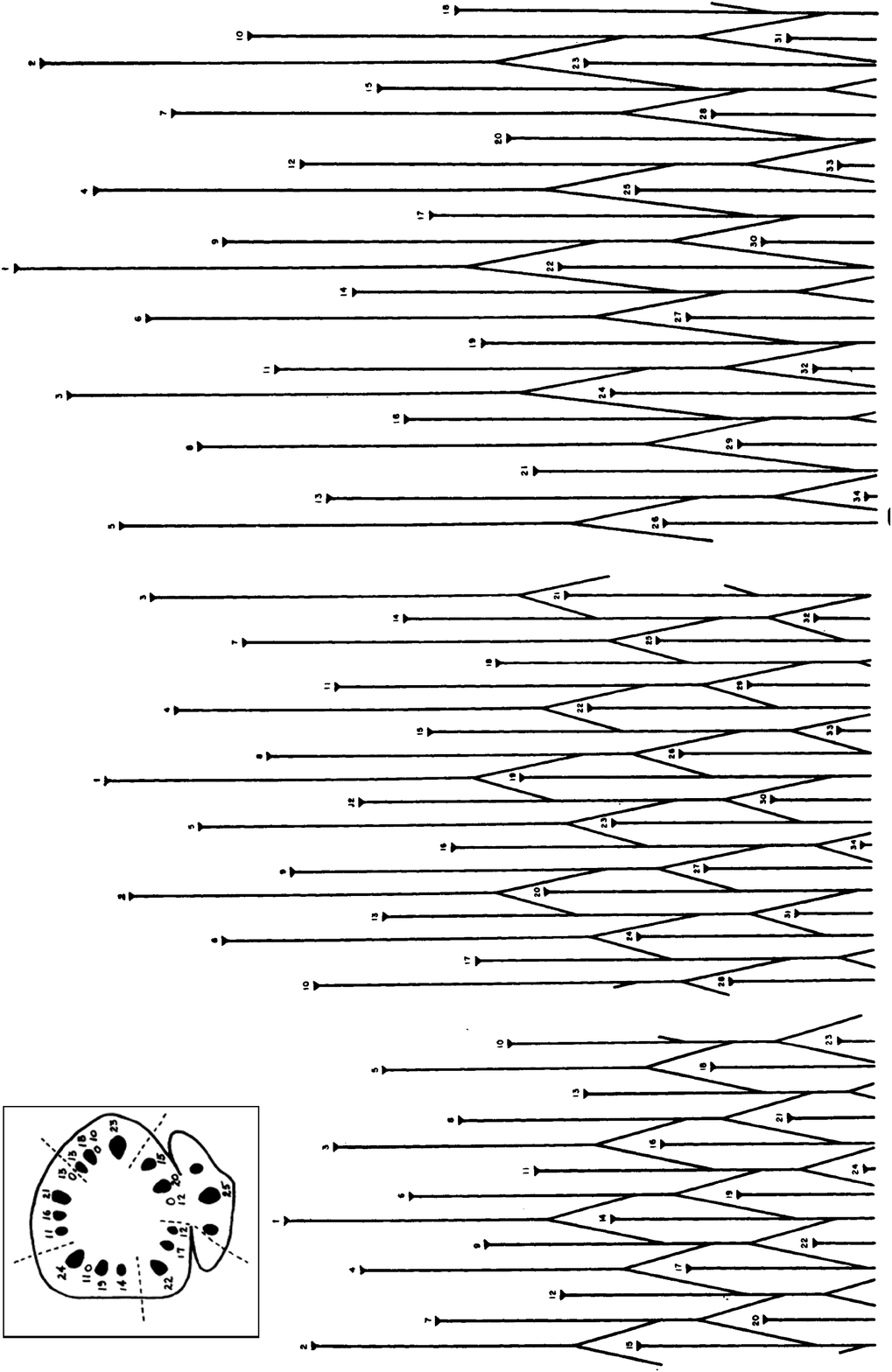}
\caption{
Diagrams of the primary vascular system of $\frac{5}{13}$ (left), $\frac{5}{18}$ (center),
$\frac{8}{21}$ (right) phyllotaxis of 
{\it Linum usitatissimum}. 
In the inset (top left), 
dashed lines mark off
five parastichy sectors in a transverse section of the $\frac{5}{13}$
 phyllotaxis stem.  
Adapted from \cite{girolami53}. 
}
\label{fig:girolami}
\end{center}
\end{figure}
\cite{girolami53} investigated 
the relation between phyllotaxis and vascular organization of  {\it Linum} (flax),  
whose vascular structures of a $\frac{5}{13}$, $\frac{5}{18}$ and $\frac{8}{21}$ phyllotaxis 
are given in the left, center and right of Fig.~\ref{fig:girolami}, respectively.
On the one hand, 
the genetic spirals 
of the $\frac{5}{13}$ and $\frac{5}{18}$ phyllotaxis 
wind up to the right (counterclockwise), 
while it goes to the left (clockwise) for the $\frac{8}{21}$ phyllotaxis.  
On the other hand,  the main parastichies of the three patterns run in
the same direction. 
That is to say,  
5-parastichies for $\frac{5}{13}$ (1-6-11-16-21, etc.), 
7-parastichies for $\frac{5}{18}$ (1-8-15-22-29, etc.) 
and  8-parastichies for $\frac{8}{21}$ (1-8-15-22-29, etc.) 
run steeply from the bottom right to top left (clockwise). 
The most direct vascular connection goes along the main parastichies.  
The vascular bundles of these parastichies
are recognized as sectioned clusters in a transverse section of the
stem,  called parastichy sectors. 
As shown in the inset of Fig.~\ref{fig:girolami}, 
 the $\frac{5}{13}$ phyllotaxis stem 
is divided into five parastichy sectors. 
In what follows, 
the following points remarked by
\cite{girolami53} are 
analyzed in terms of the model, whereby 
some useful general rules are  pointed out:

(G1)
The number of the parastichy sectors 
(5, 7 and 8 for $\frac{5}{13}$, $\frac{5}{18}$ and $\frac{8}{21}$, respectively)
agrees with the numerator of the phyllotactic fraction for 
$\frac{5}{13}$ and $\frac{8}{21}$ of the main sequence, 
but not 
for $\frac{5}{18}$ of the accessory sequence. 

(G2)
The length of leaf traces per internode 
increases with the number of parastichy sectors, namely 
12 for $\frac{5}{13}$, 17 for $\frac{5}{18}$, and 19 for $\frac{8}{21}$ 
approximately. 


(G3) 
There is no 
correlation in 
the relative directions of the genetic spiral
and the parastichies.

As noted in the first point, there is 
no easy-to-use general formula 
between the parastichy numbers and 
the phyllotactic fraction (see below however), 
but the numerical correspondence is immediately read 
from Fig.~\ref{SBtree} and Tables \ref{tab:2}$\sim$\ref{tab:2112}. 
According to Table~\ref{tab:3}, the parastichy pair of the $\frac{5}{18}$ phyllotaxis is $(7,11)$. 
The number of parastichy sectors is the small number of the parastichy pair. 
Therefore, 
the number 7 of the parastichy sectors of the $\frac{5}{18}$ phyllotaxis
is obtained.   
%
%
Unlike the numerator, the denominator satisfies simple rules. 
Most notably,  the denominator of a fraction is equal to the sum of the
contact parastichy pair corresponding to the fraction (e.g. $18=7+11$). 
Mathematical relations between various numbers in phyllotaxis 
have been investigated since early times on an empirical ground 
based on 
purely mathematical properties of a regular lattice (\cite{bravais1837,naumann45, jean94}). 




On the second point, 
Tables \ref{tab:2} and \ref{tab:3} give 
the conditions $8\le n_c<13$, $11\le n_c<18$ and $13\le n_c<21$ 
for the phyllotactic fractions $\frac{5}{13}$, $\frac{5}{18}$ and
$\frac{8}{21}$, respectively. 
The predictions of the model are supported by 
the reported values $n_c=12, 17$ and 19, 
which satisfy their respective conditions near their upper limits. 
Nevertheless, 
a close look at Fig.~\ref{fig:girolami} indicates that these figures are not accurate.
As a matter of fact, $n_c$ appears not constant but somewhat larger in the upper part of the stem. 
Changes in length of the leaf traces are revealed 
in a more sophisticated analysis of \cite{meicenheimer86}, where progressive transitions 
from $\frac{1}{3}$ through $\frac{2}{5}$ and $\frac{3}{8}$ up to
$\frac{5}{13}$ have been reported. 
Phyllotactic transition caused by changes in $n_c$ is discussed in the next section.

\begin{figure}[t]
  \begin{center}
\includegraphics[width= .6\textwidth]
{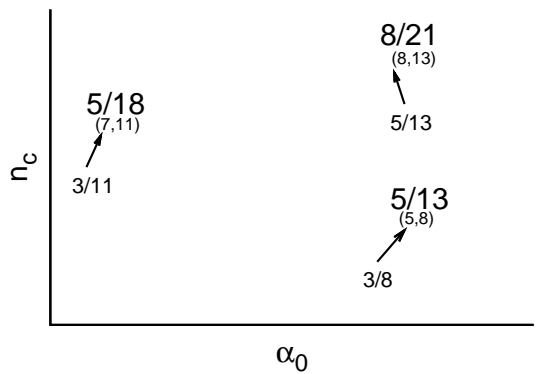}  
    \label{girofractions}
  \caption{
Three branches from Fig.~\ref{SBtree} 
with which to explain spiral directions of 
$\frac{5}{13}$, $\frac{5}{18}$, $\frac{8}{21}$ phyllotaxis. 
As the fraction $\frac{5}{18}$ and $\frac{5}{13}$ are numerically bigger than 
their `mother' fraction $\frac{3}{11}$ and $\frac{3}{8}$, 
their main parastichies of 7 and 5
are contrary in direction to 
the genetic spiral. 
On the contrary, 
8 parastichies for $\frac{8}{21}$ are in the same direction as  the genetic spiral. 
}
  \end{center}
\end{figure}
\begin{figure}[t]
  \begin{center}
\includegraphics[width= .6\textwidth]{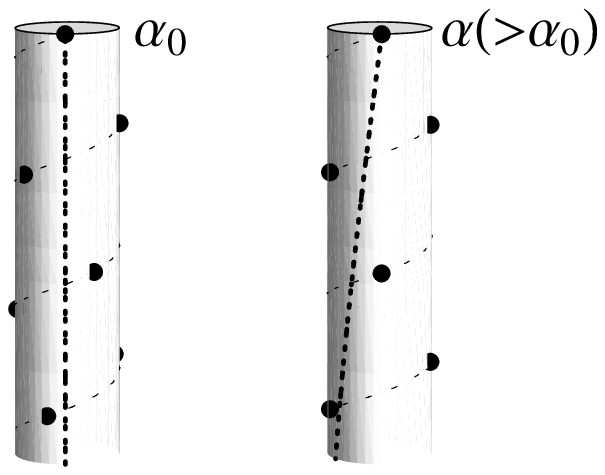}  
    \label{torsions}
  \caption{
When the final divergence $\alpha$ is numerically bigger than the initial divergence
   $\alpha_0$,  the stem is twisted in the direction of the genetic
   spiral.  N.B. Divergence angles are less than 180 degrees.
}
  \end{center}
\end{figure}

On the third point, 
a general rule holding between directions of
parastichies and the genetic spiral 
is presented based on Fig.~\ref{SBtree}. 
%
To this end, it is convenient to introduce a `mother' fraction of a fraction $\alpha$, 
which is defined as the fraction lying immediately below the fraction 
$\alpha$ in the tree of Fig.~\ref{SBtree}. 
The mother fractions of $\frac{5}{13}$, $\frac{5}{18}$ and $\frac{8}{21}$
are $\frac{3}{8}$, $\frac{3}{11}$ and $\frac{5}{13}$, respectively. 
%
It is shown that
{\it if and only if a phyllotactic fraction $\alpha$ is numerically bigger than its mother fraction, 
the main parastichies run in the direction opposite to the genetic spiral.}
(The main parastichies are gentle, long spirals 
 characterized by the small number of the contact parastichy pair.) 
The fraction $\alpha=\frac{5}{18}$ and $\frac{5}{13}$ are 
bigger than the mother fraction $\frac{3}{8}$ and $\frac{3}{11}$, respectively, 
while $\alpha=\frac{8}{21}$ is smaller than the mother fraction $\frac{5}{13}$. 
%
The magnitude relations are schematically shown in 
Fig.~\ref{girofractions} extracted from Fig.~\ref{SBtree}. 
Thus, the above rule explains Girolami's observation consistently.  
In practice, this rule may be used  
to identify the direction of the genetic spiral of a high order phyllotactic pattern for
which parastichies are far easy to follow visually. 
Some special cases of this general rule 
have been remarked
(\cite{church04}(p.~96), \cite{nb68}) 
and occasionally taken up for discussion (\cite{meicenheimer86,fhm02}). 
The directional relations between various spirals of a phyllotactic pattern 
are also mathematical consequences of the regularity of the phyllotactic pattern.



The mother fraction enables us to state general rules for 
the phyllotactic fraction and the parastichy number:  
One of the parastichy pair for a fraction $\alpha$
is equal to the denominator of the mother fraction of $\alpha$; 
The other number in the pair is determined such that the sum of
the pair is equal to the denominator of $\alpha$. 
Consider  $\alpha=\frac{5}{18}$, for instance. 
One of its parastichy pair is the denominator 11 of 
the mother fraction $\frac{3}{11}$,
while the other is the difference of the denominators, $18-11=7$. 
As a result, 
the parastichy pair $(7,11)$ is obtained for $\frac{5}{18}$.
Thus, the rules are used to relate the parastichy numbers and the
phyllotactic fraction.

The vascular systems shown in 
 Fig.~\ref{fig:girolami} 
form closed networks. 
In each system, connections between leaf traces are formed along
both the paired parastichies, 
so that the vascular bundles are divided into parastichy sectors. 
Among dicotyledons with helical phyllotaxis, however,  an open vascular
system is 
rather common (\cite{bsr82}). 
Primitive angiosperms and many gymnosperms have open vascular systems (\cite{beck10}). 
According to  \cite{bsr82}, open systems of five sympodia (a $\frac{2}{5}$ phyllotaxis) characterize
67\% of the species with helical phyllotaxy and are clearly a common type among dicotyledons.  
In an open system, leaf traces are connected along one direction. 
Although the present model determines the basic architecture of vascular phyllotaxis, 
it does not specify detailed structure of the reticulate pattern,  
whether it remains open or becomes closed. 
This is not a shortcoming of the model,  
because actual linkages between leaf traces are likely to be secondary events 
depending on circumstances (\cite{ktdd03}).

To conclude this section, let us remark another obvious correlation 
between the direction of the genetic spiral and the secondary torsion of
the stem.   
The initial divergence $\alpha_0$ is related to the fractional
divergence $\alpha$ of a mature pattern 
by 
the angle of twist $\alpha-\alpha_0$ undergone in the secondary torsion. 
The direction of the torsion is the same as the genetic spiral 
if and only if $\alpha>\alpha_0$. 
This is shown schematically in Fig.~\ref{torsions}. 
The direction of the secondary torsion 
would not be difficult to check experimentally.  
In most typical cases, the direction is reversed, or the sign of $\alpha-\alpha_0$ changes, 
as $n_c$ crosses a threshold of phyllotactic transition.  
\cite{bravais1837} evaluated the limit divergence $\alpha_0$ 
from mature shoots 
by correcting 
 the torsion angle $\alpha-\alpha_0$.

%
%
%
%
%
%
%
%




\section{Phyllotactic transition}
\label{sec:larson}

\begin{figure}
\begin{center}
\includegraphics[width=.85 \textwidth,angle=-90]{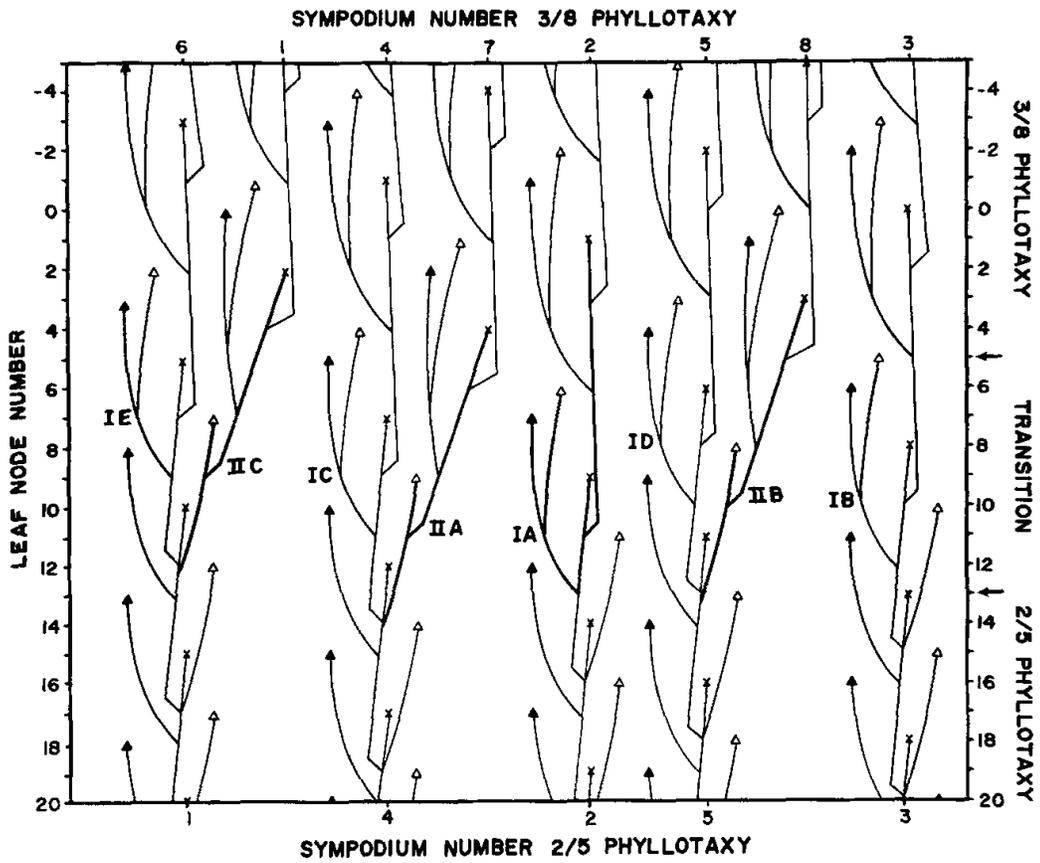}
\caption{
Transition in the primary vascular system of a cottonwood plant
from a $\frac{2}{5}$ to $\frac{3}{8}$ phyllotaxis. 
Central, right and left traces are indicated with crosses, filled and open triangles, respectively. 
After \cite{larson77}.
}
\label{fig:larson}
\end{center}
\end{figure}

\cite{larson77} has investigated phyllotactic transition 
in the vascular system of {\it Populus} (cottonwood).  
His result showing transition from a $\frac{2}{5}$ to $\frac{3}{8}$
phyllotaxis is reproduced in Fig.~\ref{fig:larson}. 
Each leaf has three traces; 
central, right and left traces are indicated with crosses, filled and
open triangles, respectively. 
The leaf traces are connected with the stem vascular bundles to make sympodia. 
The sympodia  are separated from each other, or the vascular system is open.  
The three traces leading to each leaf primordium arise on different sympodia. 
The number of the sympodia changes from five in the lower portion to eight
in the upper portion of Fig.~\ref{fig:larson}. 
The number agrees with the denominator of the phyllotactic fraction in each part. 
The region of the $\frac{2}{5}$ phyllotaxis occurs in the basal stem above some primary leaves, 
while  the $\frac{3}{8}$ phyllotaxis occurs at mid and upper stem levels, 
principally in the zone of expanding leaves (\cite{larson77}). 
In Fig.~\ref{fig:larson}, 
once the transition is initiated at a point IA on a sympodium number 2, 
it progresses through the sympodia at points IB through IE. 
Three new central traces to establish the three additional
sympodia of the $\frac{3}{8}$ system are derived from left traces in
sequence at points IIA-IIC. 
Various interrelations between phyllotaxis and leaf development have been studied (\cite{larson80}). 
In what follows, 
a correlation between phyllotactic transition and lengths of the leaf traces
is analyzed by means of the model, 
whereby supporting evidence of the model is pointed out. 



\begin{figure}[t]
\begin{center}
\includegraphics[width=.85\textwidth]{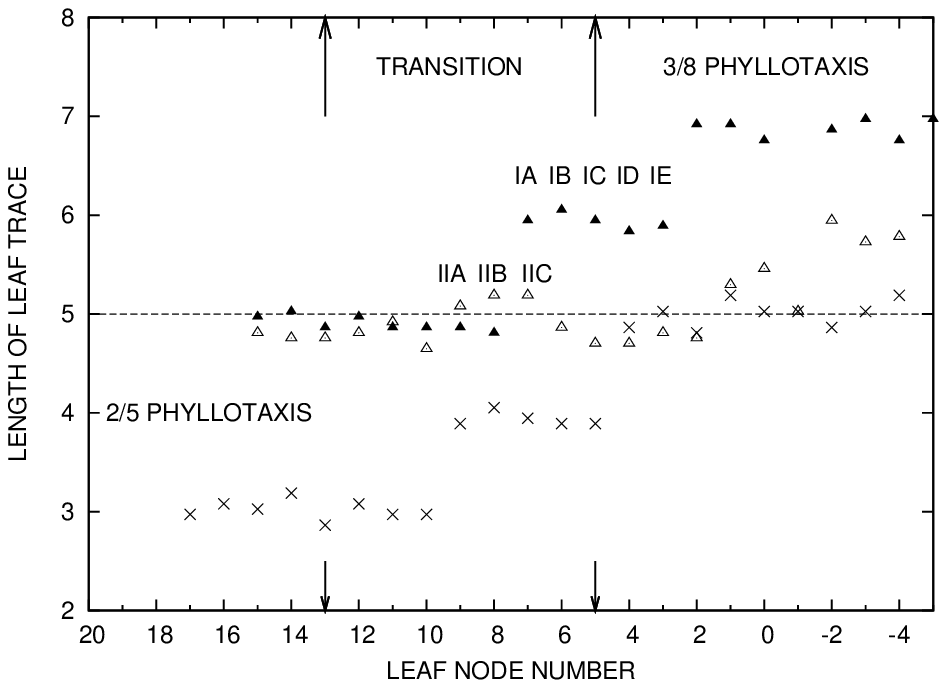}
\caption{
Length per internode of the leaf traces in 
Fig.~\ref{fig:larson} is
plotted against the leaf node number, the vertical axis of Fig.~\ref{fig:larson}. 
Arrows for a transition region between the $\frac{2}{5}$ and
 $\frac{3}{8}$ phyllotaxis and 
labels IA-IE and IIA-IIC to indicate initiation of the transition
are marked in accordance with Fig.~\ref{fig:larson}  by Larson. 
The phyllotactic transition is consistent with the threshold value of
 $n_c=5$ predicted by the model (Table \ref{tab:2}). 
}
\label{fig:larsonnc}
\end{center}
\end{figure}
The lengths per internode of 
the leaf traces are optically read from Fig.~\ref{fig:larson}
and plotted in Fig.~\ref{fig:larsonnc}. 
%
%
Arrows indicating  the transition region between 
the $\frac{2}{5}$ and
$\frac{3}{8}$ phyllotaxis in Fig.~\ref{fig:larsonnc} 
are marked 
in accordance with Fig.~\ref{fig:larson} after \cite{larson77}. 
By comparison, a dashed line at $n_c=5$ is drawn 
to indicate the theoretical threshold between the $\frac{2}{5}$ and $\frac{3}{8}$ phyllotaxis (Table \ref{tab:2}). 
In accordance with the model, 
the phyllotactic transition is triggered by the increasing length of the
leaf traces crossing a threshold value of five internodes. 



\begin{figure}[t]
  \begin{center}
\includegraphics[width= .57\textwidth]{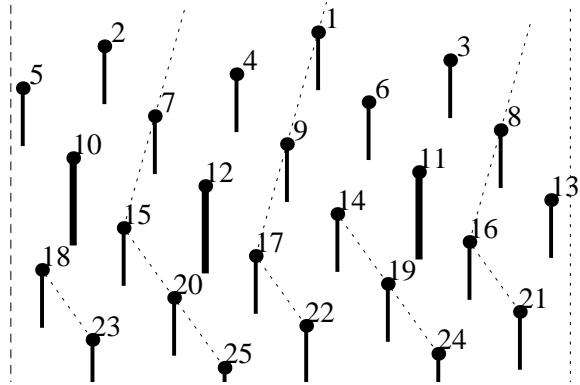}  
  \end{center}
  \caption{
A phyllotactic pattern with $\alpha_0=1/(1+\tau)$ (cf. Fig.~\ref{leftfig1}). 
Length of leaf traces is $n_c=4$ (solid bars) except for 10, 11 and 12 with $n_c=6$ (bold bars).  
The longer traces can induce a transition 
from $\alpha=\frac{2}{5}$ 
in the lower portion (cf. Fig.~\ref{rightfig1}) 
to $\alpha=\frac{3}{8}$ 
in the upper portion (cf. Fig.~\ref{rightfig}). 
At the transition, the longer traces deflect main parastichies (dotted lines), 
and the parastichy number increases from 5
to 8.
}
  \label{al2nc46hindrance}
\end{figure}

According to Table \ref{tab:2}, phyllotactic transition is predictable.  
Transitions of the main sequence 
 occur whenever the trace length $n_c$ crosses Fibonacci numbers. 
The trace length, like other parameters of the plant,  is
predictably correlated with plant vigor (\cite{larson80}). 
Therefore, in principle, the model allows us to control phyllotaxis artificially. 
%
In Sec.~\ref{sec:model}, leaf traces are assumed to have a common length.
As noted at the end of the last section, 
the direction of the secondary torsion
is reversed when $n_c$ crosses a threshold value,  
%
so that 
it may be fixed by a leaf trace of length longer than the threshold. 
Fig.~\ref{al2nc46hindrance} schematically shows that 
long leaf traces 10, 11 and 12 
trigger a transition  from $\frac{2}{5}$ to $\frac{3}{8}$. 
In the transition region of Fig.~\ref{fig:larsonnc}, 
three left traces (open triangles) of the node number 7, 8 and 9 
are the first to  cross the threshold at $n_c=5$. 
These are the very traces labeled with IIA, IIB and IIC by Larson 
as those from which the three extra sympodia branch. 
A close look at Fig.~\ref{fig:larson} reveals that 
central traces below and above the transition region 
are inclined 
in the opposite direction. 
This is consistent with the prediction of the model, for  
$\alpha=\frac{3}{8}<\alpha_0 < \frac{2}{5}$. 
Furthermore,  five right traces (filled triangles) 
striking around $n_c\simeq 6$ in Fig.~\ref{fig:larsonnc} 
agree with the special traces labeled with IA through IE.
Thus, the observation supports 
the special role of the Fibonacci number 5 for the trace length $n_c$.

\begin{figure}
\begin{center}
\includegraphics[width=.7 \textwidth,angle=-90]{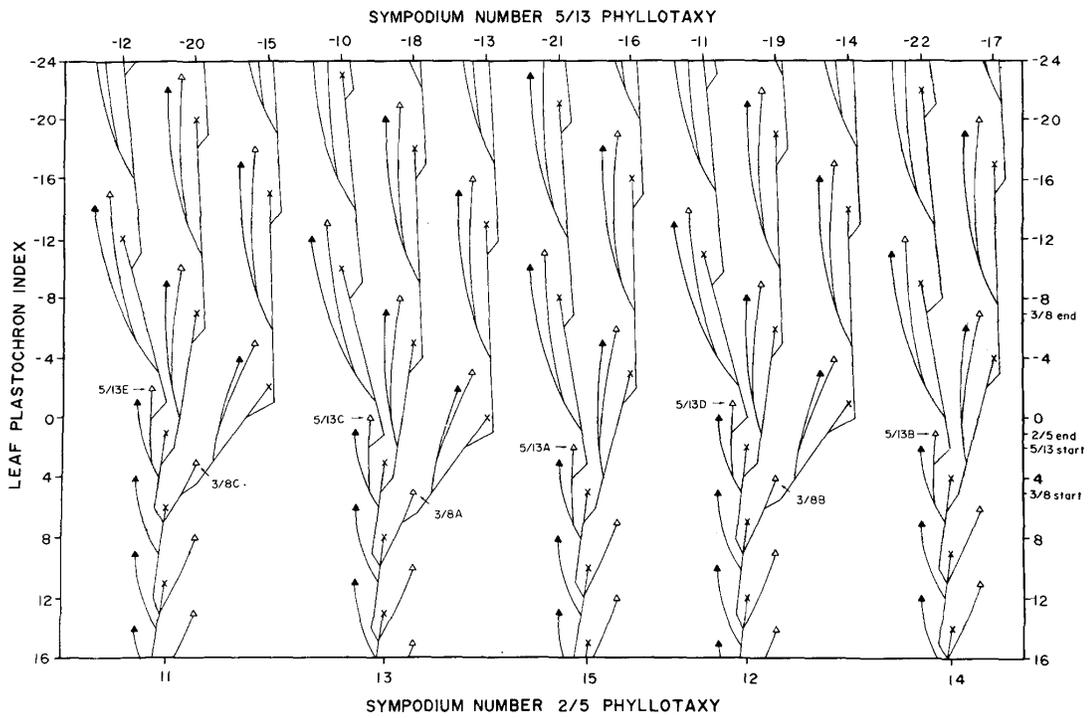}
\caption{
Reconstructed vascular system of a cottonwood plant showing 
transition from $\frac{2}{5}$ through $\frac{3}{8}$ to
 $\frac{5}{13}$ phyllotaxis by \cite{larson77}. 
See Fig.~\ref{fig:larson} for symbols.  
}
\label{fig:larsonfig4}
\end{center}
\end{figure}

\begin{figure}[t]
\begin{center}
\includegraphics[width=.9\textwidth]{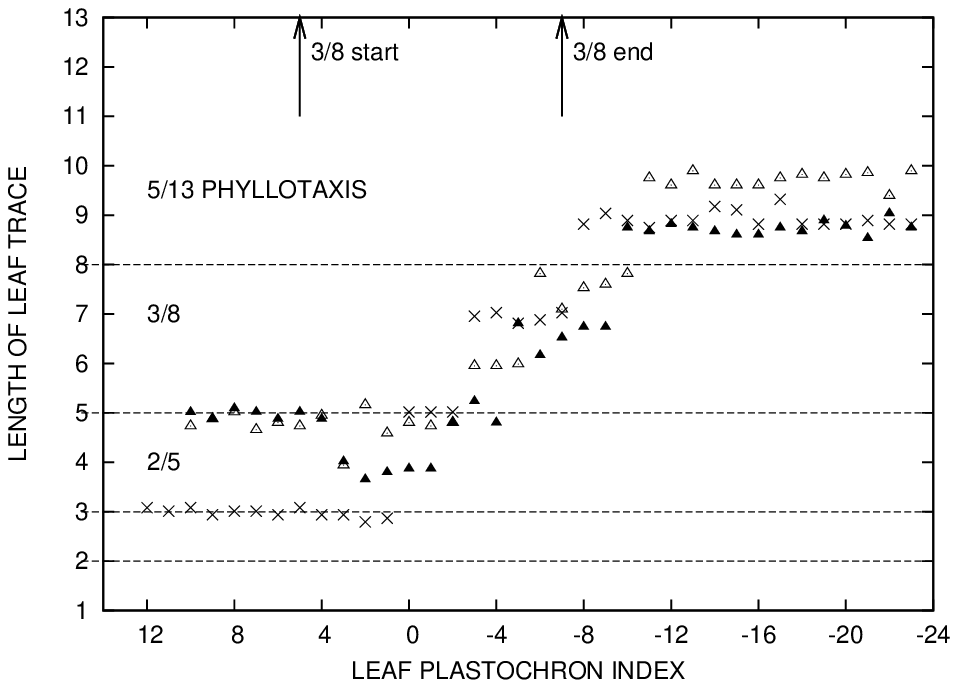}
\caption{
Length per internode of the leaf traces 
read from Fig.~\ref{fig:larsonfig4} is 
plotted against the leaf index (the vertical axis of Fig.~\ref{fig:larsonfig4}).  
Two arrows at the top indicate where the $\frac{3}{8}$ phyllotaxis starts and
 ends according to \cite{larson77} (see Fig.~\ref{fig:larsonfig4}). 
According to the theoretical model, 
stable regions for the $\frac{2}{5}$, $\frac{3}{8}$ and $\frac{5}{13}$ phyllotaxis 
are 
separated by horizontal dashed lines at Fibonacci numbers 3, 5 and 8 (cf. Table~\ref{tab:2}). 
Thus, Larson's estimate of the region of the $\frac{3}{8}$ phyllotaxis agrees with the theory. 
Left traces (open triangles) reaching 
a maximum length of  about 10
internodes is consistent with an observation that 
the highest-order phyllotactic fraction that this plant attains is $\frac{5}{13}$. 
}
\label{fig:larsonfig4nc}
\end{center}
\end{figure}
Two-step transition from a $\frac{2}{5}$ 
to $\frac{5}{13}$ phyllotaxis is shown in Fig.~\ref{fig:larsonfig4} after \cite{larson77}, 
where steady increase in length of leaf traces is more obvious than Fig.~\ref{fig:larson}. 
Fig.~\ref{fig:larsonfig4nc} is obtained from Fig.~\ref{fig:larsonfig4} 
in the same manner as Fig.~\ref{fig:larsonnc} is obtained 
from Fig.~\ref{fig:larson}. 
Leaf positions at which the $\frac{3}{8}$ phyllotaxis starts and ends
are marked on the right side of Fig.~\ref{fig:larsonfig4} by \cite{larson77}, 
according to which the transient pattern of the $\frac{3}{8}$
phyllotaxis is maintained 
for the leaves with plastochron index from 5 to $-7$. 
Accordingly, the corresponding positions are marked 
by arrows in Fig.~\ref{fig:larsonfig4nc}. 
On the other hand, 
horizontal lines at Fibonacci numbers 3, 5, and 8 in
Fig.~\ref{fig:larsonfig4nc} theoretically 
divide the regions for the $\frac{1}{3}$,
$\frac{2}{5}$, $\frac{3}{8}$ and $\frac{5}{13}$ phyllotaxis
(Table~\ref{tab:2}). 
Thus, 
it is confirmed again that continuous changes in length of leaf traces cause
discontinuous transitions in the vascular structure.  

Fig.~\ref{fig:larsonfig4nc} indicates that $n_c$ 
increases steadily up to an upper bound of about 10. 
This observation is consistent with the fact that 
the $\frac{5}{13}$ phyllotaxis was the stable pattern of the old plant (\cite{larson80}). 
According to the model, 
the $\frac{5}{13}$ phyllotaxis is stable 
insofar as $n_c$ lies between 8 and 13, i.e., 
there is a 5-internode allowance for the trace length of the $\frac{5}{13}$ phyllotaxis. 
The main sequence is special for this wide clearance
between successive threshold values. 
The interval is denoted as $\Delta n_c$ in \cite{okabe11}. 
As shown there, 
the widest clearances are achieved for Fibonacci numbers, 
and a sequence of Fibonacci numbers is realized when the limit
divergence angle is one of the special irrational numbers related to the golden ratio. 
As shown in Fig.~\ref{ntfraction}, 
the number of transitions encountered while $n_c$ grows up to above 10 
is kept to a minimum number 
insofar as  the initial divergence is restricted within 
$\frac{3}{8}<\alpha_0<\frac{3}{7}$  
(from 135$^\circ$ to 154$^\circ$, as noted at the end of
Sec.~\ref{sec:model}). 
When $n_c$ becomes larger than 12, 
the range 
is narrowed to $\frac{3}{8}<\alpha_0<\frac{2}{5}$
(from 135$^\circ$ to 144$^\circ$). 
Thus, the normal phyllotaxis of the main sequence is singled out. 
Owing to the observation that the highest-order fraction was $\alpha=\frac{5}{13}$, 
the model predicts 
that the initial divergence $\alpha_0$
should be contained within $\frac{3}{8}<\alpha_0 <\frac{2}{5}$,  
just as observed by \cite{fujita39} for other species (Sec.~\ref{sec:model}). 
Unfortunately,  
initial divergences of the cottonwood plant are not
available to us.  
To support this argument, 
\cite{pulawska65} has reported 
for {\it Actinidia arguta} (hardy kiwi) 
that initial divergence remains constant despite changes 
in the vascular organization between $\frac{3}{8}, \frac{5}{13}$
and $\frac{8}{21}$. 

When $n_c$ is increased past 8, the model predicts vascular phyllotaxis of 
either $\alpha=\frac{5}{13}$ or $\alpha=\frac{5}{12}$ 
depending on whether the initial divergence $\alpha_0$
is smaller or larger than $\frac{2}{5}$ (angle of 144$^\circ$).  
%
%
Suppose $\alpha_0=[3]$ (99.5$^\circ$),  
then one should have five threshold lines at 2, 3, 4, 7 and 11 (Table
\ref{tab:3}), 
instead of four thresholds at 2, 3, 5 and 8 for $\alpha_0=[2]$ in
Fig.~\ref{fig:larsonfig4nc}. 
If the initial divergence were $\alpha_0=[5]$ (64.1$^\circ$ in Table~\ref{tab:5}), 
one should have six threshold lines at $n_c=2$, 3, 4, 5, 6 and 11 separating patterns of 
$\alpha=\frac{1}{2}, \frac{1}{3}, \frac{1}{4}, \frac{1}{5},
\frac{1}{6},\frac{2}{11}$ and $\frac{3}{17}$
(cf. Fig.~\ref{ntfraction}). 
The vascular phyllotaxis is very unstable. 
The instability is energetically unfavorable.
Therefore, $\alpha_0=[5]$ (64.1$^\circ$) is very improbable to survive natural
selection because of the multiplicity of expected transitions. 
%
%
%
A general remark should be made when discussing multiple patterns in sequence. 
In order for a pattern with a definite value of $\alpha$ to be distinguished as such, 
the pattern should consist of more leaves than the denominator of the fraction $\alpha$. 
This holds true if $n_c$ varies sufficiently gradually; 
otherwise phyllotaxis transition may not be distinctly discernible.

%


Last but not least, whorled phyllotaxis has not been discussed in this paper. 
A $J$-jugate pattern with $J$ fundamental spirals is formed when 
$J$ leaves are borne at each node. 
%
Compared with a helical phyllotaxis with $J=1$, 
divergence angles of a $J$-jugate system are divided by $J$
and the parastichy pairs $(m,n)$ are multiplied by $J$.  
Therefore, one obtains $0<J\alpha_0<\frac{1}{2}$ and $J(m,n)$ 
for the divergence angle and parastichy pair of a $J$-jugate system.    
It is known that 
sometimes vascular structure may change 
between helical and whorled phyllotaxis during ontogeny.  
This type of `anomalous' phyllotactic transition 
also appears to be caused by a decrease in length of leaf traces 
(\cite{jensen68,bsr82,kwiatkowska95}). 
The present model gives 
$\alpha=\frac{1}{2}$ for $1\le n_c <2$ and
$\alpha=\frac{1}{3}$ for $2\le n_c <3$ irrespective of $\alpha_0$. 
Correspondingly, it seems natural to consider that 
a whorled phyllotaxis is a variation of the most primitive alternate phyllotaxis 
and that a whorled phyllotaxis is triggered as $n_c$ becomes less than 1. 
%
%
However, changes in the vascular structure 
have to be coordinated with changes in the positioning of initiated leaf
primordia while a whorled pattern is established (\cite{zm94,meicenheimer98, kc03}).  
The physiological processes involved are unlikely to be amenable to
simple mathematical analysis. 
Still, 
a similar transition rule as a helical pattern 
should hold for an established whorled pattern 
in terms of  trace length redefined with a new internode.

%

%

%
%



\section{Conclusions}


The present work puts forward an important role of 
Fibonacci numbers as critical values of the length per internode of leaf traces 
played in vascular phyllotaxis transition.  


The regular arrangement of leaves and 
the regularity in divergence angle of 137.5$^\circ$
are a result of selective pressure to reduce possible changes in the
vascular structure during growth,  
i.e., aperiodic arrangements will necessitate extra nutrients to
reconstruct the sectorial or fractional order of vascular connections.


The phyllotactic fraction $\alpha$ of mature patterns of leaf traces normally 
makes transitions through 
$\frac{1}{2}$, $\frac{1}{3}$, $\frac{2}{5}$, $\frac{3}{8}$,
 $\frac{5}{13}$,  $\frac{8}{21}$, $\cdots$, 
whenever the number of internodes traversed by the leaf
traces, $n_c$,  crosses Fibonacci numbers, 1, 2, 3, 5, 8, 13, 21, $\cdots$. 
The Fibonacci numbers make appearances because 
initial divergence angle $\alpha_0$ of leaves 
at the shoot apex is normally 
the golden angle of 
about $137.5^\circ$ with a good precision. 
The golden angle is prevalent because it is the selectively advantageous angle 
at which the number of the phyllotactic transition 
is the minimum (Fig.~\ref{ntfraction}). 
The precision of the initial  divergence is determined by the trace
length $n_c$. 


%
%

\section*{Acknowledgement} 
The author would like to thank Prof. Rolf Rutishauser for valuable
comments on {\it Picea abies} and others. 
He would like to thank Prof. Beata Zag\'orska-Marek 
for informing him about a different view on divergence angle. 

%
%

\appendix
\section{Relation between the trace length $n_c$ and the plastochron ratio $a$}
\label{appendix}

A point on a cylinder surface is located 
with 
 the angular coordinate $\varphi$ and the height $z$. 
Leaves on a stem are represented by 
a lattice of points given by $\varphi=2\pi \alpha n$ (in radians)
and $z=h n$, 
where $\alpha$ is a constant angle of divergence, 
$h$ an internode length,  and $n$ is an integer index. 
On the other hand, 
a point on a plane is located in a polar coordinate system $(r,
\varphi)$, 
where $r$ and $\varphi$ 
is the radial distance from the central axis 
and the angular coordinate about the axis, respectively. 
Leaf primordia at a shoot apex are represented by 
$r=a^n$ and $\varphi=2\pi \alpha n$, 
where $a$ is a plastochron ratio. 
In a conformal growth preserving angles, 
 the two representations are related by $ {2\pi} z= {\log r}$. 
Hence, the internode length $h$ corresponds to 
the logarithm of the plastochron ratio, $\frac{1}{2\pi}\log a$. 
The number of internodes traversed by the leaf traces is 
$n_c={Z_{\rm lt}}/{h}= 2\pi{Z_{\rm lt}}/{\log a}$, 
where $Z_{\rm lt}$ is a length of leaf traces in the stem.  
Therefore, 
$n_c$ may be regarded as inversely proportional to $\log a$,  
or 
 the relative growth rate per plastochron 
$\frac{{\rm d}r}{{\rm d}n}/ r$. 
The growth rate should depend on cell types. 
Accordingly, 
$n_c$ may change during plant growth.

The plastochron ratio may change as a result of alteration in size of
the apex and primordia. 
\cite{richards51} discussed changing phyllotaxis to the effect that 
a continuous shift in the parastichy pair of normal Fibonacci phyllotaxis
is linearly correlated with a double logarithm $\log (\log a)$. 
He defined the phyllotaxis index (P.I.) by 
\begin{equation}
 {\rm P.I.}= 0.38 - 2.39 \log_{10} \log_{10} a, 
\label{P.I.}
\end{equation}
where numerical values are chosen such that 
the index assumes an integral value whenever two sets of parastichies in
the Fibonacci system intersect orthogonally. 
The crossing angle between the contact parastichies changes continuously 
as a function of the plastochron ratio.
In this descriptive model, 
the divergence angle $\alpha_0$ is 
fixed at the golden angle. 
%
%
%
%

%
%

Changing phyllotaxis due to change in the plastochron ratio is
consistent with the present model of vascular phyllotaxis.  
In this model, 
the divergence $\alpha$ on the stem changes discontinuously, however. 
To show a correspondence between 
changes in phyllotaxis on the apex and the stem,  
let us consider 
the normal phyllotaxis with an initial divergence of the golden angle $\alpha_0=\tau^{-2}$ (Table~\ref{tab:2}). 
Let us introduce 
the Fibonacci sequence $F_n$ 
generated from initial integers $F_1=1$ and $F_2=1$ 
by the recurrence relation $F_{n+2}=F_{n+1}+F_{n}$.  
Accordingly,  $F_n=1,1,2,3,5,8$ and $13$ for $n=1,2,3,4,5,6$ and $7$, respectively. 
In terms of $F_n$, 
the phyllotactic fraction $\alpha=\frac{F_n}{F_{n+2}}$
and the parastichy pair $(F_n, F_{n+1})$ are obtained for $F_{n+1}\le n_c< F_{n+2}$, 
or for 
\[
(n+1) \log \tau- \log \sqrt{5} 
\le \log n_c< 
(n+2) \log \tau- \log \sqrt{5} 
\]
owing to an approximate formula $F_n\simeq \tau^n/\sqrt{5}$ valid for
large $n$ (see below (B.31) in \cite{okabe11}). 
Therefore, 
 $\log n_c$ is proportional to
the integer index $n$. 
%
%

To put it concretely,  we get 
$\alpha=\frac{1}{3}$ and the parastichy pair 
(1,2) 
for $2\le n_c <3$ or
\[
0.7 \le \log n_c< 1.1, 
\] 
$\alpha=\frac{2}{5}$ and $(2,3)$ 
for $3\le n_c <5$ or
\[
1.1 \le \log n_c< 1.6, 
\] 
$\alpha=\frac{3}{8}$ and $(3,5)$ 
for $5\le n_c <8$ or
\[
1.6 \le \log n_c< 2.1, 
\] 
$\alpha=\frac{5}{13}$ and $(5,8)$ 
for $8\le n_c <13$ or
\[
2.1 \le \log n_c< 2.6, 
\] 
and so on. 
Thus, the shift in the parastichy pair 
is linearly correlated with $\log n_c \propto \log (\log a)$.  
This is a general property holding also for other initial divergences found in nature. 

For the systematic study 
of the mature stem, 
the index $n_c$ is more usefully regarded 
as a developmental index than $a$,  
not only because an internode is a natural unit of length
as the plastochron is the developmental unit of time, 
but values of $n_c$ allowed for a phyllotactic pattern are delimited by 
the special integers traditionally familiar to those who are 
enchanted by phyllotaxis;  Fibonacci numbers. 
For a given initial divergence, 
the numbers comprise a sequence generated 
by the Fibonacci recurrence relation $F_{n+2}=F_{n+1}+F_{n}$ 
from a pair of different seed integers. 
The main sequence,  1, 2, 3, 5, 8, $\cdots$ in Table \ref{tab:2},  
is generated from the simplest seed pair $(1, 2)$.
The next simplest seed integers $(1, 3)$ give the accessory sequence
1, 3, 4, 7, 11, 18 $\cdots$  of Table \ref{tab:3}. 
In this manner, any phyllotactic sequence is characterized 
by a pair of seed integers,  as well as the limit divergence $\alpha_0$. 
This is in accordance with accumulated empirical wisdom of phyllotaxis.  
Traditionally,  these special integers have been remarked 
in connection with parastichy numbers
(cf. Tables~\ref{tab:2}$\sim$\ref{tab:2112}). 
The present work puts emphasis on these numbers as critical values for
the length per internode of leaf traces. 
This point has never been remarked before.

\bibliography{okabe}

\begin{thebibliography}{126}
\expandafter\ifx\csname natexlab\endcsname\relax\def\natexlab#1{#1}\fi
\expandafter\ifx\csname url\endcsname\relax
  \def\url#1{\texttt{#1}}\fi
\expandafter\ifx\csname urlprefix\endcsname\relax\def\urlprefix{URL }\fi

\bibitem[{Adler(1974)}]{adler74}
Adler, I., 1974. A model of contact pressure in phyllotaxis. Journal of
  Theoretical Biology 45, 1--79.

\bibitem[{Airy(1873)}]{airy1873}
Airy, H., 1873. On leaf-arrangement. Proc. Royal Soc. London 21, 176--179.

\bibitem[{Allard(1942)}]{allard42}
Allard, H., 1942. Some aspects of the phyllotaxy of tobacco. Journal
  Agricultural Research 64, 49--55.

\bibitem[{{Atela} et~al.(2002){Atela}, {Gol{\'e}}, and {Hotton}}]{agh02}
{Atela}, G., {Gol{\'e}}, J.~A., {Hotton}, J.~P., 2002. {A Dynamical System for
  Plant Pattern Formation: A Rigorous Analysis}. Journal of NonLinear Science
  12, 641--676.

\bibitem[{Barab{\'e} et~al.(2010)Barab{\'e}, Bourque, Yin, and
  Lacroix}]{bbyl10}
Barab{\'e}, D., Bourque, L., Yin, X., Lacroix, C., 2010. Phyllotaxis of the
  palm \textit{{E}uterpe oleracea} {M}art. at the level of the shoot apical
  meristem. Botany 88~(5), 528--536.

\bibitem[{Beck(2010)}]{beck10}
Beck, C.~B., 2010. An Introduction to Plant Structure and Development.
  Cambridge University Press.

\bibitem[{Beck et~al.(1982)Beck, Schmid, and Rothwell}]{bsr82}
Beck, C.~B., Schmid, R., Rothwell, G.~W., 1982. Stelar morphology and the
  primary vascular system of seed plants. Botanical Review 48, 691--815.

\bibitem[{Braun(1831)}]{braun31}
Braun, A., 1831. Vergleichende {U}ntersuchung \"uber die {O}rdnung der
  {S}chuppen an den {T}annenzapfen als {E}inleitung zur {U}ntersuchung der
  {B}lattstellung. Verhandlungen der Kaiserlichen Leopoldinisch-Carolinischen
  Akademie der Naturforscher 15, 195--402.

\bibitem[{Braun(1835)}]{braun1835}
Braun, A., 1835. Dr. {C}arl {S}chimper's {V}ortr{\"a}ge {\"u}ber die
  {M}{\"o}glichkeit eines wissenschaftlichen {V}erst{\"a}ndnisses der
  {B}lattstellung, nebst {A}ndeutung der haupts{\"a}chlichen
  {B}lattstellungsgesetze und insbesondere der neuentdeckten {G}esetze der
  {A}neinanderreihung von {C}yclen verschiedene maasse. Flora 18, 145--191.

\bibitem[{Bravais and Bravais(1837)}]{bravais1837}
Bravais, L., Bravais, R., 1837. Essai sur la disposition des feuilles
  curvis\'eri\'ees. Annales des Sciences Naturelles Botanique 7, 42--110.

\bibitem[{Bryntsev(2004)}]{bryntsev04}
Bryntsev, V.~A., 2004. Types of phyllotaxis and patterns of their realization.
  Russ. J. Dev. Biol. 2, 114--156.

\bibitem[{Chapman and Perry(1987)}]{cp87}
Chapman, J.~M., Perry, R., 1987. A diffusion model of phyllotaxis. Annals of
  Botany 60~(4), 377--389.

\bibitem[{Church(1904)}]{church04}
Church, A.~H., 1904. On the Relation of Phyllotaxis to Mechanical Laws. On the
  Relation of Phyllotaxis to Mechanical Laws. Williams \& Norgate, London.

\bibitem[{Church(1920)}]{church20}
Church, A.~H., 1920. On the interpretation of phenomena of phyllotaxis.
  Botanical memoirs. Hafner Pub. Co.

\bibitem[{Coxeter(1972)}]{coxeter72}
Coxeter, H. S.~M., 1972. The role of intermediate convergents in {T}ait's
  explanation for phyllotaxis. Journal of Algebra 20, 167--175.

\bibitem[{Cummings and Strickland(1998)}]{cs98}
Cummings, F., Strickland, J., 1998. A model of phyllotaxis. Journal of
  Theoretical Biology 192~(4), 531--544.

\bibitem[{Davies(1939)}]{davies39}
Davies, P.~A., 1939. Leaf position in {A}ilanthus altissima in relation to the
  {F}ibonacci series. American Journal of Botany 26, 67--74.

\bibitem[{de~Candolle(1881)}]{candolle81}
de~Candolle, C., 1881. Consid\'erations sur l'\'etude de la phyllotaxie.
  Geneva: H. Georg.

\bibitem[{Delpino(1883)}]{delpino1883teoria}
Delpino, F., 1883. Teoria generale della fillotassi. Atti della R. Universita
  di Genova. Armanino.

\bibitem[{Dickson(1871)}]{dickson71}
Dickson, A., 1871. On the phyllotaxis of \textit{{L}epidodendron} and the
  allied, if not identical, genus \textit{{K}norria}. Journal of botany,
  British and foreign 9, 166--167.

\bibitem[{Dormer(1972)}]{dormer72}
Dormer, K., 1972. Shoot organization in vascular plants. Shoot Organization in
  Vascular Plants. Syracuse University Press.

\bibitem[{Douady and Couder(1996)}]{dc96a}
Douady, S., Couder, Y., 1996. Phyllotaxis as a dynamical self organizing
  process part {I}: The spiral modes resulting from time-periodic iterations.
  Journal of Theoretical Biology 178, 255--274.

\bibitem[{Erickson(1983)}]{erickson83}
Erickson, R.~O., 1983. The geometry of phyllotaxis. In: Dale, J., Milthorpe, F.
  (Eds.), The Growth and functioning of leaves: proceedings of a symposium held
  prior to the thirteenth International Botanical Congress at the University of
  Sydney, 18-20 August 1981. Cambridge University Press, pp. 53--88.

\bibitem[{Erickson and Michelini(1957)}]{em57}
Erickson, R.~O., Michelini, F.~J., 1957. The plastochron index. American
  Journal of Botany 44, 297--305.

\bibitem[{Esau(1965)}]{esau65}
Esau, K., 1965. Vascular differentiation in plants. New York: Holt, Rinehart
  and Winston.

\bibitem[{Fredeen et~al.(2002)Fredeen, Horning, and Madill}]{fhm02}
Fredeen, A.~L., Horning, J.~A., Madill, R.~W., 2002. Spiral phyllotaxis of
  needle fascicles on branches and scales on cones in pinus contorta var.
  latifolia: Are they influenced by wood-grain spiral? Canadian Journal of
  Botany 80~(2), 166--175.

\bibitem[{Fujita(1937)}]{fujita37}
Fujita, T., 1937. {\"U}ber die {R}eihe 2,5,7,12.... in der schraubigen
  {B}lattstellung und die mathematische {B}etrachtung verschiedener
  {Z}ahlenreihensysteme. Bot.\ Mag.\ Tokyo 51, 298--307.

\bibitem[{Fujita(1938)}]{fujita38}
Fujita, T., 1938. Statistische {U}ntersuchung {\"u}ber die {Z}ahl der
  konjugierten {P}arastichen bei den schraubigen {O}rganstellungen. Bot.\ Mag.\
  Tokyo 52, 425--433.

\bibitem[{Fujita(1939)}]{fujita39}
Fujita, T., 1939. Statistische {U}ntersuchungern {\"u}ber den {D}ivergenzwinkel
  bei den schraubigen {O}rganstellungen. Bot.\ Mag.\ Tokyo 53, 194--199.

\bibitem[{Girolami(1953)}]{girolami53}
Girolami, G., 1953. Relation between phyllotaxis and primary vascular
  organization in linum. American Journal of Botany 40, 618--625.

\bibitem[{Gola et~al.(2007)Gola, Jernstedt, and Zag\'orska-Marek}]{gjz07}
Gola, E.~M., Jernstedt, J.~A., Zag\'orska-Marek, B., 2007. Vascular
  architecture in shoots of early divergent vascular plants,
  \textit{{L}ycopodium clavatum} and \textit{{L}ycopodium annotinum}. New
  Phytologist 174~(4), 774--786.

\bibitem[{Green et~al.(1996)Green, Steele, and Rennich}]{gss96}
Green, P.~B., Steele, C.~S., Rennich, S.~C., 1996. Phyllotactic {P}atterns: {A}
  {B}iophysical {M}echanism for their {O}rigin. Annals of Botany 77, 515--527.

\bibitem[{Hellwig et~al.(2006)Hellwig, Engelmann, and Deussen}]{hed06}
Hellwig, H., Engelmann, R., Deussen, O., 2006. Contact pressure models for
  spiral phyllotaxis and their computer simulation. Journal of Theoretical
  Biology 240~(3), 489--500.

\bibitem[{Hirmer(1922)}]{hirmer22}
Hirmer, M., 1922. Zur {L}{\"o}sung des {P}roblems der {B}lattstellungen. G.
  Fischer.

\bibitem[{Hirmer(1931)}]{hirmer31}
Hirmer, M., 1931. Zur {K}enntnis der {S}chraubenstellungen im {P}flanzenreich.
  Planta 14, 132--206.

\bibitem[{Hofmeister(1868)}]{hofmeister68}
Hofmeister, W., 1868. Allgemeine {M}orphologie der {G}ew{\"a}chse. In: de~Bary,
  A., Irmisch, T.~H., Sachs, J. (Eds.), Handbuch der {P}hysiologischen
  {B}otanik. Leipzig: W. Engelmann, pp. 405--664.

\bibitem[{Hotton et~al.(2006)Hotton, Johnson, Wilbarger, Zwieniecki, Atela,
  Gol\'e, and Dumais}]{hjwzagd06}
Hotton, S., Johnson, V., Wilbarger, J., Zwieniecki, K., Atela, P., Gol\'e, C.,
  Dumais, J., 2006. The possible and the actual in phyllotaxis: Bridging the
  gap between empirical observations and iterative models. Journal of Plant
  Growth Regulation 25, 313--323.

\bibitem[{Jean(1986)}]{jean84}
Jean, R., 1986. An interpretation of {F}ujita's frequency diagrams in
  phyllotaxis. Bulletin of Mathematical Biology 48, 77--86.

\bibitem[{Jean(1994)}]{jean94}
Jean, R.~V., 1994. Phyllotaxis: A Systemic Study in Plant Morphogenesis.
  Cambridge Univ. Press, Cambridge, New York.

\bibitem[{Jensen(1968)}]{jensen68}
Jensen, L. C.~W., 1968. Primary stem vascular patterns in three subfamilies of
  the {C}rassulaceae. American Journal of Botany 55, 553--563.

\bibitem[{J\"onsson et~al.(2006)J\"onsson, Heisler, Shapiro, Meyerowitz, and
  Mjolsness}]{jhsmm06}
J\"onsson, H., Heisler, M.~G., Shapiro, B.~E., Meyerowitz, E.~M., Mjolsness,
  E., 2006. An auxin-driven polarized transport model for phyllotaxis.
  Proceedings of the National Academy of Sciences of the United States of
  America 103~(5), 1633--1638.

\bibitem[{Kang et~al.(2003)Kang, Tang, Donnelly, and Dengler}]{ktdd03}
Kang, J., Tang, J., Donnelly, P., Dengler, N., 2003. Primary vascular pattern
  and expression of {ATHB-8} in shoots of arabidopsis. New Phytologist 158~(3),
  443--454.

\bibitem[{Kelly and Cooke(2003)}]{kc03}
Kelly, W.~J., Cooke, T.~J., 2003. Geometrical relationships specifying the
  phyllotactic pattern of aquatic plants. American Journal of Botany 90~(8),
  1131--1143.

\bibitem[{King et~al.(2004)King, Beck, and L\"uttge}]{kbl04}
King, S., Beck, F., L\"uttge, U., 2004. On the mystery of the golden angle in
  phyllotaxis. Plant, Cell \& Environment 27~(6), 685--695.

\bibitem[{Kirchoff(1984)}]{kirchoff84}
Kirchoff, B.~K., 1984. On the relationship between phyllotaxy and vasculature:
  a synthesis. Botanical Journal of the Linnean Society 89, 37--51.

\bibitem[{Kirchoff(2003)}]{kirchoff03}
Kirchoff, B.~K., 2003. {Shape Matters: Hofmeister's Rule, Primordium Shape, and
  Flower Orientation}. International Journal of Plant Sciences 164~(4),
  505--517.

\bibitem[{Koch et~al.(1998)Koch, Bernasconi, and Rothen}]{kbr98}
Koch, A.-J., Bernasconi, G., Rothen, F., 1998. Phyllotaxis as a geometrical and
  dynamical system. In: Jean, R.~V., Barab{\'e}, D. (Eds.), Symmetry in plants.
  World Scientific, pp. 459--486.

\bibitem[{Koch and Meinhardt(1994)}]{km94}
Koch, A.~J., Meinhardt, H., 1994. {Biological pattern formation: from basic
  mechanisms to complex structures}. Review of Modern Physics 66, 1481--1507.

\bibitem[{Kuhlemeier(2007)}]{kuhlemier07}
Kuhlemeier, C., 2007. Phyllotaxis. TRENDS in Plant Science 12, 143--150.

\bibitem[{Kumazawa and Kumazawa(1971)}]{kumazawa71}
Kumazawa, M., Kumazawa, M., 1971. Periodic variations of the divergence angle,
  internode length and leaf shape, revealed by correlogram analysis.
  Phytomorphology 21, 376--389.

\bibitem[{Kunz(2001)}]{kunz01}
Kunz, M., September 2001. Dynamical models of phyllotaxis. Phys. D 157,
  147--165.

\bibitem[{Kwiatkowska(1995)}]{kwiatkowska95}
Kwiatkowska, D., 1995. Ontogenetic changes in the shoot primary vasculature of
  \textit{Anagallis arvensis} {L.} Acta Societatis Botanicorum Poloniae 64,
  213--222.

\bibitem[{Larson(1977)}]{larson77}
Larson, P.~R., 1977. Phyllotactic transitions in the vascular system of
  \textit{{P}opulus} \textit{deltoides} {B}artr. as determined by $^{14}${C}
  labeling. Planta 134, 241--249.

\bibitem[{Larson(1980)}]{larson80}
Larson, P.~R., 1980. Interrelations between phyllotaxis, leaf development and
  the primary-secondary vascular transition in \textit{{P}opulus deltoides}.
  Annals of Botany 46, 757--769.

\bibitem[{Larson(1983)}]{larson83}
Larson, P.~R., 1983. Primary vascularization and the siting of primordia. In:
  Dale, J.~E., Milthorpe, F.~L. (Eds.), The growth and functioning of leaves.
  Cambridge, UK: Cambridge University Press, pp. 25--51.

\bibitem[{Leigh(1972)}]{leigh72}
Leigh, E.~G., 1972. The golden section and spiral leaf-arrangement.
  Transactions of the Connecticut Academy of Arts and Sciences 44, 163--176.

\bibitem[{Lestiboudois(1848)}]{lestiboudois}
Lestiboudois, M.~T., 1848. Phyllotaxie anatomique. Annales des Sciences
  Naturelles 3, 15--105, 136--189.

\bibitem[{Levitov(1991)}]{levitov91b}
Levitov, L.~S., 1991. Energetic approach to phyllotaxis. Europhys.\ Lett. 14,
  533--539.

\bibitem[{Lyndon(1990)}]{lyndon90}
Lyndon, R., 1990. Plant development: the cellular basis. Topics in plant
  physiology. Unwin Hyman.

\bibitem[{Maksymowych and Erickson(1977)}]{me77}
Maksymowych, R., Erickson, R.~O., 1977. Phyllotactic change induced by
  gibberellic acid in \textit{{X}anthium} shoot apices. American Journal of
  Botany 64, 33--44.

\bibitem[{Malygin(2006)}]{malygin06}
Malygin, A.~G., 2006. Morphodynamics of phyllotaxis. Int. J. Dev. Biol. 50,
  277--287.

\bibitem[{Marc and Hackett(1991)}]{mh91}
Marc, J., Hackett, W.~P., 1991. Gibberellin-induced reorganization of spatial
  relationships of emerging leaf primordia at the shoot apical meristem in
  \textit{Hedera helix} {L.} Planta 185, 171--178.

\bibitem[{Marzec and Kappraff(1983)}]{mk83}
Marzec, C., Kappraff, J., 1983. Properties of maximal spacing on a circle
  related to phyllotaxis and to the golden mean. Journal of Theoretical Biology
  103, 201--226.

\bibitem[{Meicenheimer(1986)}]{meicenheimer86}
Meicenheimer, R.~D., 1986. Role of parenchyma in {L}inum usitatissimum leaf
  trace patterns. American Journal of Botany 73, 1649--1664.

\bibitem[{Meicenheimer(1998)}]{meicenheimer98}
Meicenheimer, R.~D., 1998. Decussate to spiral transitions in phyllotaxis. In:
  Jean, R.~V., Barab{\'e}, D. (Eds.), Symmetry in plants. World Scientific, pp.
  125--144.

\bibitem[{Meicenheimer(2006)}]{meicenheimer06}
Meicenheimer, R.~D., 2006. Stem unit growth analysis of \textit{Linum
  usitatissimum} ({L}inaceae) internode development. American Journal of Botany
  93~(1), 55--63.

\bibitem[{Meinhardt et~al.(1998)Meinhardt, Koch, and Bernasconi}]{mkb98}
Meinhardt, H., Koch, A.-J., Bernasconi, G., 1998. Models of pattern formation
  applied to plant development. In: Jean, R.~V., Barab{\'e}, D. (Eds.),
  Symmetry in plants. World Scientific, pp. 723--758.

\bibitem[{Mitchison(1977)}]{mitchison77}
Mitchison, G.~H., 1977. {P}hyllotaxis and the {F}ibonacci series. Science 196,
  270--275.

\bibitem[{N\"ageli(1858)}]{naegeli58}
N\"ageli, C.~W., 1858. Das {W}achsthum des {S}tammes und der {W}urzel bei den
  {G}ef\"asspflanzen und die anordnung der {G}ef\"asstr\"ange im {S}tengel.
  Beitrage Zur Wissenschaftlichen Botanik 1, 1--156.

\bibitem[{Namboodiri and Beck(1968)}]{nb68}
Namboodiri, K.~K., Beck, C.~B., 1968. A comparative study of the primary
  vascular system of conifers. {I.} genera with helical phyllotaxis. American
  Journal of Botany 55, 447--457.

\bibitem[{Naumann(1845)}]{naumann45}
Naumann, C., 1845. Ueber den {Q}uincunx als {G}rundgesetz der {B}lattstellung
  vieler {P}flanzen. Arnold.

\bibitem[{Nelson and Dengler(1997)}]{nd97}
Nelson, T., Dengler, N., 1997. Leaf vascular pattern formation. The Plant Cell
  Online 9~(7), 1121--1135.

\bibitem[{Newell et~al.(2008)Newell, Shipman, and Sun}]{nss08}
Newell, A.~C., Shipman, P.~D., Sun, Z., 2008. Phyllotaxis: cooperation and
  competition between mechanical and biochemical processes. Journal of
  Theoretical Biology 251~(3), 421--439.

\bibitem[{Niklas(1988)}]{niklas88}
Niklas, K.~J., 1988. The role of phyllotatic pattern as a "developmental
  constraint" on the interception of light by leaf surfaces. Evolution 42,
  1--16.

\bibitem[{Niklas(1997)}]{niklas97}
Niklas, K.~J., 1997. The evolutionary biology of plants. University of Chicago
  Press.

\bibitem[{Niklas(1998)}]{niklas98}
Niklas, K.~J., 1998. Light harvesting "fitness landscapes" for vertical shoots
  with different phyllotactic patterns. In: Jean, R.~V., Barab{\'e}, D. (Eds.),
  Symmetry in plants. World Scientific, pp. 759--773.

\bibitem[{Okabe(2011)}]{okabe11}
Okabe, T., 2011. Physical phenomenology of phyllotaxis. Journal of Theoretical
  Biology 280, 63--75.

\bibitem[{Pearcy and Yang(1998)}]{py98}
Pearcy, R.~W., Yang, W., 1998. The functional morphology of light capture and
  carbon gain in the redwood forest understorey plant \textit{{A}denocaulon
  bicolor} hook. Functional Ecology 12~(4), 543--552.

\bibitem[{Priestley and Scott(1933)}]{ps33}
Priestley, J.~H., Scott, L.~I., 1933. Phyllotaxis in the dicotyledon from the
  standpoint of developmental anatomy. Biological Reviews 8~(3), 241--268.

\bibitem[{Priestley and Scott(1936)}]{ps36}
Priestley, J.~H., Scott, L.~I., 1936. The vascular anatomy of
  \textit{{H}elianthus annuus} \textit{L.} Proc. Leeds Phil. Soc. 3, 159--173.

\bibitem[{Prusinkiewicz and Lindenmayer(1991)}]{pl91}
Prusinkiewicz, P., Lindenmayer, A., 1991. The Algorithmic Beauty of Plants (The
  Virtual Laboratory). Springer.

\bibitem[{Pu{\l}awska(1965)}]{pulawska65}
Pu{\l}awska, Z., 1965. Correlations in the development of the leaves and leaf
  traces in the shoot of \textit{{A}ctinidia arguta} {P}lanch. Acta Societatis
  Botanicorum Poloniae 34, 697--712.

\bibitem[{Reinhardt(2005)}]{reinhardt05}
Reinhardt, D., 2005. Regulation of phyllotaxis. Int. J. Dev. Biol. 49,
  539--546.

\bibitem[{Richards(1948)}]{richards48}
Richards, F.~J., 1948. The geometry of phyllotaxis and its origin. Symp. Soc.
  Exp. Biol 2, 217--245.

\bibitem[{Richards(1951)}]{richards51}
Richards, F.~J., 1951. Phyllotaxis: Its quantitative expression and relation to
  growth in the apex. Philos. Trans. R. Soc. B 225, 509--564.

\bibitem[{Ridley(1982)}]{ridley82a}
Ridley, J.~N., 1982. Packing efficiency in sunflower heads. Math.\ Biosci. 58,
  129--139.

\bibitem[{Rivier et~al.(1984)Rivier, Occelli, Pantaloni, and
  Lissowski}]{ropl84}
Rivier, N., Occelli, R., Pantaloni, J., Lissowski, A., 1984. Structure of
  {B}{\'e}nard convection cells, phyllotaxis and crystallography in cylindrical
  symmetry. J.\ Phys. (Paris) 45, 49--63.

\bibitem[{Roberts(1984)}]{roberts84}
Roberts, D.~W., 1984. A chemical contact pressure model for phyllotaxis.
  Journal of Theoretical Biology 108, 481--490.

\bibitem[{Roberts(1987)}]{roberts87}
Roberts, D.~W., 1987. The chemical contact pressure model for phyllotaxis --
  application to phyllotaxis changes in seedlings and to anomalous phyllotaxis
  systems. Journal of Theoretical Biology 125, 141--161.

\bibitem[{Rothen and Koch(1989)}]{rk89b}
Rothen, F., Koch, A.~J., 1989. Phyllotaxis or the properties of spiral
  lattices. {II}. packing of circles along logarithmic spirals. J.\ Phys.
  (Paris) 50, 1603--1621.

\bibitem[{Rutishauser(1998)}]{rutishauser98}
Rutishauser, R., 1998. Plastochrone ratio and leaf arc as parameters of a
  quantitative phyllotaxis analysis in vascular plants. In: Jean, R.~V.,
  Barab{\'e}, D. (Eds.), Symmetry in plants. World Scientific, pp. 171--212.

\bibitem[{Schimper(1835)}]{schimper1835beschreibung}
Schimper, K.~F., 1835. {Beschreibung des Symphytum Zeyheri und seiner zwei
  deutschen verwandten der S. bulbosum Schimper und S. tuberosum Jacq}. Winter.

\bibitem[{Schoute(1913)}]{schoute13}
Schoute, J.~C., 1913. {B}eitr\"age zur {B}lattstellungslehre. Rec. Trav. Bot.
  N\'eerl 10, 153--324.

\bibitem[{Schwabe and Clewer(1984)}]{sc84}
Schwabe, W., Clewer, A., 1984. Phyllotaxis -- a simple computer model based on
  the theory of a polarly-translocated inhibitor. Journal of Theoretical
  Biology 109, 595--619.

\bibitem[{Schwendener(1878)}]{schwendener78}
Schwendener, S., 1878. {Mechanische Theorie der Blattstellungen}. Leipzig:
  Engelmann.

\bibitem[{Schwendener(1883)}]{schwendener83}
Schwendener, S., 1883. Zur {T}heorie der {B}lattstellungen. Sitzungsber. d.
  Berl. Akad. d. Wissensch XXXII, 741--773.

\bibitem[{Shipman et~al.(2011)Shipman, Sun, Pennybacker, and Newell}]{sspn11}
Shipman, P., Sun, Z., Pennybacker, M., Newell, A., 2011. How universal are
  {F}ibonacci patterns? The European Physical Journal D - Atomic, Molecular,
  Optical and Plasma Physics 62, 5--17.

\bibitem[{Skutch(1927)}]{skutch27}
Skutch, A.~F., 1927. Anatomy of leaf of banana, \textit{{M}usa sapientum} {L.}
  var. {H}ort. {G}ros {M}ichel. Botanical Gazette 84, 337--391.

\bibitem[{Smith et~al.(2006{\natexlab{a}})Smith, Guyomarc'h, Mandel, Reinhardt,
  Kuhlemeier, and Prusinkiewicz}]{smith06}
Smith, R.~S., Guyomarc'h, S., Mandel, T., Reinhardt, D., Kuhlemeier, C.,
  Prusinkiewicz, P., 2006{\natexlab{a}}. A plausible model of phyllotaxis.
  Proceedings of the National Academy of Sciences of the United States of
  America 103~(5), 1301--1306.

\bibitem[{Smith et~al.(2006{\natexlab{b}})Smith, Kuhlemeier, and
  Prusinkiewicz}]{smith06b}
Smith, R.~S., Kuhlemeier, C., Prusinkiewicz, P., 2006{\natexlab{b}}. Inhibition
  fields for phyllotactic pattern formation: a simulation study. Canadian
  Journal of Botany 84~(11), 1635--1649.

\bibitem[{Snow and Snow(1934)}]{snows34}
Snow, M., Snow, R., 1934. The interpretation of phyllotaxis. Biological Reviews
  9, 132--137.

\bibitem[{Snow and Snow(1962)}]{snow62}
Snow, M., Snow, R., 1962. A theory of the regulation of phyllotaxis based on
  \textit{{L}upinus albus}. Philos. Trans. Roy. Soc. London B 244, 483--513.

\bibitem[{Snow(1955)}]{snow55}
Snow, R., 1955. Problems of phyllotaxis and leaf determination. Endeavour 14,
  190--199.

\bibitem[{Steeves and Sussex(1989)}]{ss89}
Steeves, T., Sussex, I., 1989. Patterns in plant development. Cambridge
  University Press.

\bibitem[{Sterling(1945)}]{sterling45b}
Sterling, C., 1945. Growth and vascular development in the shoot apex of
  \textit{{S}equoia sempervirens} ({L}amb.) {E}ndl. {II.} vascular development
  in relation to phyllotaxis. American Journal of Botany 32, 380--386.

\bibitem[{Takenaka(1994)}]{takenaka94}
Takenaka, A., 1994. Effects of leaf blade narrowness and petiole length on the
  light capture efficiency of a shoot. Ecological Research 9, 109--114.

\bibitem[{Teitz(1888)}]{teitz88}
Teitz, P., 1888. Ueber definitive {F}ixirung der {B}lattstellung durch die
  {T}orsionswirkung der {L}eitstrange. Flora 71. Jahrgang, 419--439.

\bibitem[{Thomas(1975)}]{thomas75}
Thomas, R.~L., 1975. Orthostichy, parastichy and plastochrone ratio in a
  central theory of phyllotaxis. Annals of Botany 39~(3), 455--489.

\bibitem[{Thompson(1917)}]{dt17}
Thompson, D.~W., 1917. On Growth and Form. Oxford. Clarendon Press.

\bibitem[{Thornley(1975)}]{thornley75}
Thornley, J. H.~M., 1975. Phyllotaxis. {I}. {A} {M}echanistic {M}odel. Annals
  of Botany 39, 491--507.

\bibitem[{Unruh(1950)}]{unruh50}
Unruh, M., 1950. Neue {B}eobachtungen uber die {R}hythmik der {S}ymmetrie am
  zerstreut beblatterten {S}pross. Berichte der Deutschen Botanischen
  Gesellschaft 63, 88--96.

\bibitem[{Valladares and Brites(2004)}]{vb04}
Valladares, F., Brites, D., 2004. Leaf phyllotaxis: Does it really affect light
  capture? Plant Ecology 174, 11--17.

\bibitem[{van Iterson(1907)}]{vaniterson07}
van Iterson, G., 1907. {Mathematische und Mikroskopisch-Anatomische Studien
  \"uber Blattstellungen}. G. Fischer, Jena.

\bibitem[{Veen and Lindenmayer(1977)}]{veenlindenmayer77}
Veen, A.~H., Lindenmayer, A., 1977. {D}iffusion mechanism for phyllotaxy. Plant
  Physiol. 60, 127--139.

\bibitem[{Vogel(1979)}]{vogel79}
Vogel, H., 1979. A better way to construct the sunflower head. Mathematical
  Biosciences 44~(3-4), 179--189.

\bibitem[{Watson and Casper(1984)}]{wc84}
Watson, M.~A., Casper, B.~B., 1984. Morphogenetic constraints on patterns of
  carbon distribution in plants. Annual Review of Ecology and Systematics 15,
  233--258.

\bibitem[{Wiesner(1875)}]{wiesner75}
Wiesner, J., 1875. Bemerkungen \"uber rationale und irrationale {D}ivergenzen.
  Flora 58, 113--115, 139--143.

\bibitem[{Wiesner(1907)}]{wiesner07}
Wiesner, J., 1907. Der {L}ichtgenuss der {P}flanzen: photometrische und
  physiologische {U}ntersuchungen mit besonderer {R}\"ucksichtnahme auf
  {L}ebensweise, geographische {V}erbreitung ung {K}ultur der {P}flanzen. W.
  Engelmann.

\bibitem[{Williams(1974)}]{williams74}
Williams, R., 1974. The shoot apex and leaf growth: a study in quantitative
  biology. Cambridge University Press.

\bibitem[{Williams and Brittain(1984)}]{wb84}
Williams, R.~F., Brittain, E.~G., 1984. A geometrical model of phyllotaxis.
  Australian Journal of Botany 32, 43--72.

\bibitem[{Wright(1873)}]{wright1873}
Wright, C., 1873. The uses and origin of the arrangements of leaves in plants.
  Memoirs of the American Academy of Arts and Sciences 9~(2), 379--415.

\bibitem[{Yotsumoto(1993)}]{yotsumoto93}
Yotsumoto, A., 1993. A diffusion model for phyllotaxis. Journal of Theoretical
  Biology 162, 131--151.

\bibitem[{Young(1978)}]{young78}
Young, D.~A., 1978. On the diffusion theory of phyllotaxis. Journal of
  Theoretical Biology 71~(3), 421--432.

\bibitem[{Zag\'orska-Marek(1985)}]{zm85}
Zag\'orska-Marek, B., 1985. Phyllotactic patterns and transitions in
  \textit{Abies balsamea}. Canadian Journal of Botany 63~(10), 1844--1854.

\bibitem[{Zag\'orska-Marek(1994)}]{zm94}
Zag\'orska-Marek, B., 1994. Phyllotaxic diversity in \textit{{M}agnolia}
  flowers. Acta Societatis Botanicorum Poloniae 63, 117--137.

\bibitem[{Zag\'orska-Marek and Szpak(2008)}]{zms08}
Zag\'orska-Marek, B., Szpak, M., 2008. Virtual phyllotaxis and real plant model
  cases. Functional Plant Biology 35, 1025--1033.

\end{thebibliography}

\end{document}